%% file: Cinar_et_al.tex
\documentclass[twocolumn,trackchanges]{aastex7}

\usepackage{natbib}[1999/02/25]
\usepackage{rotating}
\usepackage{booktabs}

\usepackage{SIunits}
\usepackage{paralist}
\usepackage{amsmath}
\usepackage{psfrag,color}
\usepackage{multirow}
\usepackage{hyperref}
\usepackage{nameref}

\shorttitle{Origin of Four G-type Stars}
\shortauthors{\c{c}ınar et al.}
\graphicspath{{./}{figures/}}
\begin{document}
\title{Tracing the Galactic Origins of Selected Four G-type Stars in the Solar Neighbourhood}

\author[0000-0001-7940-3731]{Deniz Cennet \c{C}ınar}
\affiliation{Programme of Astronomy and Space Sciences, Institute of Graduate Studies in Science, Istanbul University,  34116, Istanbul, T\"{u}rkiye}
\email[show]{denizcdursun@gmail.com}  

\author[0000-0003-3510-1509]{Sel\c{c}uk Bilir}
\affiliation{Department of Astronomy and Space Sciences, Faculty of Science, Istanbul University, 34119, Istanbul, T\"{u}rkiye}
\email{sbilir@istanbul.edu.tr}

\author[0000-0002-0296-233X]{Timur \c Sahin}
\affiliation{Department of Space Sciences and Technologies, Faculty of Science, Akdeniz University, 07058, Antalya, T\"{u}rkiye}
\email{timursahin@akdeniz.edu.tr}

\author[0000-0002-0435-4493]{Olcay Plevne}
\affiliation{Department of Astronomy and Space Sciences, Faculty of Science, Istanbul University, 34119, Istanbul, T\"{u}rkiye} 
\email{olcayplevne@istanbul.edu.tr}

\begin{abstract}
\noindent
We present a multi-method investigation of four metal-poor G-type main-sequence stars to resolve their Galactic origins. By combining high-resolution spectroscopy from PolarBase, photometric/astrometric data from {\it Gaia} DR3, and spectral energy distribution (SED) modelling, we derive precise stellar parameters, chemical abundances, kinematics and Galactic orbital parameters. The stars HD 22879, HD 144579, HD 188510, and HD 201891 show effective temperatures of 5855 $\pm$ 110, 5300 $\pm$ 160, 5370 $\pm$ 60, and 5880 $\pm$ 90 K; surface gravities of 4.40 $\pm$ 0.18, 4.52 $\pm$ 0.37, 4.57 $\pm$ 0.13, and 4.48 $\pm$ 0.18 cgs; and metallicities of $-0.86 \pm 0.08$, $-0.55 \pm 0.12$, $-1.60 \pm 0.07$, and $-1.15 \pm 0.07$ dex, respectively. Kinematic analysis suggests that HD 22879, HD 144579, and HD 201891 are potential bulge-origin escapees, possibly ejected by the Galactic bar or spiral arm perturbations. HD 188510, however, shows halo-like dynamics, including a retrograde orbit. Chemical abundance trends ([$\alpha$/Fe] vs. [Fe/H]) reveal mixed origins, challenging kinematic classifications. This discrepancy highlights the importance of integrative methodologies in Galactic archaeology. We associate HD 22879, HD 144579, and HD 201891 with the bulge globular cluster NGC 6441, NGC 5927, and NGC 6544, respectively. HD 188510’s retrograde motion and low metallicity align with ejection from the halo globular cluster NGC 5139 ($\omega$ Cen). These results illustrate the complex interplay of dynamical processes—including bar resonances, spiral arm perturbations, and tidal stripping—in depositing metal-poor stars into the solar neighborhood.
\end{abstract}
	
	%\keywords{Galaxy: Solar Neighborhood, Abundances, Star: Metal Poor, Kinematics and Dynamics}

    \keywords{\uat{Solar neighborhood}{1509} --- \uat{Chemical abundances}{224} --- \uat{Stellar kinematics}{1608} --- \uat{Stellar dynamics}{1596}} 

    %\keywords{\uat{Galaxies}{573} --- \uat{Cosmology}{343} --- \uat{High Energy astrophysics}{739} --- \uat{Interstellar medium}{847} --- \uat{Stellar astronomy}{1583} --- \uat{Solar physics}{1476}}
	
	
\section{Introduction} \label{sec:intro}

Older stellar populations in the Milky Way retained traces of metal enrichment from the early universe. Analyzing stellar elemental abundances, ages, and kinematics offers insights into the sources of metal enrichment, notably Pop III stars. Studying the chemical composition of stars with varying metal levels is crucial for understanding Galactic evolution \citep{Reddy2003, Frebel2010, Yong2013, Hayden2015, Battistini2016}. Examining the chemical abundances of metal-poor stars in the atmosphere provides observational data that can be compared with the theoretical models of nucleosynthesis in the first metal-poor massive stars that ended as supernovae. Analyzing trends in element abundance ratios relative to metal content can help trace the chemical history of the Milky Way \citep{Beers2005, Frebel2015}.

The chemical compositions and kinematic characteristics of these stars can be compared with models of Galactic chemical evolution to elucidate the early chemical enrichment processes of galaxies. \citet{Chamberlain1951} disclosed the presence of metal-poor stars like HD\,19445 and HD\,140283, initiating the investigation. Today, highly metal-poor stars like SMSS 0313-6708 have also been identified. \citet{Beers2005} classified stars based on their metal content, from super metal-rich (SMR) stars to extremely metal-poor (EMP) stars. Stars with metal content exceeding 0.5 dex are termed super metal-rich (SMR), while those below -6 dex are classified as ultra metal-poor (UMP). Within the intermediate range, stars fall into categories such as metal-poor (MP, $\le$ -1 dex), very metal-poor (VMP, $\le$ -2 dex), extremely metal-poor (EMP, $\le$ -3 dex), ultra-metal-poor (UMP, $\le$ -4 dex), and hypermetal-poor (HMP, $\le$ -5 dex).

The evolution of the Milky Way can be further understood by studying metal-poor stars to provide insights into early star formation and galaxy mergers. \citet{Freeman2002} suggest that these stars reveal the formation history of our Galaxy. The chemical composition of metal-poor stars reflects the conditions of the early universe and the effects of initial supernovae. \citet{Bonaca2017} traced the past mergers of the Milky Way by examining the Galactic orbits of metal-poor stars, showing radial migrations from the center to distant regions. Additionally, \citet{Salvadori2009} emphasized the role of these stars in clarifying the chemical enrichment history of the Galaxy and the effects of early supernovae. \citet{Tumlinson2010} highlights the importance of metal-poor stars in the outer Milky Way for Galactic formation processes. Analyzing the chemical and dynamic characteristics of these stars is essential for reconstructing the early stages of Galactic evolution. \citet{Hirano2014} investigate the mass distribution of Pop III stars and the factors influencing their formation, revealing a wide mass spectrum ranging from 10 to 1,000 $M/M_{\odot}$.

The classification of stellar populations based on fundamental astrophysical parameters has transitioned from simple kinematic criteria to sophisticated multidimensional analysis. \citet{Eggen1962} introduced the first systematic framework by correlating stellar kinematics and metallicities, proposing a dichotomy between the halo and disk populations. This was later augmented by \citet{Gilmore1983}, who identified the thick disk as a separate Galactic component. Recent spectroscopic campaigns such as GALAH \citep{Buder2018} and APOGEE \citep{Majewski2017} have employed chemical tagging techniques, leveraging [X/Fe] abundance ratios to trace star formation sites. In recent years, datasets from \textit{Gaia} \citep{GaiaDR1, GaiaDR2, GaiaDR3} and complementary ground-based spectroscopic programs have revolutionized the field, offering unprecedented resolution to trace the hierarchical assembly of the Galaxy and the history of star formation.

Metal-poor stars serve as critical tracers of the chemical evolution of the Milky Way because their atmosphere preserves the nucleosynthetic signatures of their progenitor populations. Stars with intermediate iron abundances (\(-2.5 < \text{[Fe/H] (dex)} < -1\)) are likely formed from gas enriched by the ejecta of double-stability supernovae, retaining the chemical imprints of the first stellar generation \citep[140–260 M$_{\odot}$;][]{Salvadori2019}. Numerous studies have focused on the halo regions of the Galaxy, where the mean metallicity ranges from \(\text{[Fe/H]} = -1.6\) dex in the inner halo to \(\text{[Fe/H]} = -2.2\) dex in the outer halo \citep{Bekki2001}; however, the origin and evolutionary pathways of metal-poor stars remain incompletely understood. Recent investigations combining stellar orbits, kinematics, and chemical abundance \citep[for example,][]{Bonaca2019, DiMatteo2019} have advanced our knowledge of their Galactic distribution and dynamic histories.

However, a review of the literature revealed that detailed chemical abundance analyses of metal-poor stars are scarce \citep{Pereira2019, Marismak2024, Senturk2024, Holanda2024}. Notably, except for \cite{Pereira2019}, the cited studies represent comprehensive work conducted by the project team members as part of this research initiative.  \cite{Pereira2019} analyzed HD 55496, a metal-poor star enriched in slow neutron-capture (s-process) elements, proposing its potential origin as a second-generation GC star. However, their conclusion favoring a dwarf galaxy progenitor, derived solely from abundance trends without comparative kinematic or dynamical evidence, warrants critical scrutiny.

In contrast, the foundational work of \cite{Sahin2020} established a robust framework for studying metal-poor stars. Their high-resolution spectroscopic analysis of selected F-type stars in the northern Galactic Hemisphere, combined with Bayesian age recalibrations and kinematic-orbit dynamic modeling, provided novel insights into the physical nature (population type) and Galactic origins of metal-poor stars. The updated atmospheric parameters and age determination for HD 84937, a $Gaia$ benchmark star, are the key outcomes. Crucially, the Galactic orbital parameters calculated for these stars, mirroring the methodology of the current project, were employed to verify population classifications. A persistent challenge in metal-poor star studies is the significant discrepancies in reported atmospheric parameters and derived abundances across the literature.

These inconsistencies have been emphasized in recent studies by project teams \citep{Sahin2020, Marismak2024, Senturk2024}, hinders definitive conclusions about their characteristic properties (kinematic, dynamic, and composition-based) and origins. For instance, simulations by \cite{DiMatteo2019} demonstrated how orbital dynamics and chemical evolution jointly shape the distribution of metal-poor stars near the Galactic center; however, observational validation remains limited. Systematic spectral, kinematic, and dynamic orbital analyses of highly proper motion (HPM) stars offer a pathway for resolving these uncertainties. By correlating precise abundance patterns with orbital histories, future studies may disentangle the contributions of ancient supernovae, accretion events, and in-situ formation to the metal-poor stellar inventory.

In this study, we determined the fundamental astrophysical parameters and Galactic origins of four G spectral-type main-sequence stars, HD\,22879, HD\,144579, HD\,188510, and HD\,201891, in different metal abundance ranges selected from the solar neighborhood. The remainder of this paper is organized as follows: Section 2 introduces the data used in this study, including spectroscopic, photometric, and astrometric observations. Section 3 focuses on the results obtained from spectral energy distribution analysis, photometric and astrometric analyses, spectral analysis, and kinematic and dynamic orbit analyses. Finally, Section 4 summarizes the findings and discusses the membership of the studied stars to Galactic populations and their possible Galactic origins.

\section{Data} \label{sec:data}

\subsection{Ground-based: High-Resolution Spectroscopy} \label{sec:specdata}

To identify suitable target stars, we initially selected G-type stars from the ELODIE Stellar Library \citep{Soubiran2003}. Of the 545 G-type stars in this catalog, 173 were cross-matched with the entries in $Gaia$ Data Release 3 \citep[DR3;][]{GaiaEDR3}. We subsequently refined this subset by focusing on the HPM stars. Potential spectroscopic binaries were systematically excluded via cross-examination with the SIMBAD astronomical database \citep{Wenger2000}. This process yielded a preliminary sample of 90 G-type HPM stars. To ensure data quality, we imposed a threshold of renormalized unit weight error (RUWE) of 1.4, following the recommendations of \cite{Lindegren2021}. RUWE values below this threshold indicate robust astrometric solutions as they quantify the consistency between the observed stellar positions and model predictions. The application of these criteria produced a metallicity-constrained sample of 74 G-type HPM stars. The final target prioritization considered three key factors: (1) confinement to main-sequence stars (evidenced by surface gravities $\log g > 4.0$), (2) availability of high-quality observational data (particularly high-resolution spectra), and (3) precise {\it Gaia} DR3 astrometric measurements (trigonometric parallax and proper motion components), which are essential for deriving reliable stellar parameters. Through this multi-stage selection process, we identified four metal-poor stars in the solar neighborhood that satisfied all criteria. The complete target sample is presented in Table \ref{tab:spectroscopic_data}.

The spectra, which are downloaded from the {\sc PolarBase} library\footnote{\url{http://polarbase.irap.omp.eu/}}, consist of approximately 4,700 stellar spectra across different luminosity classes, ranging from spectral types O4 to M8 \citep{Petit2014}. Stellar spectra were obtained from the region of the electromagnetic spectrum between 3,700 and 10,000 \AA. The library also contains the spectra of a large number of metal-poor stars for the study of the early universe and the formation of the Milky Way. The high-resolution and high-($S/N$) spectra of the four stars were obtained using the ESPaDOnS and Narval high-resolution spectrometers, and their basic data are listed in Table \ref{tab:spectroscopic_data}.

We employed LIME code \citep{Sahin2017} to normalize the spectra within the spectral library. The {\sc LIME} code facilitated line identification in the continuum-normalized spectrum, providing associated atomic data such as the Rowland multiplet number (RMT), $\log gf$, and lower-level excitation energy (L.E.P.). Atomic data were sourced from the NIST\footnote{NIST Atomic Spectra Database \url{http://physics.nist.gov/PhysRefData/ASD}} and VALD\footnote{VALD Atomic Spectra Database \url{http://vald.astro.uu.se}}. The majority of the identified lines exhibited good isolation, making them suitable for equivalent width (EW) analysis performed using the {\sc LIME} code. The {\sc LIME} code was instrumental in performing equivalent width (EW) analysis on the majority of well-isolated identified lines.

\begin{table}[hb]%Table 01
\setlength{\tabcolsep}{3pt}
\centering
\footnotesize
\caption{Name, spectral type, spectrograph, $S/N$, radial velocity and observation date of four stars from the {\sc PolarBase} database. The spectral types of the program stars were obtained from SIMBAD. The corrected {\sc PolarBase} spectra accounted for the reported heliocentric radial velocities, which were cross-checked by the Solar spectrum.}
\label{tab:spectroscopic_data}%
\begin{tabular}{l|cccc}
\hline
Star   & HD\,22879   & HD\,144579  & HD\,188510  & HD\,201891   \\
\hline
Spectral Type   & G0VmF2     & G8V        & G5V:       & G5V         \\
Spectrograph & Narval     & ESPaDOnS   & Narval     & Narval      \\
 $S/N$   & 226.96     & 149.52     & 105.67     & 143.99      \\
 $V_{\rm R}$ (km s$^{-1}$) & 120.18     & -59.61     & -192.72    & -44.69      \\
Obs. Date  & 12.11.2010 & 12.03.2017 & 11.08.2010 & 16.08.2010 \\

\hline
\end{tabular}
\end{table}%

\subsection{Space-based: Comprehensive Data from {\it Gaia} DR3} \label{sec:photodata}

\begin{table*} %%%% Table 02
\centering
\caption{Astrometric and spectroscopic data of the four programme stars from the {\it Gaia} DR3 catalogue.}
\label{tab:Gaia_data}
\begin{tabular}{l|cccc}
\hline
Star & HD ~22879              & HD\,144579              & HD\,188510               & HD\,201891                \\
\hline
$\alpha$ (hh:mm:ss)    & ~~03 40 22.06          & ~~16 04 56.79          & 19 55 09.67             & 21 11 59.03              \\
$\delta$ (dd:mm:ss)    & $-$03 13 01.12         & +39 09 23.43           & +10 44 27.39            & +17 43 39.89             \\
$l$ (°)           & 189.7665                 & ~62.3693                 & 49.9314                   & 66.7067                    \\
$b$ (°)           & $-$43.1192               & ~~48.2726                & ~$-$8.9179                & $-$20.4288                 \\
$\mu_\alpha\cos\delta$ (mas yr$^{-1}$) & ~~690.795 $\pm$ 0.032  & $-$570.872 $\pm$ 0.016 & ~~$-$38.395 $\pm$ 0.022 & $-$122.133 $\pm$ 0.019   \\
$\mu_\delta$ (mas yr$^{-1}$) & $-$213.443 $\pm$ 0.028 & ~~~52.633 $\pm$ 0.017  & ~~290.612 $\pm$ 0.014   & ~$-$899.404 $\pm$ 0.015  \\
$\varpi$ (mas)         & 38.325 $\pm$ 0.031     & 69.641 $\pm$ 0.014     & 26.278 $\pm$ 0.020      & 29.877 $\pm$ 0.021\\
$V_{\rm R}$ (km s$^{-1}$)   & ~~120.58 $\pm$ 0.13    & ~$-$59.44 $\pm$ 0.12   & $-$192.22 $\pm$ 0.17    & ~$-$44.33 $\pm$ 0.12  \\
$T_{\rm eff}$ (K) & $5926^{+1}_{-2}$   & $5300^{+4}_{-14}$  & -- & -- \\
$\log g$ (cgs) & 4.32$\pm$0.01 & $4.57^{+0.03}_{-0.02}$& -- & -- \\
$[$Fe/H] (dex) & $-1.61^{+0.07}_{-0.03}$ & -0.20$\pm$0.02 & -- & -- \\
\hline
\end{tabular}
\end{table*}

The {\it Gaia} satellite, launched in 2013 by the European Space Agency (ESA), provides photometric, astrometric, and spectroscopic observations of Galactic and extragalactic sources to create a three-dimensional spatial and velocity map of the Milky Way, aiming to measure the spatial and velocity distributions of stars and determine their astrophysical properties, such as effective temperature and surface gravity, to understand the structure, formation, and evolution of the Galaxy \citep{Gaia_Mission}. The first data release (DR1) was published in 2016, followed by DR1, DR2, EDR3, and DR3, to date \citet{GaiaDR1, GaiaDR2, GaiaEDR3, GaiaDR3}. The third release of data the {\it Gaia} mission \citep[{\it Gaia} DR3;][]{GaiaDR3} offers unprecedented precision in astrometry, photometry, and radial velocity measurements for more than 1.8 billion objects. The photometric data collected in the $G$, $G_{\rm BP}$, and $G_{\rm RP}$ bands provide insights into various stellar properties. The effective temperature, luminosity, radius, surface gravity, and metallicity are interconnected properties that provide a holistic view of stellar physics. The three-band photometric data ($G, G_{\rm BP}, G_{\rm RP}$), combined with astrometric data ($\alpha$, $\delta$, $\mu_\alpha\cos\delta$, $\mu_\delta$, $\varpi$) and spectroscopic measurements ($V_{\rm R}$), allow for the accurate and detailed derivation of the fundamental stellar parameters. In this study, we used {\it Gaia} DR3 data to determine the basic astrophysical parameters and Galactic origins of the four stars. The astrometric and spectroscopic data of the four stars analyzed by querying the database and model atmospheric parameters calculated by the {\it Gaia} consortium are listed in Table \ref{tab:Gaia_data}.
%\newpage
\section{Results} \label{sec:results}
\subsection{Spectral Energy Distribution Analysis} \label{sec:sedmethod}

%%%%%%%%%%%%%%%%%% FIGURE 1
\begin{figure}   
\centerline
{\includegraphics[width=0.45\textwidth,clip=]{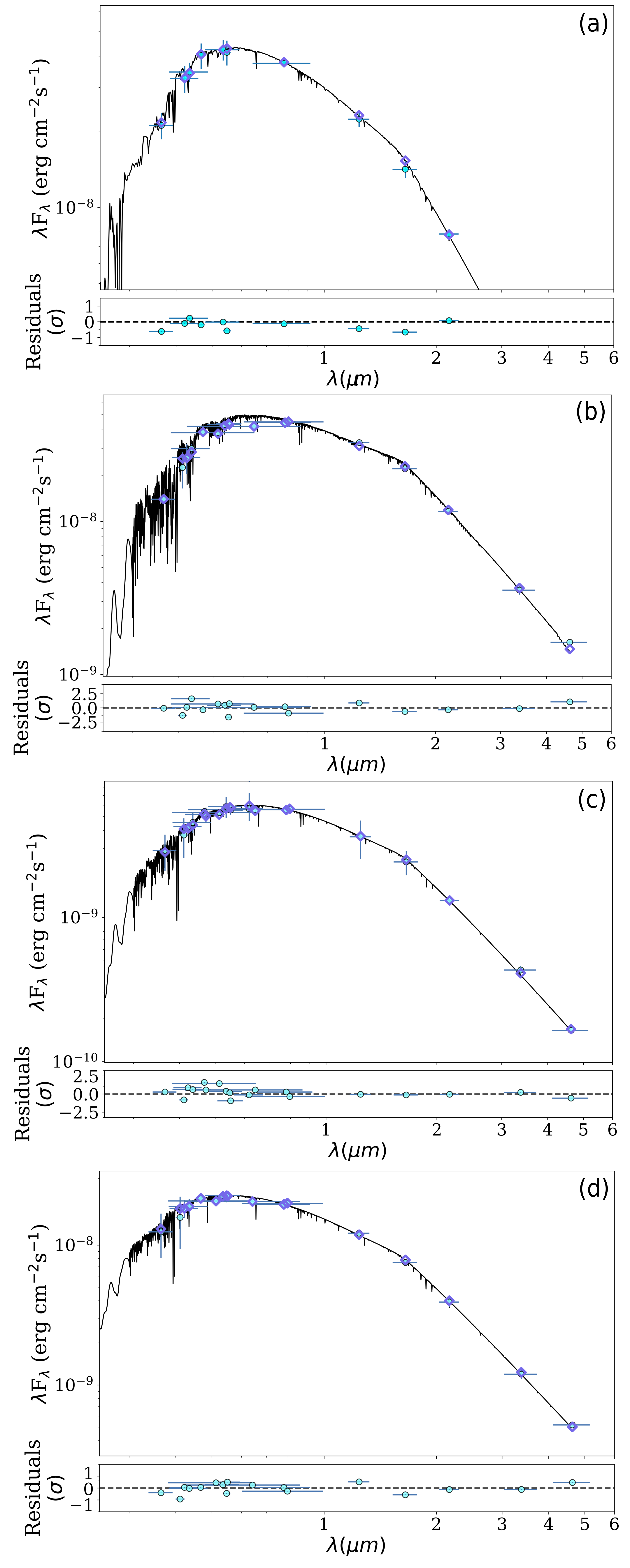}}
\caption{Best-fitting SED model for HD\,22879 (a), HD\,144579 (b), HD\,188510 (c), and HD\,201891 (d). The black curve represents the best-fitting model, whereas the cyan pluses and circles denote the retrieved photometric measurements. The blue diamonds correspond to synthetic photometry.}
\label{fourstar_sed}
\end{figure}

%%%%%%%%%%%%%%%%%% FIGURE 2
\begin{figure*}   
\centerline
{\includegraphics[width=0.8\textwidth,clip=]{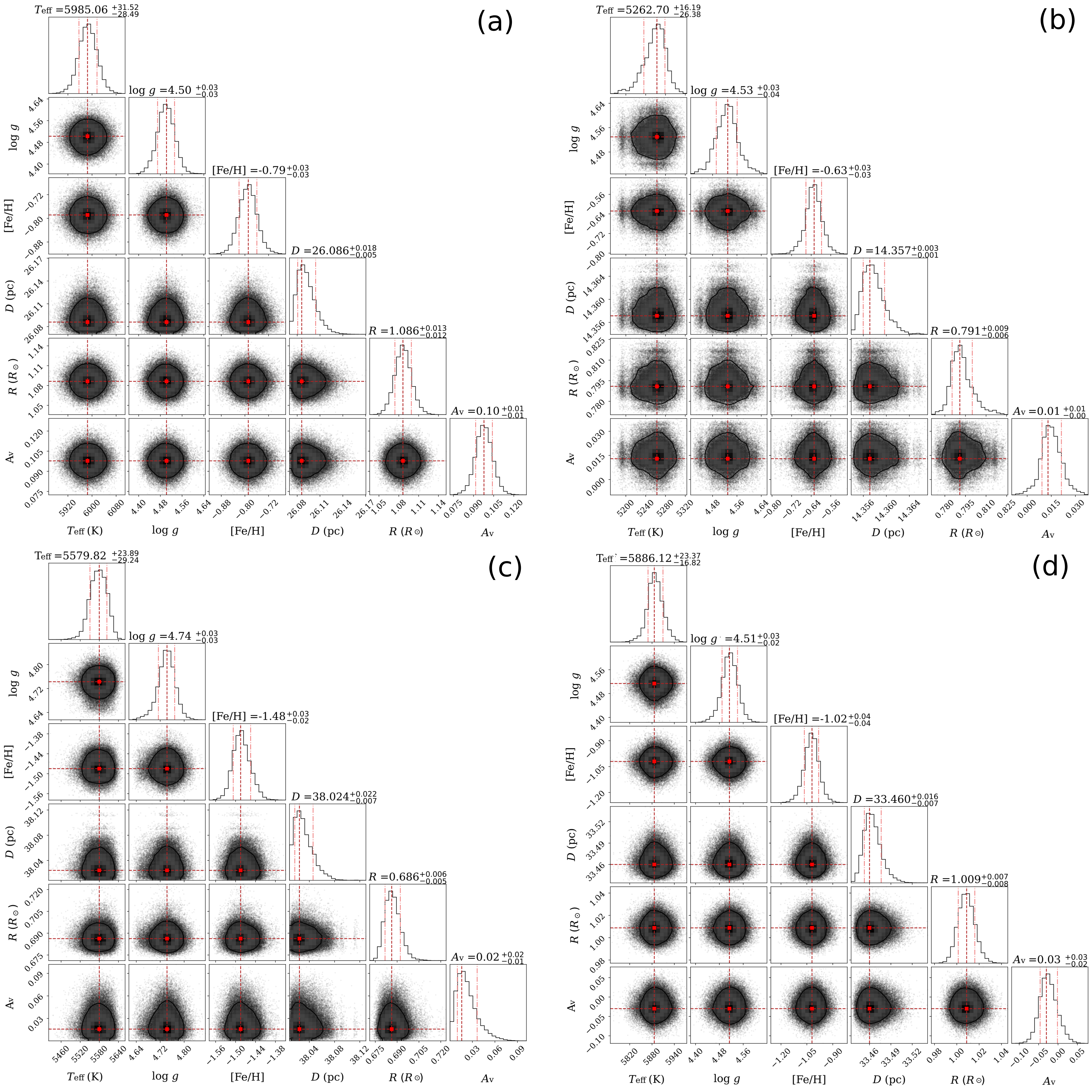}}
\caption{Posterior distributions from the best-fit SED model for HD\,22879 (a), HD\,144579 (b), HD\,188510 (c), and HD\,201891 (d) are shown. Red points indicate mean posteriors, and red vertical lines represent 16, 50 and 84 percentiles, respectively.}
\label{fourstar_corner}
\end{figure*}

The spectral energy distribution (SED) fitting was performed using the Python package SpectrAl eneRgy dIstribution bAyesian moDel averagiNg fittEr ({\sc ARIADNE}), which aims to use Bayesian model averaging to incorporate information from as many different atmospheric model grids to arrive at accurate and precise stellar parameters \citep{Vines2022}. {\sc ARIADNE} has been used with synthetic spectral model libraries, such as PHOENIX v2 \citep{Husser2013} and \citep{Castelli2003}, which are frequently referenced in the literature, to determine the values of the effective temperature ($T_{\rm eff}$), surface gravity ($\log g$), iron abundance [Fe/H], $V$-band extinction ($A_{\rm V}$), and distance ($d$). In addition, MIST evolution models \citep{Morton2015, Dotter2016} have been used to calculate the mass ($M$), stellar radius ($R$), and age ($\tau$) of stars. While fitting the SEDs, the program first used the current astrometric and photometric data of the stars \emph{Gaia} DR3 \citep{GaiaDR3} catalogue. In general, luminosity data in the wavelength range $0.1 \le \lambda~(\mu {\rm m})\le  5$ of the electromagnetic spectrum were used to generate SED fitting of the stars.

We collected broadband photometric data from various catalogues to fit the SED, including TYCHO $B$ and $V$ \citep{ESA1997}, {\it Gaia} DR3 $G$, $G_{\rm BP}$, and $G_{\rm RP}$ \citep{GaiaDR3}, \textit{TESS} $T$ \citep{Ricker2014}, 2MASS $J$, $H$, and $K_{\rm s}$ \citep{Skrutskie2006}, and \textit{WISE~W1} and \textit{W2} \citep{Wright2010}. We noticed that incorporating \textit{GALEX} FUV/NUV and Stromgren $ubv$ photometry significantly degrades the quality of our fits. This results in unreliable posterior constraints and frequent large discrepancies between the models and data \citep{Harada2024}. Consequently, we excluded \textit{GALEX} FUV/NUV and Stromgren $ubv$ photometry from SED analyses. In {\sc ARIADNE}, we used the extinction law of \citet{Fitzpatrick1999}.

We determined that using 1,000 live points and a stopping criterion of $d \log z$ = 0.5 produced stable and well-converged posterior distributions. Figure \ref{fourstar_sed} shows the SED fitting results \citep[e.g.][]{Dursun2024}. \citet{Castelli2003} model can accurately represent the SED of HD\,22879 and HD\,188510. The best-fitting parameters with their 68\% confidence intervals for HD\,22879 and HD\,188510, respectively, are as follows: $T_{\rm eff}$ = $5985\pm30$ K, $T_{\rm eff}$ = $5580\pm26$ K, $\log g$ = $4.5\pm0.03$ cgs, $\log g$ = $4.74\pm0.03$ cgs, and $R$ = $1.08\pm0.01$ $R_{\odot}$, $R$ = $0.69\pm0.01$ $R_{\odot}$. The PHOENIX v2 \citet{Husser2013} model accurately represented the SED of HD\,144579 and HD\,201891. The best-fitting parameters for these stars, respectively, are as follows: $T_{\rm eff}$ = $5263\pm21$ K, $T_{\rm eff}$ = $5886\pm20$ K, $\log g$ = $4.53\pm0.04$ cgs, $\log g$ = $4.51\pm0.03$ cgs, and $R$ = $0.79\pm0.01$ $R_\odot$, $R$ = $1.01\pm0.01$ $R_\odot$. We then estimated the stellar masses of stars based on the derived surface gravity and radius.

\subsection{Photometric and Astrometric Analyses} \label{sec:photastromethod}

\begin{table*}
\centering
\caption{The photometric ($G$, $G_{\rm BP}$, $G_{\rm RP}$) and astrometric ($\varpi$) data from \textit{Gaia} DR3, along with the extinction values in the $G$ band ($A_{\rm G}$), the derived absolute magnitudes ($M_{\rm G}$), and the de-reddening colors ($(G_{\rm BP} - G_{\rm RP})_0$), are reported for the four metal-poor stars. The parameters are provided in mag.}
\label{tab:gaia_ext}
\begin{tabular}{lccccccl}
\hline
Star       & $G$             & $G_{\rm BP}$          & $G_{\rm RP}$          & $A_{\rm G}$           & $M_{\rm G}$           & $(G_{\rm BP} - G_{\rm RP})_0$   \\
\hline
HD\,22879 & 6.536 $\pm$ 0.003 & 6.822 $\pm$ 0.003 & 6.075 $\pm$ 0.004 & 0.079 $\pm$ 0.005 & 4.374 $\pm$ 0.002 & 0.704 $\pm$ 0.005 \\
HD\,144579  & 6.461 $\pm$ 0.003  & 6.843 $\pm$ 0.003 & 5.898 $\pm$ 0.004 & 0.012 $\pm$ 0.001 & 5.675 $\pm$ 0.001 & 0.944 $\pm$ 0.005  \\
HD\,188510  & 8.647 $\pm$ 0.003 & 8.973 $\pm$ 0.003 & 8.138 $\pm$ 0.004 & 0.017 $\pm$ 0.002 & 5.744 $\pm$ 0.002 & 0.834 $\pm$ 0.005  \\
HD\,201891  & 7.231 $\pm$ 0.003  & 7.501 $\pm$ 0.003 & 6.781 $\pm$ 0.004 & 0.017 $\pm$ 0.001 & 4.607 $\pm$ 0.001 & 0.719 $\pm$ 0.005 \\
\hline
\end{tabular}
\end{table*}

The age of stars is often investigated using color magnitude diagrams (CMDs), which provide insights into stellar populations. With the advent of the \textit{Gaia} era, precise proper motions and trigonometric parallaxes of stars in the solar neighborhood are now available with milli-arcsecond (mas) accuracy, whereas \textit{Gaia} DR3 photometric data are delivered with milli-magnitude precision. These advancements have enabled the placement of the four stars studied herein on the $M_G \times (G_{\rm BP}-G_{\rm RP})_0$ plane. In addition, by estimating the photometric metallicities of these stars, it was possible to determine their ages.

The \textit{Gaia} DR3 provided photometric ($G$, $G_{\rm BP}$, and $G_{\rm RP}$) and astrometric ($\varpi$) data that were used for age determination (see Table \ref{tab:Gaia_data}). The extinction coefficients ($A_{\lambda}/A_{\rm V}$) for the \textit{Gaia} photometric system bands $G$, $G_{\rm BP}$, and $G_{\rm RP}$ were obtained from \citet{Cardelli1989}, with values of 0.83627, 1.08337, and 0.63439, respectively. The following relationships were used to correct each filter for interstellar extinction \citep[see also][]{Canbay2023, iyisan2025}.
\begin{eqnarray}
G_0 &=& G - A_{\rm G} = G - 0.83627 \times A_{\rm V} \\
(G_{\rm BP})_0 &=& G_{\rm BP} - A_{\rm G(BP)} = G_{\rm BP} - 1.08337 \times A_{\rm V} \\
(G_{\rm RP})_0 &=& G_{\rm RP} - A_{\rm G(RP)} = G_{\rm RP} - 0.63439 \times A_{\rm V}
\end{eqnarray}
Thus, the \textit{Gaia} DR3 \citep{GaiaDR3} photometric data of stars were corrected for interstellar extinction, and the errors are listed in Table \ref{tab:gaia_ext}. Subsequently, the $G$ absolute magnitudes of the stars were calculated using the following distance relation:
\begin{equation}
M_{\rm G} = G - 5 \times \log d + 5 - A_{\rm G}
\end{equation}
Here, $M_{\rm G}$ is the absolute magnitude of the star, $G$ is the apparent magnitude, $A_{\rm G}$ is the extinction in the $G$ band, and $\varpi$ is the trigonometric parallax of the star. Given a star's distance $d({\rm pc})=1000/\varpi$ (mas), by differentiating both sides, the error in the absolute magnitude of the $G$ band ($\Delta M_{\rm G}$) was calculated using the following relation:
\begin{equation}
\Delta M_{\rm G} = \frac{5}{d \times \ln 10} \times \Delta d + \Delta A_{\rm G}
\end{equation}
Here, $\Delta A_{\rm G}$ represents the error in the $G$ band extinction, whereas $\Delta d$ corresponds to the distance error arising from the trigonometric parallax measurements. The distance error was calculated using the following equation:
\begin{equation}
\Delta d = \frac{1000}{\varpi^2} \times \Delta \varpi
\end{equation}
The obtained $G$-band absolute magnitudes ($M_{\rm G}$), along with their associated errors, are listed in Table \ref{tab:gaia_ext}. To determine the ages of stars from their astrometric and photometric data, $M_{\rm G}$ and $(G_{\rm BP} - G_{\rm RP})_0$ are provided in Table \ref{tab:gaia_ext}. The CMDs were constructed for each star.

\subsection{Spectral Analysis} \label{sec:spectralmethod}

This study utilized the local thermodynamic equilibrium (LTE) {\sc ATLAS9} model atmospheres (ODFNEW) from \citet{Castelli2004} to conduct an abundance analysis on a sample of G-type stars. We used the MOOG {\sc LTE} line analysis code\footnote{\url{http://www.as.utexas.edu/~chris/moog.html}} \citep{Sneden1974} to calculate the elemental abundances of the program stars. The methodology adopted for the abundance analysis and atomic data sources are consistent with those used in previous studies by \citet{Sahin2009}, \citet{Sahin2011}, \citet{Sahin2016}, \citet{Sahin2020}, \citet{Sahin2023a}, \citet{Marismak2024}, \citet{Senturk2024}, and \citet{Sahin2024}. The preceding sections discuss the employed line list, atomic data, and derivation of stellar parameters.

The determination of the effective temperature ($T_{\rm eff}$) and other stellar parameters, such as the microturbulence ($\xi$) and surface gravity ($\log g$), are critical aspects of the analysis of the stellar atmosphere. $T_{\rm eff}$ was derived using the excitation balance method, which relies on neutral spectral lines with varying excitation potentials, specifically focusing on Fe\,{\sc i} lines. The selected $T_{\rm eff}$ ensures that the derived abundance exhibits no dependence on the lower excitation potential (L.E.P.) of the spectral lines.

Microturbulence, which characterizes small-scale gas motion within the stellar atmosphere, is determined by applying the traditional criterion that the abundance of neutral or singly ionized species (e.g., Fe\,{\sc i}) remains independent of the reduced equivalent width ($EW/ \lambda$). This analysis was conducted under the assumption of local thermodynamic equilibrium (LTE). For the Fe\,{\sc i} lines, the conditions of independence from both the L.E.P. and $EW/ \lambda$ were simultaneously enforced.

Furthermore, microturbulence was individually determined using the Fe\,{\sc i},  Fe\,{\sc ii}, Ti\,{\sc i}, Ti\,{\sc ii}, Cr\,{\sc i}, and Cr\,{\sc ii} lines. For each model, the dispersion in the abundances of Fe, Ti, and Cr was computed over a range of  $\xi$ values from 0 to 1.5 km~s$^{\rm -1}$. As shown in Figure \ref{fig:fourstar_abundances}, this approach was applied to all stars analyzed in this study. For example, in the case of HD\,22879 (top panel), the Fe\,{\sc i}, Fe\,{\sc ii}, and Ti\,{\sc i} lines yielded a micro-turbulent velocity of $\approx$1.0 km~s$^{-1}$. For the same star, the Cr\,{\sc i}, and Ti\,{\sc ii} lines indicate a value of $\approx$0.8 km~s$^{-1}$, which is consistent with the results of classical spectroscopic analysis (Table \ref{tab:atmosphere_parameters}). For HD\,144579, the Fe\,{\sc i}, Fe\,{\sc ii}, and Ti\,{\sc i} lines exhibited minimal sensitivity to $\xi$ across the range of 0.0$\approx$0.8 km~s$^{\rm -1}$, whereas the Cr\,{\sc i} lines suggested a microturbulent velocity of 0.8 km~s$^{\rm -1}$. For HD\,188510, the Ti\,{\sc ii} lines indicated microturbulence values between 0.5 and 0.6 km~s$^{\rm -1}$, whereas the Fe\,{\sc i}, Ti\,{\sc i} and Cr\,{\sc i} lines suggested values ranging from 0.0 to 0.5 km~s$^{\rm -1}$. For HD\,201891, the Fe\,{\sc i}, Fe\,{\sc ii}, Ti\,{\sc i}, and Ti\,{\sc ii} lines indicated a microturbulent velocity range of 0.7 to 1.2 km~s$^{-1}$, highlighting the uncertainty in the microturbulence value derived from classical spectroscopic analysis. A comparison of the two methods demonstrates that the estimated uncertainties in the microturbulence for all stars are consistent with those obtained from the classical spectroscopic analysis.

$\log g$ was constrained by enforcing the ionization equilibrium, ensuring that the Fe\,{\sc i} and Fe\,{\sc ii} lines yielded consistent iron abundances. This analysis was performed using MOOG software. The metallicity ([Fe/H]) was iteratively adjusted to obtain agreement between the derived iron abundance and the initial abundance used to construct the model atmosphere. Convergence of the derived abundances was achieved by iteratively varying $T_{\rm eff}$, $\log g$, and $\xi$ within the model.

\begin{figure*} 
\centering
\centerline{\includegraphics[width=0.95\textwidth,clip=]{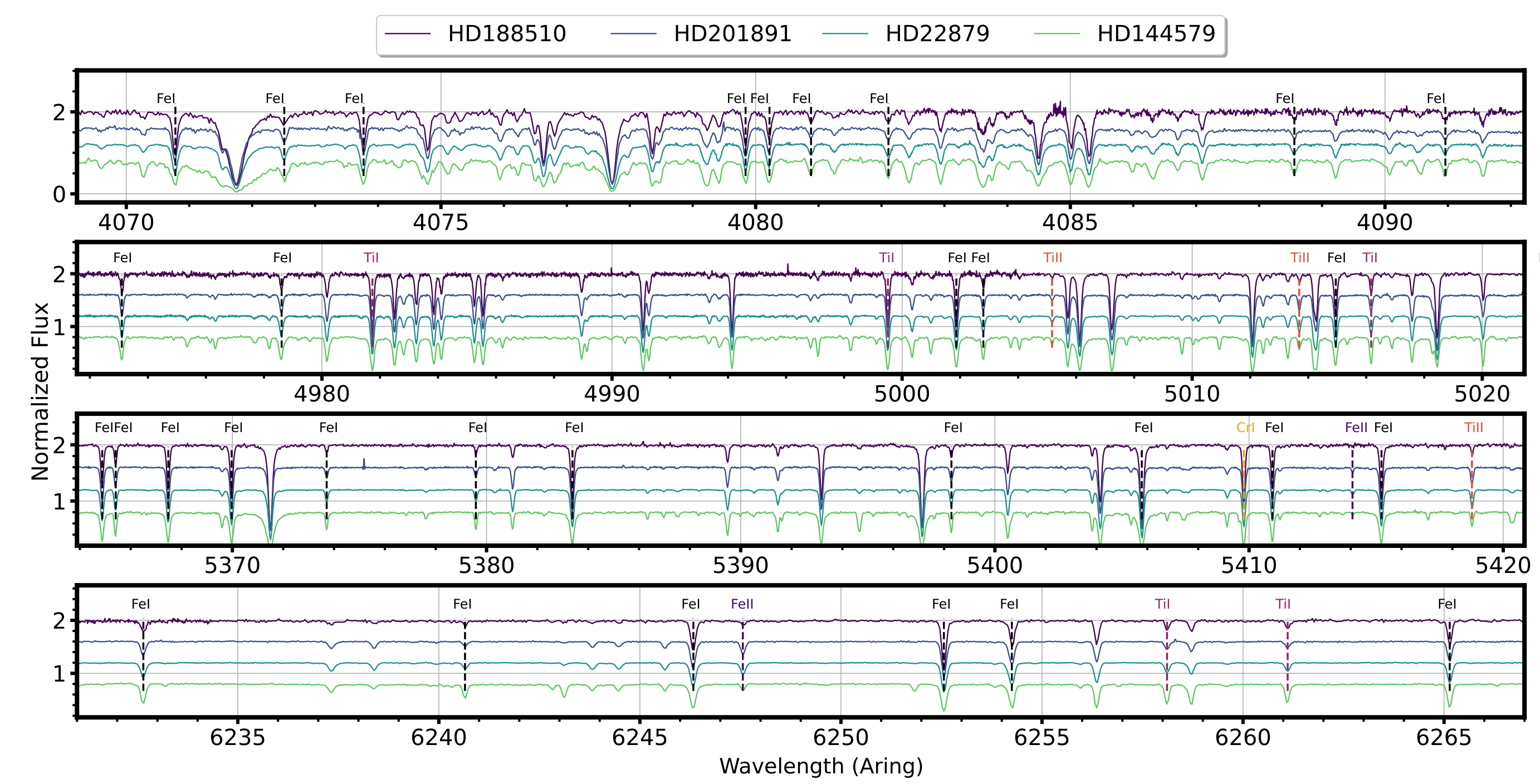}}
\caption{Small regions of the spectra for HD\,188510, \textbf{HD\,201891}, HD\,22879, and HD\,144579. For convenience, the spectra were shifted vertically. A selection of neutral and ionized lines used in the analysis are indicated by a vertical dashed line.}
\label{fourstar_spectra}
\end{figure*}

The uncertainty in the derived effective temperature arises from the error in the slope of the relationship between the Fe\,{\sc i} abundance and L.E.P. of the lines. For HD\,22879, a temperature variation of  $\pm$ 110 K  results in a significant change in the slope. Similarly, a 1$\sigma$ discrepancy in [X/H] abundance between neutral and ionized Fe lines corresponded to a shift of approximately 0.2 dex in $\log g$. For HD\,144579, this approach yielded uncertainties of $\pm$160 K for $T_{\rm eff}$ and $\pm$0.37 dex for $\log g$. The atmospheric parameters derived for all the stars are summarized in Table \ref{tab:atmosphere_parameters}.

The determination of elemental abundance may be subject to various uncertainties. These include factors associated with the accuracy of $gf$ values and potential deviations from the local thermodynamic equilibrium (LTE) in the formation processes of atomic spectral lines. To ensure the accuracy of $gf$values and mitigate systematic errors, we derived solar abundances using stellar line measurements. Equivalent widths were measured from the solar flux atlas of \citet{Kurucz1984} and analyzed with a solar model atmosphere selected from the \citet{Castelli2004} grid, adopting the parameters $T_{\rm eff}$ = 5790 K and $\log g\,=$ 4.40 cgs, [Fe/H]= 0 dex, and a microturbulence of $\xi$=0.66 km s$^{-1}$. The derived solar abundance values are listed in Table \ref{tab:solar_abund}. Our results satisfactorily reproduced the solar abundances and aligned closely with the values reported in the critical review by \citet{Asplund2009}. These solar abundances were used to reference stellar abundances to solar values, thereby performing the analysis differentially with respect to the Sun. In Table \ref{tab:solar_abund}, log$\epsilon$ is the logarithm of abundance. The number of lines used in this analysis has also been provided. The errors reported in the log $\epsilon$ abundances are represented by a 1$\sigma$ line-to-line scatter in abundance.

In this study, the abundances of Fe\,{\sc i} and Fe\,{\sc ii} were used to derive the atmospheric model parameters. For the Fe\,{\sc i} lines, we incorporated 1D non-LTE corrections using the methodology of \,{\sc i} and the INSPECT program (v1.0). Previous studies \citep{Bergemann2012, Lind2012, Bensby2014} have suggested that non-LTE effects are negligible for Fe\,{\sc ii} lines. Specifically,  \citet{Lind2012} demonstrated that departures from the LTE for Fe\,{\sc ii} lines with low excitation potentials ($<$8 eV) in stars with metallicities [Fe/H] $>$ -3 dex were minimal. To assess the influence of non-LTE corrections on atmospheric parameter determination, we adopted the approach outlined in \citet[][see Figures 4, 5, and 7]{Lind2012}. For most metal-poor stars in our sample, the resulting corrections to the atmospheric parameters were small, with temperature adjustments not exceeding the reported errors on $T_{\rm eff}$ listed in Table \ref{tab:atmosphere_parameters}.

\begin{table}
\small
\centering
\setlength{\tabcolsep}{2pt}
\caption{Model atmospheric parameters of the four program stars.}
\label{tab:atmosphere_parameters}
\centering
\begin{tabular}{lcccc}
\hline
Star	  &	$T_{\rm eff}$	      &	$\log g$ 	           &	[Fe/H]	             &  $\xi$               \\
\cline{2-5}
          & (K)                   & (cgs)                  &     (dex)               & (km s$^{-1}$)        \\
\hline
HD\,22879  & $5855 \pm 110$      & $4.40 \pm 0.18$  & $-0.86 \pm 0.08$ & $0.95 \pm 0.50$ \\
HD\,144579  & $5300 \pm 160$     & $4.52 \pm 0.37$  & $-0.55 \pm 0.12$ & $0.00 \pm 0.80$ \\
HD\,188510  & $5370 \pm 60$      & $4.57 \pm 0.13$  & $-1.60 \pm 0.07$ & $0.17 \pm 0.50$ \\
HD\,201891  & $5880 \pm 90$      & $4.48 \pm 0.18$  & $-1.15 \pm 0.07$ & $1.03 \pm 0.50$ \\
\hline
\end{tabular}

\end{table}

For HD\,22879, Si\,{\sc i} shows a minor NLTE correction of -0.005 dex, derived from six spectral lines, suggesting a minimal deviation from the LTE assumptions. Ti\,{\sc i} exhibits a significant NLTE effect with an average correction of 0.221 dex based on 21 spectral lines, whereas Ti\,{\sc ii} shows a slight negative correction of -0.002 dex from seven lines. Cr\,{\sc i} and Mn\,{\sc i} display positive NLTE corrections of 0.145 and 0.201 dex, respectively. In addition, Fe\,{\sc i} and Fe\,{\sc ii} had NLTE corrections of 0.017 dex and 0.006 dex, calculated from 51 and six spectral lines, respectively. The highest NLTE correction was observed for Co\,{\sc i} (0.275 dex) although it was derived from only two spectral lines.

%PAR1 yerine G-stars ile aynı sadece Mg ve Al PAR2den revize edildi--Timur
\begin{table}
\scriptsize
\centering
\setlength{\tabcolsep}{4pt}
\renewcommand{\arraystretch}{1.2}
\caption{Solar abundances obtained by employing the solar model atmosphere from \citet{Castelli2004} compared to the photospheric abundances from \citet{Asplund2009}. The abundances presented in bold typeface were measured by synthesis, while the remaining elemental abundances were calculated using the line EWs. $\Delta {\rm log} \epsilon_{\rm \odot}{\rm (X) = log} \epsilon_{\rm \odot} {\rm (X)}_{\rm This\,study} - {\rm log} \epsilon_{\rm \odot} {\rm (X)}_{\rm Asplund}$}
\label{tab:solar_abund}
\centering
\begin{tabular}{l|c|c|c|c}
\hline
     	&   This study &  &   \citet{Asplund2009} \\
\cline{2-2}
\cline{4-4}
Species   &  $\log\epsilon_{\rm \odot}$(X) & $N$ &   $\log\epsilon_{\rm \odot}$(X) & $\Delta\log\epsilon_{\rm \odot}$(X) \\
\cline{2-2}
\cline{4-4}
		&  (dex) &&  (dex)&  (dex)\\
 \hline
%Li\,{\sc i}	&    ---	    &	1     &   1.05 $\pm$0.10  & 	    \\
Na\,{\sc i}           & 6.16 $\pm$ 0.07 & 2	  & 6.24 $\pm$ 0.04 & -0.08 \\
\textbf{Mg\,{\sc i}}  & 7.62 $\pm$ 0.03 & 5	  & 7.60 $\pm$ 0.04 & 0.02  \\
\textbf{Al\, {\sc i}} & 6.45 $\pm$ 0.03 & 8   & 6.45 $\pm$ 0.03 & 0.00 \\
Si\,{\sc i}  	      & 7.50 $\pm$ 0.07 & 12  & 7.51 $\pm$ 0.03 & -0.01 \\
Ca\,{\sc i}  	      & 6.34 $\pm$ 0.08 & 18  & 6.34 $\pm$ 0.04 & 0.00  \\
Sc\,{\sc i} 	      & 3.12 $\pm$ 0.00 & 1	  & 3.15 $\pm$ 0.04 & -0.03 \\
Sc\,{\sc ii} 	      & 3.23 $\pm$ 0.08 & 7	  & 3.15 $\pm$ 0.04 & 0.08  \\
Ti\,{\sc i}  	      & 4.96 $\pm$ 0.09 & 43  & 4.95 $\pm$ 0.05 & 0.01  \\
Ti\,{\sc ii} 	      & 4.99 $\pm$ 0.08 & 12  & 4.95 $\pm$ 0.05 & 0.04  \\
\textbf{V\,{\sc i}}   & 3.90 $\pm$ 0.03 & 5	  & 3.93 $\pm$ 0.08 & -0.03 \\
Cr\,{\sc i}  	      & 5.71 $\pm$ 0.07 & 19  & 5.64 $\pm$ 0.04 & 0.07  \\
Cr\,{\sc ii} 	      & 5.64 $\pm$ 0.14 & 3	  & 5.64 $\pm$ 0.04 & 0.00  \\
Mn\,{\sc i}  	      & 5.62 $\pm$ 0.13 & 13  & 5.43 $\pm$ 0.05 & 0.19  \\
Fe\,{\sc i}  	      & 7.54 $\pm$ 0.09 & 132 & 7.50 $\pm$ 0.04 & 0.04  \\
Fe\,{\sc ii}          & 7.51 $\pm$ 0.04 & 17  & 7.50 $\pm$ 0.04 & 0.01  \\
\textbf{Co\,{\sc i}}  & 4.90 $\pm$ 0.14 & 6	  & 4.99 $\pm$ 0.07 & -0.05 \\
Ni\,{\sc i}  	      & 6.28 $\pm$ 0.09 & 54  & 6.22 $\pm$ 0.04 & 0.06  \\
\textbf{Zn\,{\sc i}}  & 4.60 $\pm$ 0.00 & 2	  & 4.56 $\pm$ 0.05 & 0.04  \\
\textbf{Sr\,{\sc i}}  & 2.86 $\pm$ 0.00 & 1   & 2.87 $\pm$ 0.07 & -0.01 \\
\textbf{Y\,{\sc ii}}  & 2.20 $\pm$ 0.04 & 2	  & 2.21 $\pm$ 0.05 & -0.01 \\
\textbf{Zr\,{\sc ii}} & 2.70 $\pm$ 0.00 & 1	  & 2.58 $\pm$ 0.04 & 0.12  \\
\textbf{Ba\,{\sc ii}} & 2.52 $\pm$ 0.19 & 4	  & 2.18 $\pm$ 0.09 & 0.32  \\
\textbf{Ce\,{\sc ii}} & 1.58 $\pm$ 0.02 & 2	  & 1.58 $\pm$ 0.04 & 0.00  \\
\textbf{Nd\,{\sc ii}} & 1.33 $\pm$ 0.09 & 3	  & 1.42 $\pm$ 0.04 & -0.09 \\
\textbf{Sm\,{\sc ii}} & 0.95 $\pm$ 0.00 & 1	  & 0.96 $\pm$ 0.04 & -0.01 \\
\hline
\end{tabular}
\end{table}

For HD 144579, the NLTE corrections range from 0.00 dex (for Si\,{\sc i}) to 0.14 dex (for Ti\,{\sc i}). Among the analyzed elements, Mg\,{\sc i} exhibited a minor NLTE correction of 0.02 dex, derived from a single spectral line. Ti\,{\sc i} demonstrates a more substantial correction of 0.14 dex, based on 26 spectral lines, whereas Ti\,{\sc ii} presents a slightly lower correction of 0.06 dex from five lines. Cr\,{\sc i} and Mn\,{\sc i} displayed moderate NLTE corrections of 0.08 dex and 0.12 dex, respectively. Iron-group elements exhibit relatively small NLTE effects, with Fe\,{\sc i} and Fe\,{\sc ii} having corrections of 0.02 dex and 0.00 dex, based on 61 and 6 spectral lines, respectively. Co\,{\sc i}, derived from five spectral lines, exhibits an NLTE correction of 0.13 dex, indicating a modest deviation from the LTE.

For HD 188510, NLTE corrections varied significantly across different elements, with values ranging from -0.001 dex (Fe\,{\sc ii}) to 0.342 dex (Co\,{\sc i}). Among the species studied, Ca\,{\sc i} exhibited a small NLTE correction of 0.027 dex, based on two spectral lines. Ti\,{\sc i} showed a more pronounced correction of 0.223 dex (from eight lines), whereas Ti \,{\sc ii} had a much smaller correction of 0.001 dex (from four lines), indicating a stronger NLTE effect in neutral titanium than in its singly ionized counterpart. Cr\,{\sc i} and Mn\,{\sc i} displayed moderate-to-significant NLTE effects, with corrections of 0.209 and 0.304 dex, respectively. Iron-group elements show minor NLTE effects, with Fe\,{\sc i} and Fe\,{\sc ii} having corrections of 0.008 dex (from 17 lines) and -0.001 dex (from one line), respectively. However, Co\,{\sc i}, which is based on a single spectral line, exhibited the highest NLTE correction (0.342 dex).

\begin{table}
\scriptsize
\centering
\setlength{\tabcolsep}{1pt}
\renewcommand{\arraystretch}{1.0}
\caption{The abundances of the observed species for HD\,22879, HD\,144579, HD\,188510, and HD\,201891 are presented. The abundances presented in bold typeface are measured by synthesis.}
\label{tab:abundances}
\centering
    \begin{tabular}{l|ccc|ccc|ccc|ccc}
    \hline

Species                & \multicolumn{3}{c|}{HD\,22879} & \multicolumn{3}{c|}{HD\,144579} & \multicolumn{3}{c|}{HD\,188510} & \multicolumn{3}{c}{HD\,201891} \\ \cline{2-13} 
~                      & [X/Fe]    & $\sigma$    & N    & [X/Fe]    & $\sigma$    & N     & [X/Fe]    & $\sigma$    & N     & [X/Fe]    & $\sigma$    & N    \\ \hline
Na \,{\sc i}           & 0.07      & 0.08        & 2    & 0.21      & 0.06        & 1     & -         & -           & -     & -         & -           & -    \\
\textbf{Mg \,{\sc i}}  & 0.16      & 0.09        & 3    & 0.17      & 0.10        & 2     & 0.25      & 0.07        & 2     & 0.23      & 0.09        & 2    \\
\textbf{Al \,{\sc i}}  & 0.07      & 0.02        & 2    & 0.33      & 0.04        & 2     & -         & -           & -     & 0.17      & 0.03        & 2    \\
Si \,{\sc i}           & 0.27      & 0.03        & 9    & 0.27      & 0.05        & 6     & -         & -           & -     & -         & -           & -    \\
Ca \,{\sc i}           & 0.26      & 0.03        & 12   & 0.32      & 0.04        & 10    & 0.24      & 0.05        & 3     & 0.18      & 0.03        & 8    \\
Sc \,{\sc ii}          & 0.15      & 0.03        & 7    & 0.29      & 0.06        & 6     & -0.17     & 0.05        & 1     & 0.17      & 0.04        & 4    \\
Ti \,{\sc i}           & 0.23      & 0.03        & 27   & 0.42      & 0.04        & 29    & 0.20      & 0.05        & 10    & 0.17      & 0.03        & 7    \\
Ti \,{\sc ii}          & 0.34      & 0.04        & 7    & 0.39      & 0.06        & 5     & 0.32      & 0.05        & 4     & 0.32      & 0.04        & 10   \\
\textbf{V \,{\sc i}}   & -         & -           & -    & 0.52      & 0.06        & 4     & -         & -           & -     & -         & -           & -    \\
Cr \,{\sc i}           & -0.09     & 0.03        & 16   & 0.06      & 0.05        & 15    & -0.06     & 0.05        & 5     & -0.13     & 0.04        & 10   \\
Cr \,{\sc ii}          & -0.10     & 0.10        & 2    & 0.13      & 0.11        & 2     & -         & -           & -     & 0.08      & 0.09        & 1    \\
Mn \,{\sc i}           & -0.40     & 0.09        & 4    & -0.05     & 0.06        & 9     & -0.30     & 0.08        & 2     & -0.46     & 0.05        & 3    \\
Fe \,{\sc i}           & -0.03     & 0.02        & 92   & 0.07      & 0.04        & 104   & 0.02      & 0.04        & 36    & -0.03     & 0.03        & 51   \\
Fe \,{\sc ii}          & 0.00      & 0.03        & 14   & 0.00      & 0.05        & 11    & 0.00      & 0.06        & 3     & 0.00      & 0.04        & 8    \\
\textbf{Co \,{\sc i}}  & -0.11     & 0.08        & 2    & 0.23      & 0.10        & 6     & 0.03      & 0.07        & 1     & 0.09      & 0.06        & 1    \\
Ni \,{\sc i}           & -0.02     & 0.03        & 27   & 0.08      & 0.04        & 32    & -0.17     & 0.06        & 2     & -0.05     & 0.05        & 10   \\
\textbf{Zn \,{\sc i}}  & 0.03      & 0.07        & 2    & 0.15      & 0.04        & 2     & 0.04      & 0.04        & 1     & 0.00      & 0.08        & 2    \\
\textbf{Sr \,{\sc i}}  & -0.15     & 0.02        & 1    & -0.12     & 0.04        & 1     & -         & -           & -     & -         & -           & -    \\
\textbf{Y \,{\sc ii}}  & -0.03     & 0.06        & 2    & -0.06     & 0.05        & 2     & -         & -           & -     & -0.08     & 0.04        & 2    \\
\textbf{Zr \,{\sc ii}} & 0.32      & 0.02        & 1    & -0.12     & 0.04        & 1     & -         & -           & -     & -         & -           & -    \\
\textbf{Ba \,{\sc ii}} & -0.36     & 0.10        & 2    & -0.44     & 0.12        & 2     & -0.36     & 0.11        & 2     & -0.27     & 0.10        & 2    \\
\textbf{Ce \,{\sc ii}} & -0.05     & 0.03        & 1    & -         & -           & -     & -         & -           & -     & -         & -           & -    \\
\textbf{Nd \,{\sc ii}} & 0.13      & 0.06        & 1    & -         & -           & -     & -         & -           & -     & -         & -           & -    \\ 
     \hline
 \end{tabular}
\end{table}

In the case of HD 201891, the analysis revealed varying degrees of non-LTE corrections across different elements. Mg\,{\sc i} shows a modest correction of 0.029 dex, as determined from a single spectral line. Ca\,{\sc i} exhibited an average correction of 0.023 dex, based on the analysis of the six lines. The most pronounced corrections were evident in Fe\,{\sc i} and Co\,{\sc i}, with non-LTE differences of 0.202 dex (derived from 27 lines) and 0.415 dex (based on one line), respectively. Conversely, elements such as Ti\,{\sc ii} and Fe\,{\sc ii} showed negligible corrections. 

The abundances obtained for all stars using their PolarBase spectra are listed in Table~\ref{tab:abundances}, where [X/Fe] represents logarithmic abundance considering the abundance of Fe\,{\sc ii}. The error in [X/Fe] is the square root of the sum of the squares of the errors in [X/H]\footnote{[X/H] is the logarithmic abundance ratio of hydrogen to the corresponding solar value.} and [Fe/H]. The formal abundance errors resulting from uncertainties in atmospheric parameters $T_{\rm eff}$, $\log g$, and $\xi$ are summarized in Table \ref{tab:abundances_errs} for changes relative to the model.

\begin{table}
\tiny
\centering
\setlength{\tabcolsep}{1pt}
\renewcommand{\arraystretch}{0.9}
\caption{The influence of model atmospheric parameter uncertainties on the precision of the derived elemental abundances for four metal-poor stars.}
\label{tab:abundances_errs}
\centering
\begin{tabular}{l|cccc|l|cccc}
\hline
\multicolumn{5}{c|}{\bf HD\,22879} & \multicolumn{5}{c}{\bf HD\,144579} \\
\cline{1-10}

Species & $\Delta T_{\rm eff}$	&	$\Delta \log g$ 	&	$\Delta$[Fe/H] &  $\Delta \xi$ &Species & $\Delta T_{\rm eff}$	&	$\Delta \log g$ 	&	$\Delta$[Fe/H] &  $\Delta \xi$ \\
\cline{2-5}
\cline{7-10}
 ~    & (+75) & (+0.17) & (+0.08) & (+0.50)&  &(+165) & (+0.37) & (+0.12) & (+0.80) \\
  \hline
Na \,{\sc i} & +0.03 & -0.02 & +0.00 & -0.02 & Na \,{\sc i} & 0.10 & -0.06 & 0.02 & -0.02 \\
Mg \,{\sc i} & +0.07 & +0.00 & +0.00 & -0.08 & Mg \,{\sc i} & 0.22 & -0.13 & 0.03 & -0.11 \\
Al \,{\sc i} & +0.05 & +0.02 & +0.01 & -0.07 & Al \,{\sc i} & 0.08 &  0.02 & 0.00 & -0.01 \\
Si \,{\sc i} & +0.02 & +0.01 & +0.00 & -0.01 & Si \,{\sc i} & 0.00 &  0.04 & 0.02 & -0.01 \\
Ca \,{\sc i} & +0.05 & -0.03 & +0.00 & -0.06 & Ca \,{\sc i} & 0.13 & -0.13 & 0.02 & -0.06 \\
Sc\,{\sc ii} & +0.02 & +0.06 & +0.02 & -0.06 & Sc \,{\sc ii}& 0.00 &  0.13 & 0.04 & -0.06 \\
Ti \,{\sc i} & +0.07 & +0.00 & +0.00 & -0.05 & Ti \,{\sc i} & 0.19 & -0.10 & 0.01 & -0.10 \\
Ti\,{\sc ii} & +0.02 & +0.05 & +0.02 & -0.09 & Ti \,{\sc ii}& 0.00 &  0.13 & 0.04 & -0.07 \\
V \,{\sc i}  &  --   &  --   &  --   &  --   & V \,{\sc i}  & 0.22 &  0.00 & 0.00 & -0.09 \\
Cr \,{\sc i} & +0.05 & -0.01 & +0.00 & -0.07 & Cr \,{\sc i} & 0.17 & -0.12 & 0.02 & -0.10 \\
Cr\,{\sc ii} & -0.03 & +0.07 & +0.01 & -0.02 & Cr\,{\sc ii} &-0.06 &  0.12 & 0.03 & -0.07 \\
Mn\,{\sc i}  & +0.06 & -0.01 & +0.00 & -0.08 & Mn\,{\sc i}  & 0.16 & -0.11 & 0.03 & -0.11 \\
Fe \,{\sc i} & +0.06 & -0.01 & +0.01 & -0.06 & Fe \,{\sc i} & 0.13 & -0.07 & 0.03 & -0.08 \\
Fe\,{\sc ii} & -0.01 & +0.06 & +0.01 & -0.07 & Fe \,{\sc ii}& -0.07&  0.16 & 0.04 & -0.06 \\
Co \,{\sc i} & +0.04 & -0.02 & +0.00 & -0.11 & Co \,{\sc i} & 0.12 & -0.01 & 0.03 & -0.06 \\
Ni \,{\sc i} & +0.04 & +0.00 & +0.00 & -0.03 & Ni \,{\sc i} & 0.08 &  0.00 & 0.03 & -0.05 \\
Zn \,{\sc i} & +0.03 & -0.02 & +0.01 & -0.10 & Zn \,{\sc i} & 0.00 &  0.00 &-0.01 & -0.02 \\
Sr \,{\sc i} & +0.04 & +0.01 & +0.01 & -0.03 & Sr \,{\sc i} & 0.16 & -0.07 & 0.02 & -0.16 \\
Y \,{\sc ii} & +0.03 & +0.06 & +0.01 & -0.11 & Y \,{\sc ii} & 0.03 &  0.10 & 0.02 & -0.13 \\
Zr\,{\sc ii} & +0.04 & +0.07 & +0.02 & -0.10 & Zr\,{\sc ii} & 0.03 &  0.11 & 0.05 & -0.13 \\
Ba\,{\sc ii} & +0.03 & +0.02 & +0.02 & -0.12 & Ba\,{\sc ii} & 0.04 &  0.08 & 0.05 & -0.12 \\
Ce\,{\sc ii} & +0.03 & +0.05 & +0.02 & -0.02 & Ce\,{\sc ii} &  --  &  --   &  --  &  --   \\
Nd\,{\sc ii} & +0.04 & +0.07 & +0.02 & -0.03 & Nd\,{\sc ii} &  --  &  --   &  --  &  --   \\
\multicolumn{10}{c}{}\\
\cline{1-10}
\multicolumn{5}{c|}{\bf HD\,188510} & \multicolumn{5}{c}{\bf HD\,201891} \\
\cline{1-10}
	Species& $\Delta T_{\rm eff}$	&	$\Delta \log g$ 	&	$\Delta$[Fe/H] &  $\Delta \xi$ & Species& $\Delta T_{\rm eff}$	&	$\Delta \log g$ 	&	$\Delta$[Fe/H] &  $\Delta \xi$ \\
\cline{2-5}
\cline{7-10}
 ~    & (+58) & (+0.13) & (+0.08) & (+0.50) &  &(+90 K) & (+0.16) & (+0.08) & (+0.50) \\
  \hline
Na \,{\sc i} &   --    & --    &  --   &  --   & Na \,{\sc i} & --    &   --  & --    & --    \\
Mg \,{\sc i} &   +0.10 & +0.01 & +0.03 & -0.03 & Mg \,{\sc i} & +0.09 & +0.02 & +0.01 & -0.05 \\
Al \,{\sc i} &    --   &  --   &  --   &  --   & Al \,{\sc i} & +0.03 & +0.05 & +0.03 & -0.01 \\
Si \,{\sc i} &    --   &  --   &  --   &  --   & Si \,{\sc i} & --    &   --  & --    &    -- \\
Ca \,{\sc i} &   +0.04 & -0.02 & +0.01 & -0.01 & Ca \,{\sc i} & +0.06 & -0.01 & +0.01 & -0.03 \\
Sc\,{\sc ii} &   +0.01 & +0.03 & +0.01 & -0.01 & Sc \,{\sc ii}& +0.03 & +0.05 & +0.01 & -0.08 \\
Ti \,{\sc i} &   +0.06 & -0.02 & +0.00 & -0.03 & Ti \,{\sc i} & +0.09 & -0.01 & +0.01 & -0.05 \\
Ti\,{\sc ii} &   +0.02 & +0.02 & +0.02 & -0.04 & Ti \,{\sc ii}& +0.03 & +0.06 & +0.01 & -0.06 \\
V \,{\sc i}  &    --   &  --   &  --   &  --   & V \,{\sc i}  & --    &  --   & --    &   --  \\
Cr \,{\sc i} &   +0.07 & -0.02 & +0.00 & -0.04 & Cr \,{\sc i} & +0.08 & -0.01 & +0.01 & -0.06 \\
Cr\,{\sc ii} &    --   &  --   &  --   &  --   & Cr \,{\sc ii}& -0.01 & +0.04 & +0.01 & -0.06 \\
Mn \,{\sc i} &   +0.03 & -0.01 & +0.00 & -0.04 & Mn \,{\sc i} & +0.07 & -0.04 & +0.02 & -0.09 \\
Fe \,{\sc i} &   +0.06 & -0.02 & +0.01 & -0.03 & Fe \,{\sc i} & +0.07 & -0.01 & +0.01 & -0.06 \\
Fe\,{\sc ii} &   -0.01 & +0.06 & +0.01 & -0.01 & Fe \,{\sc ii}& +0.00 & +0.06 & +0.01 & -0.08 \\
Co \,{\sc i} &   +0.07 & -0.05 & +0.02 & -0.06 & Co \,{\sc i} & +0.11 & -0.04 & +0.01 & -0.21 \\
Ni \,{\sc i} &   +0.03 & +0.00 & +0.01 & -0.02 & Ni I         & +0.06 & +0.00 & +0.01 & -0.03 \\
Zn \,{\sc i} &   +0.00 & +0.02 & +0.00 & -0.02 & Zn I         & +0.03 & -0.02 & +0.01 & -0.05 \\
Sr\,{\sc i} &    --   &  --   &  --   &  --   & Sr\,{\sc i} &  --   & --    & --    &    -- \\
Y\,{\sc ii} &    --   &  --   &  --   &  --   & Y II         & +0.04 & +0.04 & +0.01 & -0.04 \\
Zr\,{\sc ii} &    --   &  --   &  --   &  --   & Zr\,{\sc ii} &  --   & --    & --    &    -- \\
Ba\,{\sc ii} &   +0.04 & +0.00 & +0.03 & -0.06 & Ba II        & +0.05 & +0.02 & +0.04 & -0.11 \\
\hline
\end{tabular}
\end{table}

\begin{figure} 
\includegraphics[width=0.47\textwidth]{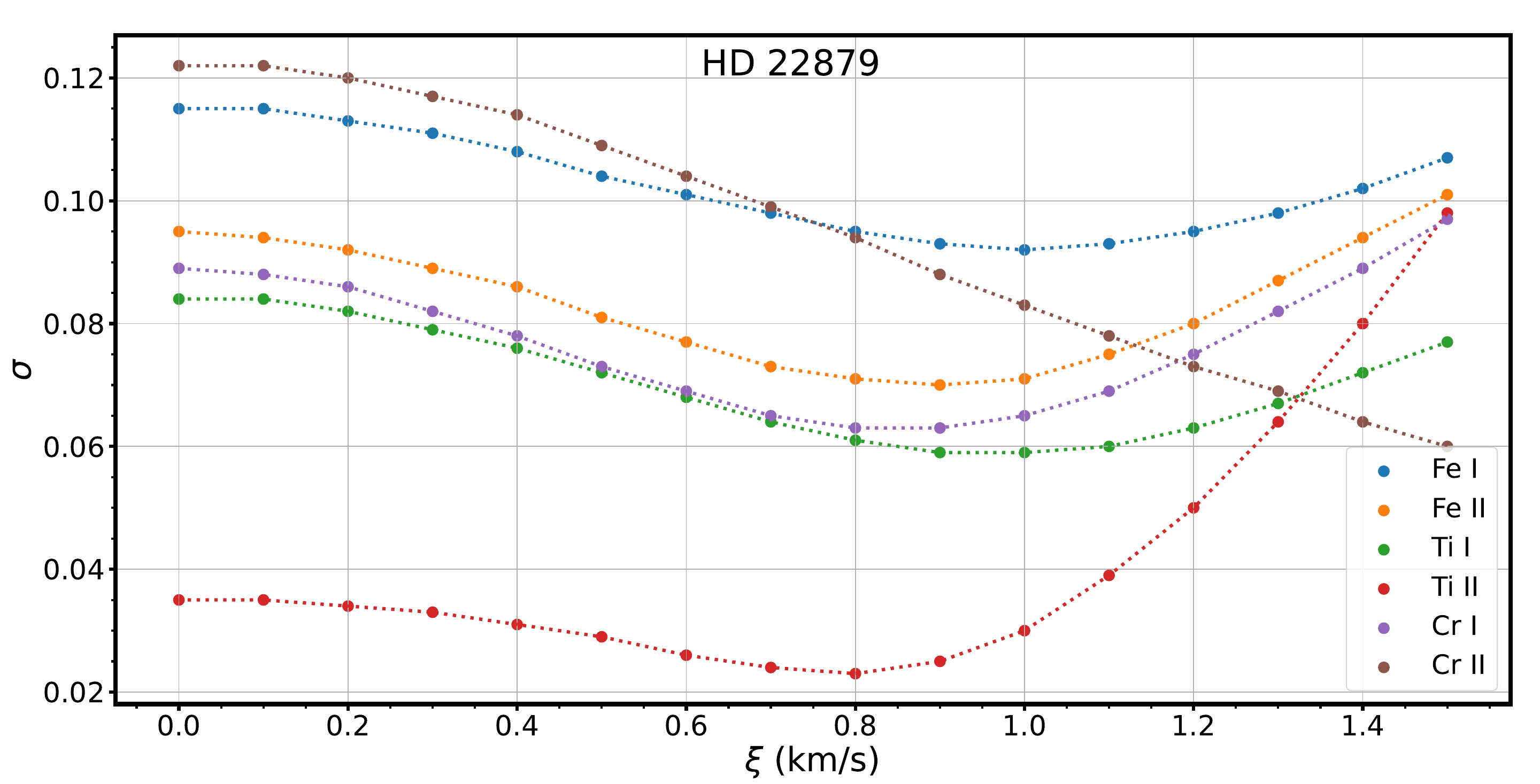}
\includegraphics[width=0.47\textwidth]{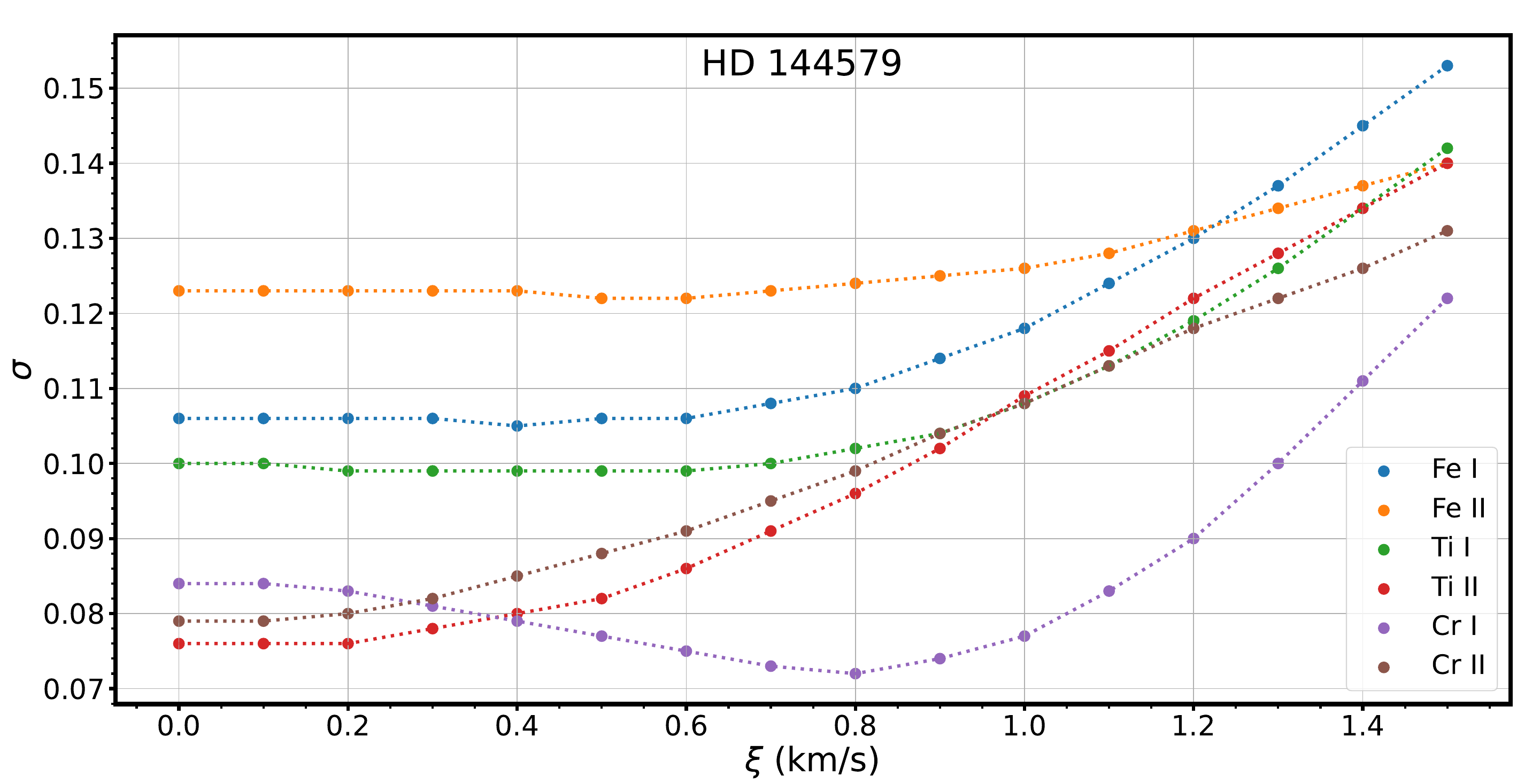}
\includegraphics[width=0.47\textwidth]{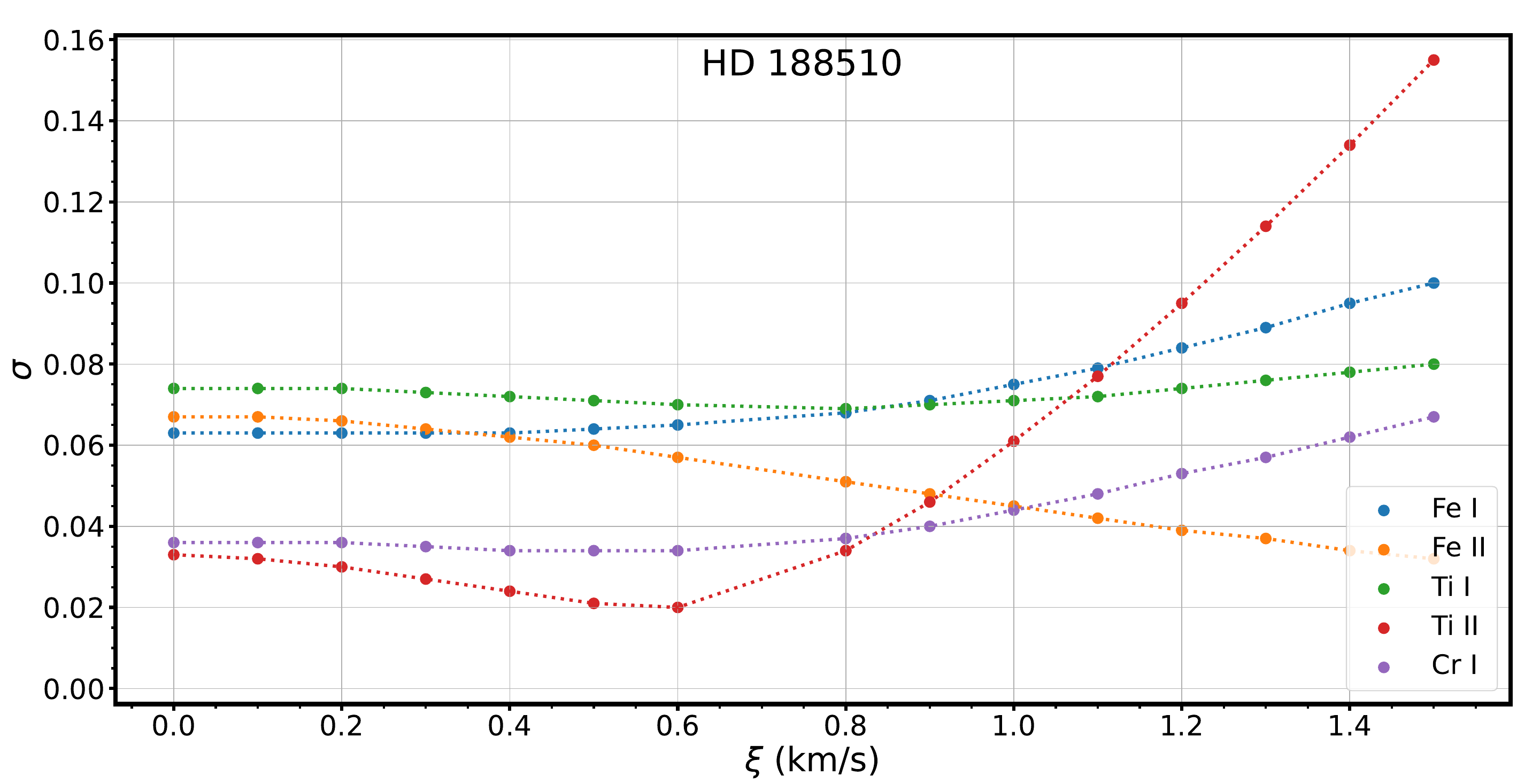}
\includegraphics[width=0.47\textwidth]{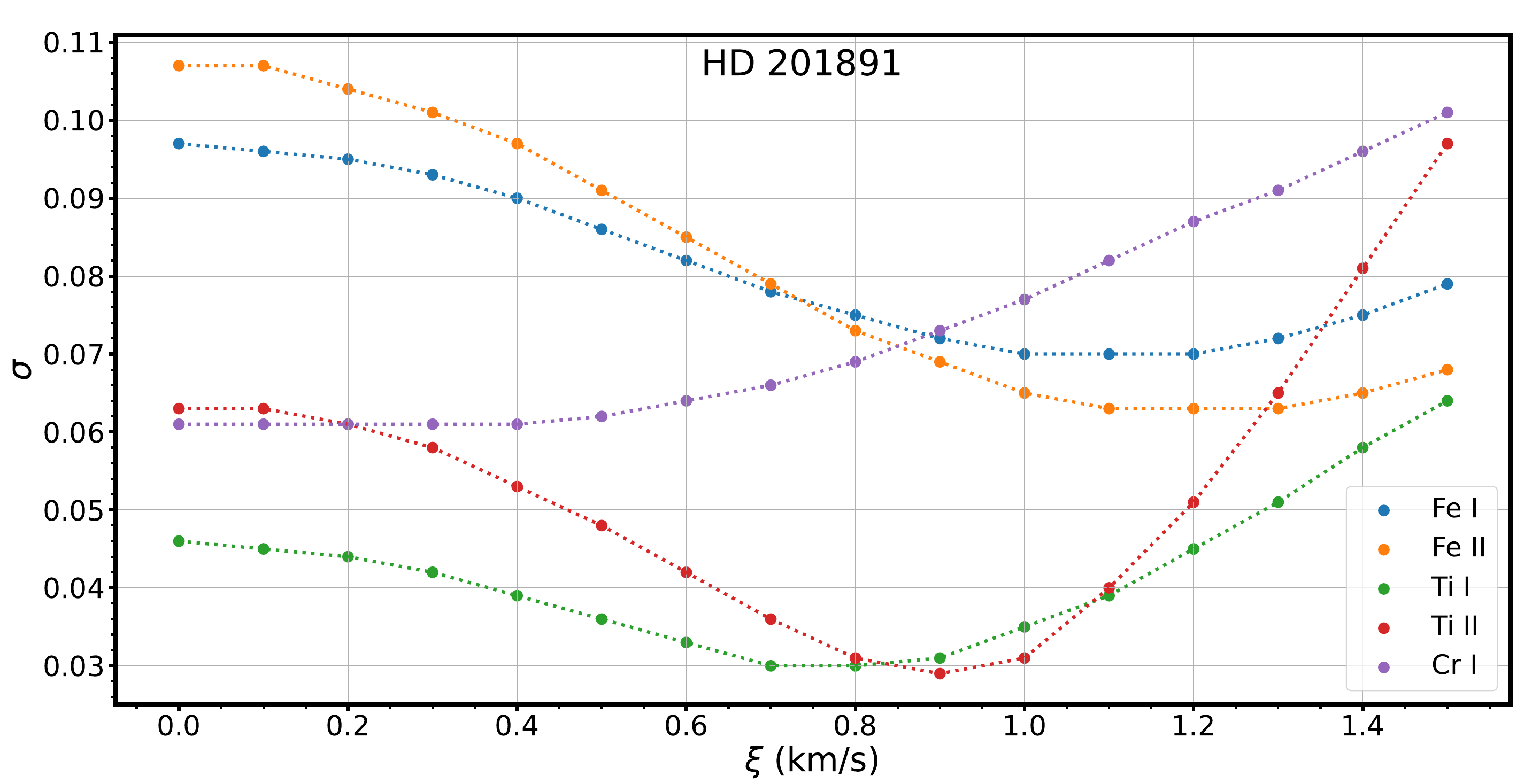}
\caption{The standard deviation of the Ti, Cr, and Fe abundances from the suite of Ti\,{\sc i}, Ti\,{\sc ii}, Cr\,{\sc i}, Cr\,{\sc ii}, Fe\,{\sc i}, and Fe\,{\sc ii} lines as a function of $\xi$.}
\label{fig:fourstar_abundances}
\end{figure}

%\newpage

\subsection{Stellar Age Estimation} \label{sec:stellarage}

In this study, determining the ages of the analyzed stars is important for understanding their origins. Therefore, stellar ages were estimated using a Monte Carlo Markov Chain (MCMC) method that employs maximum likelihood estimation and an isochrone grid based on two different input sets.

The isochrone grid used for this method was derived from version CMD 3.8\footnote{\url{http://stev.oapd.inaf.it/cgi-bin/cmd}} of the {\sc parsec} stellar isochrone library \citep{Bressan2012}. The properties of the obtained grid are as follows: in the $\log \tau$ space, steps of 0.05 were employed within the range of $6 \leq \log \tau \leq 10.13$, while in the metallicity space $Z$, an isochrone grid was created with steps of 0.0005 dex in the range of $0 \leq Z \leq 0.03$. Although the constructed grid is highly precise, it does not fully represent a continuous space. This limitation introduces uncertainties in age estimates and poses challenges for the walkers used in the MCMC method. To mitigate these issues, interpolation of intermediate points within the grid space is necessary. To achieve this, the Delaunay function from Python \texttt{Scipy} library \citep{Scipy} was used. This function enables continuous space within the input grid by performing three-dimensional interpolation among the desired input parameters. Consequently, the obtained solution is free from the uncertainties introduced by the step size of the input grid and provides the continuous space required for the Monte Carlo method.

In this process, a function that represents the desired observational parameters in the grid space was obtained using the Delaunay function, as expressed $I (Z, \tau, M) \xrightarrow[\text{Delaunay}]{ } \theta$. Using this function, stellar ages were determined for two different $\theta$ input parameter sets using the MCMC method and the maximum likelihood approach. The first consists of photometric parameters, where $Gaia$ is the photometry and spectroscopic iron abundance, and the second includes spectroscopic parameters, specifically the stellar atmospheric model parameters ($T_{\text{eff}}$, $\log g$, and [Fe/H]). These parameters were used to determine the most probable age on the isochrone grid by applying MCMC using the following maximum likelihood function:
\begin{equation}
\log \mathcal{L} = -\frac{1}{2} \sum_{i} \left( \frac{\theta_i^{\text{obs}} - \theta_i^{\text{model}}}{\sigma_i} \right)^2  
\end{equation}
where $\mathcal{L}$ represents the maximum likelihood function. The term $\theta_i^{\text{obs}}$ denotes the observed parameters, which include $M_{\rm G}$ and the $(G_{\rm BP} - G_{\rm RP})_0$ for the photometric input set, as well as the stellar atmospheric model parameters, $T_{\text{eff}}$, $\log g$, and [Fe/H], for the spectroscopic input set. The parameter $\theta_i^{\text{model}}$ corresponds to the model parameters derived from the interpolated isochrone grid, while $\sigma_i$ represents the observational uncertainties associated with each parameter. This function minimizes the squared differences between the observed and model parameters, weighted by their uncertainties, ensuring an optimal age estimation. The MCMC method explores the parameter space using this log-likelihood function to determine the most probable stellar age.

\begin{figure*}  
\centering
\includegraphics[width=0.8\textwidth]{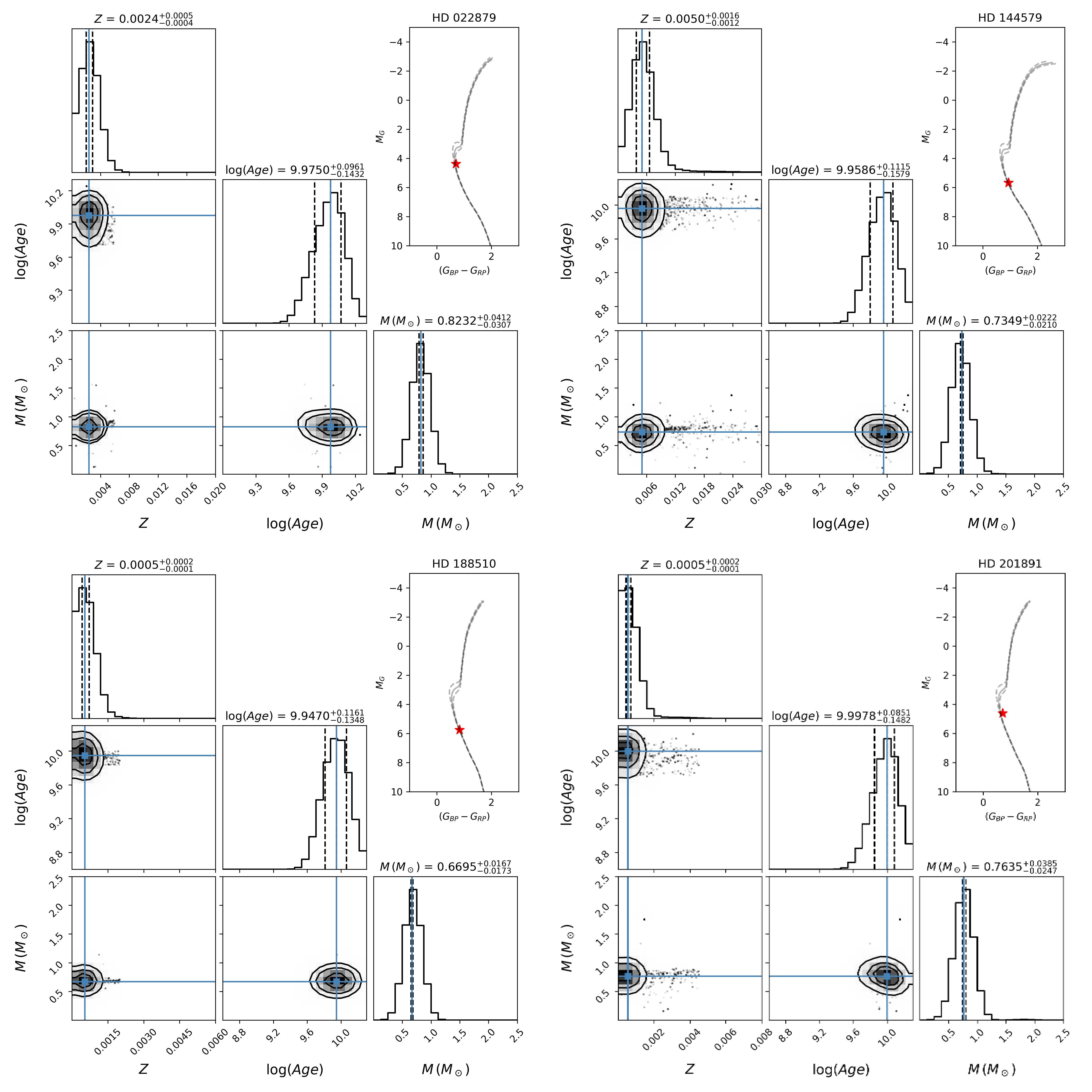}
\caption{Corner plot illustrating the two-dimensional joint and one-dimensional marginalized posterior distributions for the properties of four stars. The contours in the two-dimensional joint posterior distributions represent the confidence levels of 68\%, 90\%, and 95\%. In the one-dimensional posterior distributions, the vertical lines indicate the median and the 16th and 84th percentile confidence intervals. The $M_{\rm G} \times (G_{\rm BP} - G_{\rm RP})_0$ CMDs of the stars (red stars) are shown on the right side of the corner plots. The black solid lines are the {\sc parsec} isochrones derived from the median values of the posterior distributions, and black dashed lines indicate the calculated age errors of the stars.}

\label{fig:fourstar_photoastro}
\end{figure*}

To derive the final age estimates, the input parameters for each star were processed through 22 walkers and 5,000 iterations. The resulting posterior distributions were used to compute the median age, along with the associated uncertainties, defined as the range within one standard deviation above and below the median. This approach ensures a statistically robust estimation of stellar ages while accounting for the uncertainties inherent in the observational parameters.

\begin{figure*}    
\centering
\includegraphics[width=0.8\textwidth]{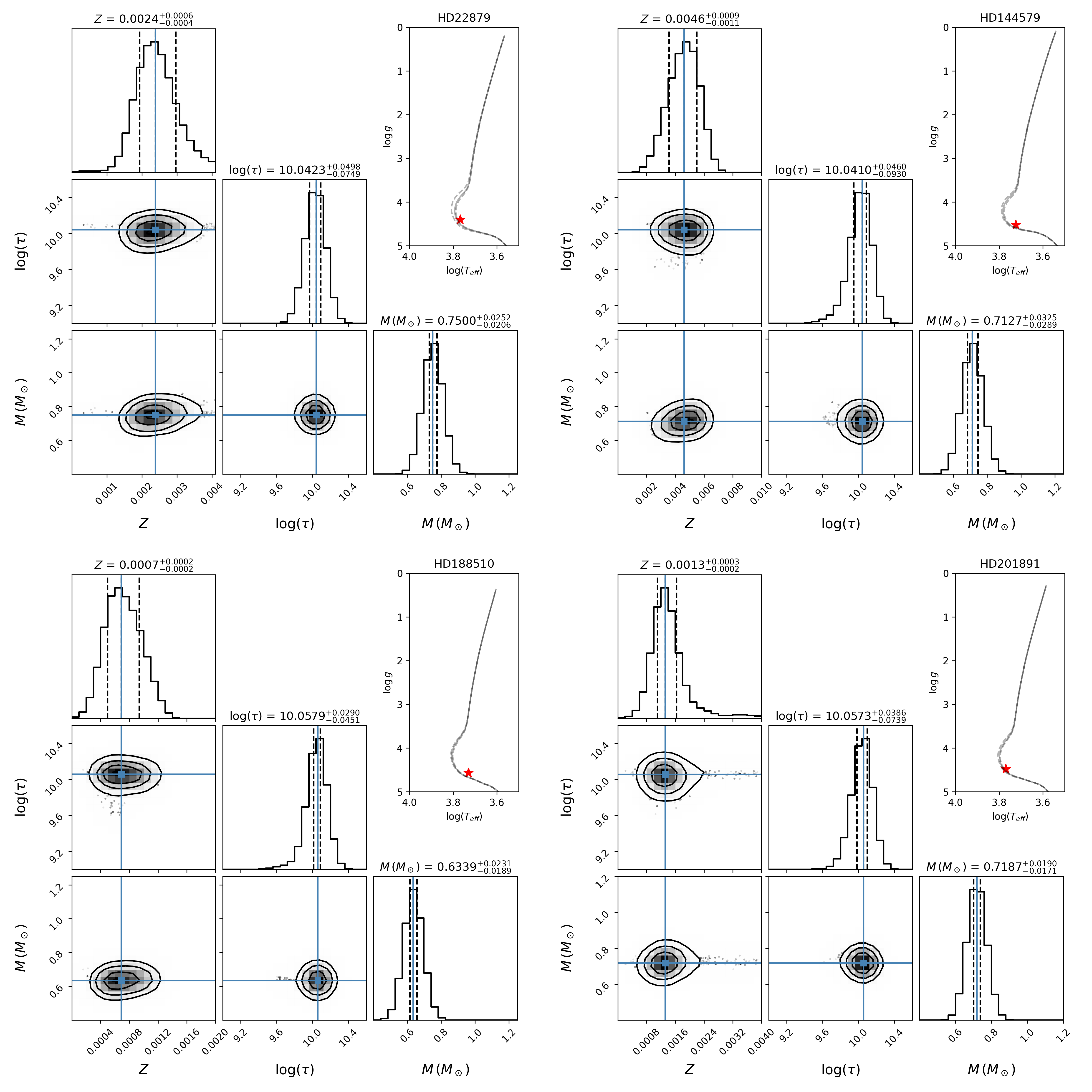}
\caption{Corner plot illustrating the posterior distributions for four stars, with confidence contours at 68\%, 90\%, and 95\%. One-dimensional distributions indicate the median and 16th/84th percentile intervals. The positions of the stars in the Kiel diagram are shown on the right-hand side of the panels. Definitions are as provided in the previous Figure \ref{fig:fourstar_photoastro}.}
\label{fig:SpecAge}
\end{figure*}

The results are summarized in Table~\ref{tab:main_results}. The posterior distributions derived from the photometric input parameter and best-fitting age results are visualized on the color-magnitude diagram in Figure~\ref{fig:fourstar_photoastro}, whereas the age estimates obtained from the atmospheric model parameters are presented in the same format as in Figure~\ref{fig:SpecAge}. The other parameters listed in Table~\ref{tab:main_results} were detected using interpolation on the isochrone grid based on the best-fitting solutions.

\begin{table*}
\centering
\caption{Model atmosphere parameters ($T_{\rm eff}$, $\log g$, [Fe/H]), mass ($M$), radius ($R$), and age ($\tau$) values of the four stars analyzed in this study, obtained using spectral energy distribution (SED), photometric and astrometric analysis (Phot. \& Astro), and spectral analysis   (Spec.) methods. Literature median values (Lit.) are included for comparison.}
\label{tab:main_results}
\begin{tabular}{c|c|cccc}
\hline
Parameter                            & Method & HD\,22879     & HD\,144579    & HD\,188510    & HD\,201891    \\ \hline
\multirow{4}{*}{$T_\text{eff}$ (K)}  & SED    & 5985 $\pm$ 30    & 5263 $\pm$ 21    & 5580 $\pm$ 26    & 5886 $\pm$ 20    \\
                                     & Phot. \& Astro. & 6332 $\pm$ 50        & 5501 $\pm$ 40         & 5766  $\pm$ 45       & 6139  $\pm$ 40       \\
                                     & Spec.      & 5855 $\pm$ 110   & 5300 $\pm$ 160   & 5370 $\pm$ 60   & 5880 $\pm$ 90   \\
                                     & Lit.    & 5848 $\pm$ 90    & 5276 $\pm$ 64    & 5489 $\pm$ 143   & 5902 $\pm$ 80    \\ \hline
\multirow{4}{*}{$\log g$ (cgs)} & SED    & 4.50 $\pm$ 0.03  & 4.53 $\pm$ 0.04  & 4.74 $\pm$ 0.03  & 4.51 $\pm$ 0.03  \\
                                     & Phot. \& Astro.     & 4.30 $\pm$ 0.08        & 4.58 $\pm$ 0.10        & 4.65  $\pm$ 0.05       & 4.60 $\pm$ 0.08        \\
                                     & Spec.      & 4.40 $\pm$ 0.18  & 4.52 $\pm$ 0.37  & 4.57 $\pm$ 0.13  & 4.48 $\pm$ 0.18  \\
                                     & Lit    & 4.24 $\pm$ 0.22  & 4.41 $\pm$ 0.24  & 4.44 $\pm$ 0.32  & 4.34 $\pm$ 0.17  \\ \hline
\multirow{4}{*}{[Fe/H] (dex)}        & SED    & -0.78 $\pm$ 0.03 & -0.63 $\pm$ 0.03 & -1.48 $\pm$ 0.03 & -1.02 $\pm$ 0.04 \\
                                     & Phot. \& Astro.     & -0.81  $\pm$ 0.07      & -0.48 $\pm$ 0.08        & -1.44  $\pm$ 0.07      & -1.45  $\pm$ 0.07      \\
                                     & Spec.      & -0.86 $\pm$ 0.08 & -0.55 $\pm$ 0.12 & -1.60 $\pm$ 0.07 & -1.15 $\pm$ 0.07 \\
                                     & Lit    & -0.84 $\pm$ 0.08 & -0.47 $\pm$ 0.05 & -1.60 $\pm$ 0.19 & -1.04 $\pm$ 0.12 \\ \hline
\multirow{3}{*}{$M$ ($M_\odot$)}     & SED    & 0.79 $\pm$ 0.02  & 0.72 $\pm$ 0.02  & 0.66 $\pm$ 0.02  & 0.87 $\pm$ 0.01  \\
                                     & Phot. \& Astro.     & 0.82 $\pm$ 0.04        & 0.73 $\pm$ 0.02         & 0.67   $\pm$ 0.02       & 0.76 $\pm$ 0.04         \\
                                     & Spec.      & 0.75 $\pm$ 0.02  & 0.71 $\pm$ 0.03  & 0.63 $\pm$ 0.02        & 0.72  $\pm$ 0.02     \\ \hline
\multirow{3}{*}{$R$ ($R_\odot$)}     & SED    & 1.08 $\pm$ 0.01  & 0.79 $\pm$ 0.01  & 0.69 $\pm$ 0.01  & 1.01 $\pm$ 0.01  \\
                                     & Phot. \& Astro.     & 1.05  $\pm$ 0.02       & 0.72 $\pm$ 0.15         & 0.64  $\pm$ 0.06       & 0.85  $\pm$ 0.07       \\
                                     & Spec.      & 0.82  $\pm$ 0.03 & 0.71 $\pm$ 0.11 & 0.61 $\pm$ 0.05 & 0.76 $\pm$ 0.08       \\ \hline
\multirow{3}{*}{$\tau$ (Gyr)}           & SED    & 13.08 $\pm$ 1.95 & 12.63 $\pm$ 2.79 & 12.54 $\pm$ 3.57 & 13.15 $\pm$ 2.13 \\
                                     & Phot. \& Astro.     & 9.44 $\pm$ 1.25       & 9.09 $\pm$ 1.35         & 8.85  $\pm$ 1.20      & 9.95 $\pm$ 1.30       \\
                                     & Spec.      & 11.02 $\pm$ 1.15 & 10.99 $\pm$ 1.15 & 11.42 $\pm$ 1.09 & 11.41 $\pm$ 1.15 \\ \hline 
\end{tabular}%
\end{table*}

\subsection{Kinematic and Dynamic Orbit Analyses} \label{sec:kinematicmethod}

The space velocity components of the four selected metal-poor stars, HD\,22879, HD\,144579, HD\,188510, and HD\,201891, were computed using the algorithm proposed by \citet{Johnson1987} for the J2000 epoch. $U$, $V$, and $W$ represent the space velocity components of each star relative to the sun. The transformations adopt a right-handed coordinate system, where $U$ is positive towards the Galactic center, $V$ is positive in the direction of Galactic rotation, and $W$ is positive towards the North Galactic Pole.

The input parameters for the calculations included equatorial coordinates ($\alpha$, $\delta$), radial velocities ($V_{\rm R}$), proper motion components ($\mu_{\alpha} \cos \delta$, $\mu_{\delta}$), and trigonometric parallaxes ($\varpi$), as shown in Table \ref{tab:Gaia_data}. Proper-motion components, trigonometric parallaxes, and radial velocities were obtained from the {\it Gaia} DR3 catalogue \citep{GaiaDR3}. Notably, the data quality from {\it Gaia} DR3 provides improved precision compared to earlier catalogues, ensuring robust input for velocity computations. A significant factor in determining the uncertainties in the derived space velocity components is the accuracy of the distance estimates, which are inversely related to the trigonometric parallax measurements. For the stars analyzed, the relative parallax errors from {\it Gaia} DR3 were uniformly low, typically below 0.001. This precision minimizes the propagation of distance errors in the space velocity components, yielding reliable kinematic results. 

To achieve precise space velocity components, the first-order correction for the Galactic differential rotation, as described by \citet{Mihalas1981}, was applied. The corrections for the $U$ and $V$ components of the space velocity were calculated as $-0.91 \le \Delta U~{\rm (km~s^{\rm -1})}\le 3.73$  and $-0.23 \le \Delta V~{\rm (km~s^{\rm -1})} \le 0.08$, respectively, whereas the $W$ component remained unaffected by this first-order approximation. Following differential rotation correction, the space velocity components were adjusted for the peculiar motion of the Sun relative to the local standard of rest (LSR). The adopted values for the Sun’s peculiar velocity are ($U_{\rm \odot}$, $V_{\rm \odot}$, $W_{\rm \odot})_{\rm LSR}$ = (8.50$\pm$0.29, 13.38$\pm$0.43, 6.49$\pm$0.26) km s$^{\rm -1}$, as reported by \citet{Coskunoglu2011}.

\begin{figure}   
\centerline{\includegraphics[width=0.45\textwidth,clip=]{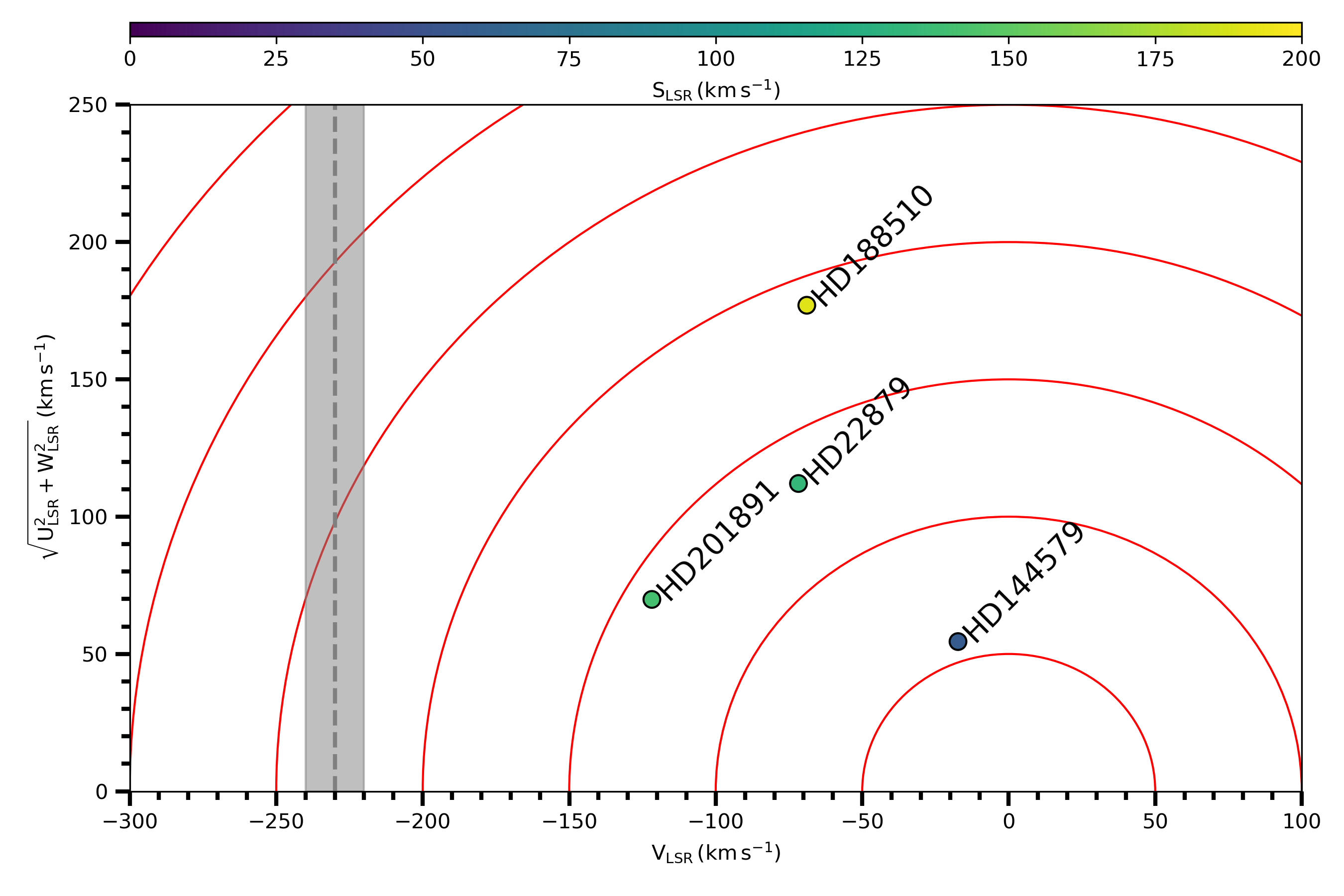}}
\caption{The positions of the stars on the Toomre diagram. The stars were color-coded according to their space velocity errors. The color scale in the upper panel shows the errors in total space velocity. The red curves in the figure represent velocity values of 50 km s$^{-1}$, whereas the dashed line in the gray region corresponds to a velocity of $S_{\rm LSR}$ = -230 km s$^{-1}$ \citep{Necib2022}.}
\label{fig:toomre}
\end{figure}

The total space velocity ($S$) of each star was calculated as the square root of the sum of the squares of its space velocity components. The uncertainties in the space velocity components ($U_{\rm err}$, $V_{\rm err}$, $W_{\rm err}$) were derived by propagating the errors in the proper-motion components, trigonometric parallaxes, and radial velocities using the methodology of \citet{Johnson1987}. The corrected space velocity components, total space velocities, and associated uncertainties for the four HPM stars—HD\,22879, HD\,144579, HD\,188510, and HD\,201891—are provided in Table \ref{tab:pop_parameters}. These results offer critical insights into the Galactic kinematics of these stars, and we applied the kinematic method outlined by \citet{Bensby2003}. This method assumes that the Galactic space velocity components follow a Gaussian distribution, expressed as

\begin{align}
P_i(U, V, W) =& k \exp \left( -\frac{U_{\text{LSR}}^2}{2\sigma_{i,U}^2} - \frac{(V_{\text{LSR}} - v_{i,a})^2}{2\sigma_{i,V}^2} - \frac{W_{\text{LSR}}^2}{2\sigma_{i,W}^2} \right)  \label{eq:P_i}, \nonumber \\  
k =&\frac{1}{(2\pi)^{3/2} \sigma_{i,U} \sigma_{i,V} \sigma_{i,W}},
\end{align}

where $\sigma_{\rm U}$, $\sigma_{\rm V}$, and $\sigma_{\rm W}$ represent the characteristic velocity distributions of distinct Galactic populations. These dispersions were as follows: 35, 20, and 16 km s$^{-1}$ for the thin disk (D); 67, 38, and 35 km s$^{-1}$ for the thick disk (TD); and 160, 90, and 90 km s$^{-1}$ for the halo (H). The term $V_{\rm asym}$ denotes the asymmetric drift velocity, with values of $-15$, $-46$, and $-220$ km s$^{-1}$ for the thin disk, thick disk, and halo, respectively, \citep{Bensby2003, Bensby2005}.
The probability that a star belongs to one Galactic population relative to another is calculated as the ratio of their Gaussian distribution functions (Equation \ref{eq:P_i}) scaled by the local space density ratios of the respective populations. For each star, the relative probabilities of belonging to a specific Galactic population were computed using the following equations:
\begin{equation} TD/D = \frac{X_{\rm TD}}{X_{\rm D}}\times\frac{f_{\rm TD}}{f_{\rm D}},~~~TD/H = \frac{X_{\rm TD}}{X_{\rm H}} \times \frac{f_{\rm TD}}{f_{\rm H}}, \label{eq:TD} \end{equation}
where $X_{\rm D}$, $X_{\rm TD}$, and $X_{\rm H}$ are the local space densities of the thin disk, thick disk, and halo, respectively. The values of 0.94, 0.06, and 0.0015 were determined based on the methodology described by \citet{Karaali2004, Bilir2006a, Bilir2006b, Bilir2006c, Cabrera-Lavers2007, Bilir2008, iyisan2025}, which utilizes the $TD/D$ ratio to categorize Galactic population membership. According to this classification, stars are identified as high-probability members of a thin disk when $TD/D \leq 0.1$. Stars with $0.1 < TD/D \leq 1$ are considered low-probability thin-disk members, whereas $1 < TD/D \leq 10$ indicates a low-probability membership in the thick disk. Finally, stars with $TD/D > 10$ were classified as high-probability members of a thick disk. The computed $TD/D$ and $TD/H$ values for the stars analyzed in this study, summarized in Table \ref{tab:pop_parameters}, provide insights into their Galactic population classifications. A lower $TD/D$ ratio corresponds to a lower likelihood of thick-disk membership than thin-disk membership.

\begin{figure}      
\centering
\centerline{\includegraphics[width=0.45\textwidth,clip=]{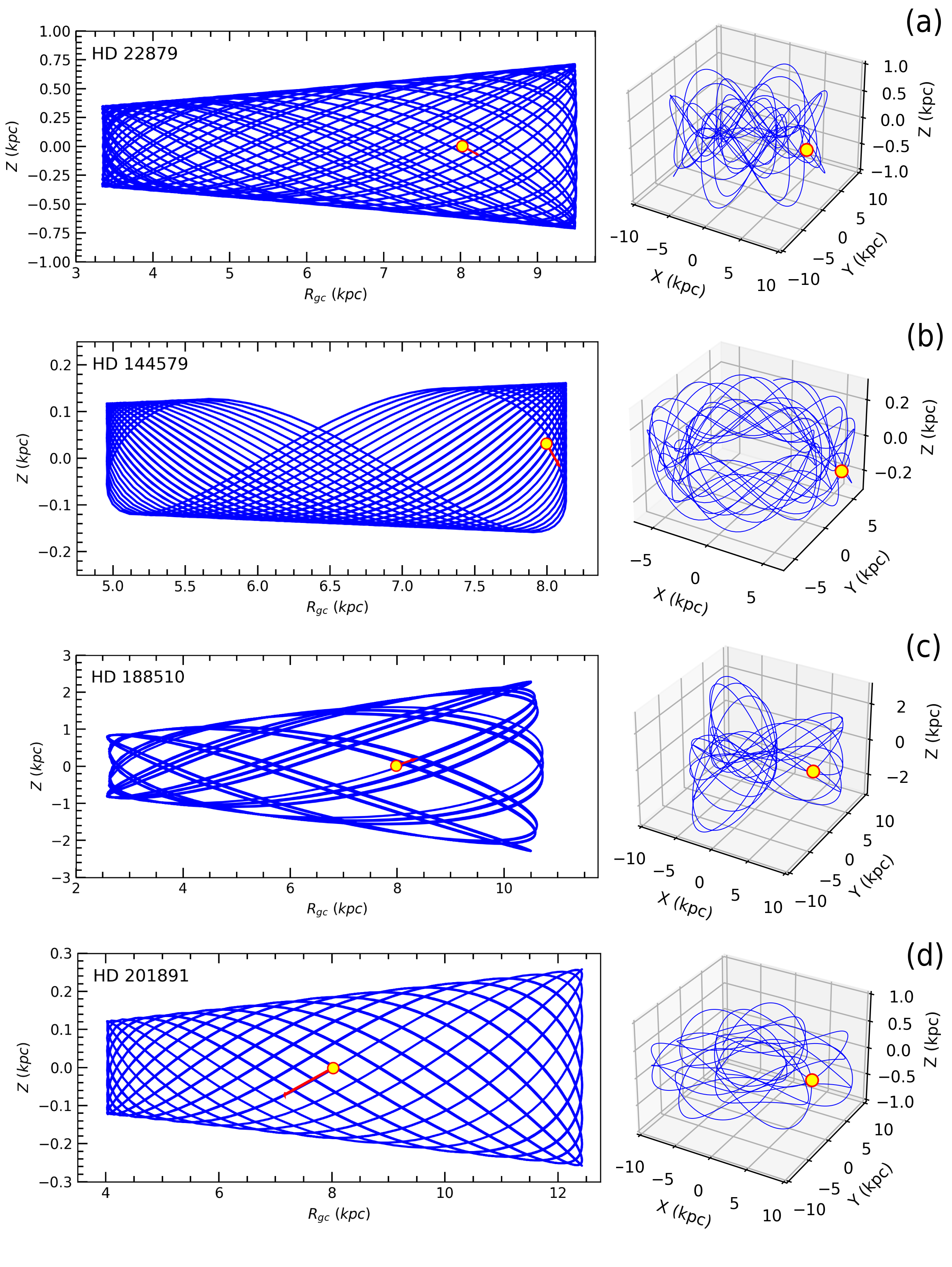}}
\caption{The orbits of the HD\,22879 (a), HD\,144579 (b), HD\,188510 (c), and HD\,201891 (d) around the Galactic center. The red-ringed yellow circles in the panels indicate the current positions of the stars, and the red arrows show their direction of motion.}
\label{fig:orbit}
\end{figure}
   
To calculate the Galactic orbital parameters of four metal-poor stars, we utilized the {\sc galpy} library, a Python-based tool for Galactic dynamics developed by \citet{Bovy2015}. In this study, we adopted a Galactic radius $R_{\rm gc}=$ 8 kpc \citep{Majewski1993} and a vertical distance of the Sun from the Galactic plane $Z_{\rm \odot}=$ 27 $\pm$ 4 pc following \citep{Chen2001}. The Milky Way was modeled using the {\sc galpy} potential {\it MWPotential2014}, which incorporates the gravitational influences of three components: the bulge, disk, and halo.
For the bulge component, we employed a spherical power-law density profile, as described by \citet{Bovy2015} and given by

\begin{equation} \rho (r) = A \left( \frac{r_{\rm 1}}{r} \right) ^{\alpha} \exp \left[-\left(\frac{r}{r_{\rm c}}\right)^2 \right] \label{eq:rho}. \end{equation} 

In this expression, $r_{\rm 1}$ and $r_{\rm c}$ are the reference and cut-off radii, respectively. Parameter $A$ denotes the amplitude of the potential in the mass density units, while $\alpha$ represents the inner power of the profile. For the Galactic disk, we adopted the potential model proposed by \citet{Miyamoto1975}, which is expressed as:
\begin{equation} \Phi_{\rm disk} (R_{\rm gc}, z) = - \frac{G M_{\rm d}}{\sqrt{R_{\rm gc}^2 + \left(a_{\rm d} + \sqrt{z^2 + b_{\rm d}^2 } \right)^2}} \label{eq:disc} \end{equation} where $z$ is the vertical distance from the Galactic plane, $ R_{\rm gc}$ is the radial distance from the Galactic center, and $ M_{\rm d}$ is the mass of the is the mass of the Galactic disk. The constants $a_{\rm d}$ and $b_{\rm d}$ represent the scale length and height of the disk, respectively. For the halo potential, we use the model proposed by \citet{Navarro1996}: 
\begin{equation} \Phi {\rm halo} (r) = - \frac{G M{\rm s}}{R_{\rm gc}} \ln \left(1+\frac{R_{\rm gc}}{r_{\rm s}}\right) \label{eq:halo} \end{equation} 
In this equation, $r_{\rm s}$ and $M_{\rm s}$ correspond to the scale radius and mass of the Galactic dark matter halo, respectively. The orbital trajectories of the four metal-poor stars around the Galactic center were computed using time steps of 1 Myr, with a total integration period of 13 Gyr. For the orbital parameter calculations, we used the same input data as those used to determine the space-velocity components. The apo- and peri-galactic distances ($R_{\rm a}$ and $R_{\rm p}$), the mean galactocentric distance, defined as $R_{\rm m}=(R_{\rm a}+R_{\rm p})/2$, along with eccentricity ($e_{\rm p}$), and the maximum distance from the Galactic plane ($Z_{\rm max}$) were obtained for each star. Eccentricity of the Galactic orbit was calculated using the following equation: $e_{\rm p}=(R_{\rm a} - R_{\rm p})/(R_{\rm a} + R_{\rm p})$. The calculated $Z_{\rm max}$ values were nearly identical to the axisymmetric model applied to the Galactic potential solutions. The resulting orbital parameters for the stars computed using the {\sc galpy} code are listed in Table \ref{tab:pop_parameters}. In addition, the Galactic orbits of the four stars projected onto the $X-Y$ and $X-Z$ planes are illustrated in Figure \ref{fig:orbit}, illustrating their vertical distance from the Galactic plane ($Z$) and their distance from the Galactic center ($R_{\rm gc}$) \citep[e.g.][]{ Yontan_2022, Tasdemir2023, Yontan2023a, Yontan2023b, Yucel2024}. The closed orbits were reconstructed by integrating the star’s present-day astrometric and radial velocity data backward in time along with the estimated age. Left panels display the motion of these stars within the Galaxy as a function of their distance from $R_{\rm gc}$ and $Z$, respectively. The right panels illustrate the variation in the stars' Galactic positions over time in terms of their $X$, $Y$, and $Z$ coordinates \citep[e.g.][]{Elsanhoury2024, Tasdemir2025, Haroon2025}.

\section{Summary and Discussion} \label{sec:summary}

This study investigated the astrophysical and chemical properties of four metal-poor main-sequence stars in the solar neighborhood using several independent methods: SED, photometric, astrometric, and high-resolution spectroscopic analyses. The model atmosphere parameters, fundamental stellar properties, and Galactic kinematics were derived using data from {\it Gaia} DR3 and the PolarBase library. The consistency of these methods was evaluated, and population classifications and potential escape scenarios from GCs were examined to determine the Galactic origins of the stars.

\begin{figure} 
\centering
\centerline{\includegraphics[width=0.45\textwidth]{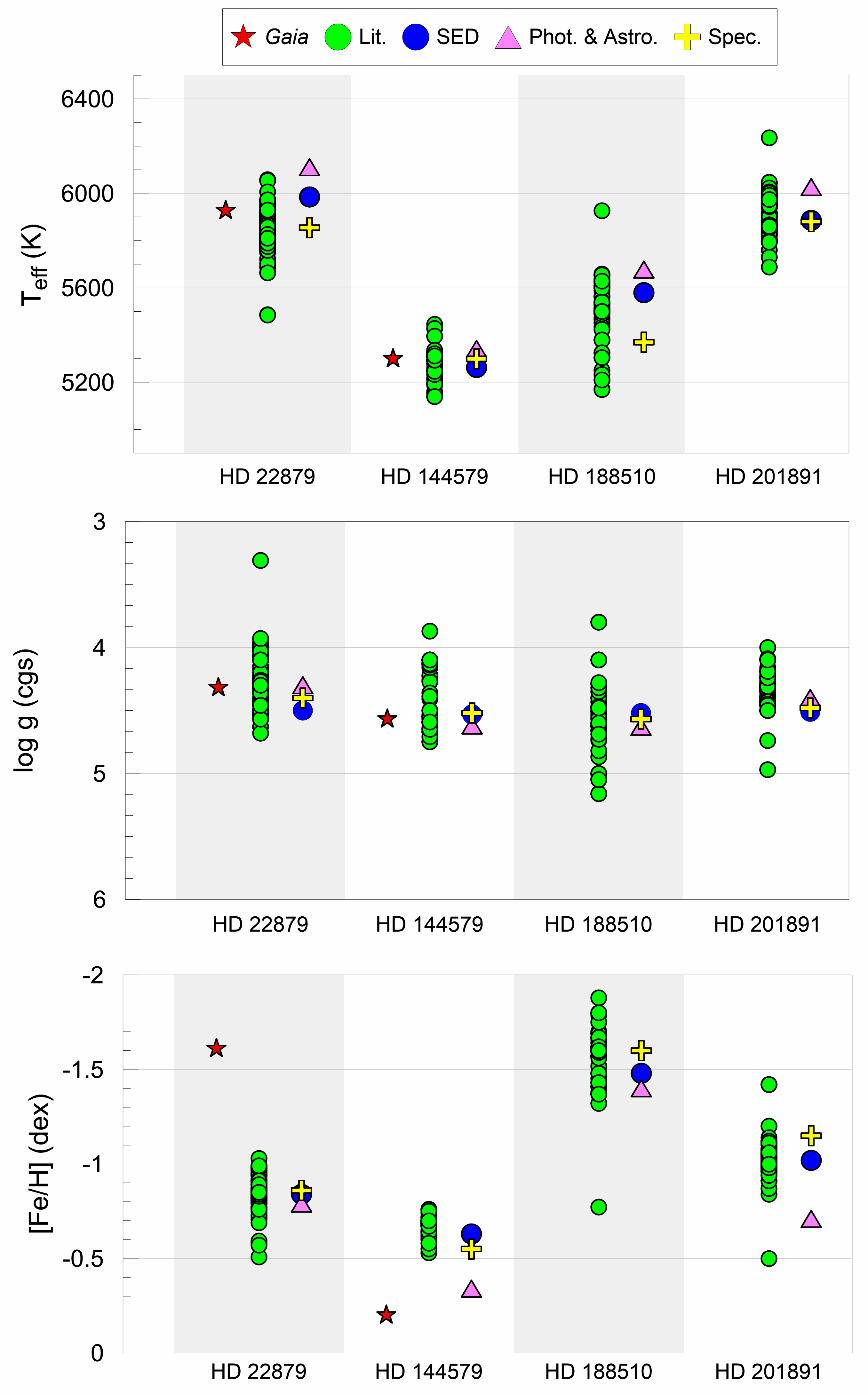}}
\caption{Comparison of stellar atmospheric parameters derived from \textit{Gaia} catalogue data and three different methods with literature values. The red star symbols represent \textit{Gaia} results, pink triangles indicate photometric and astrometric analysis, green circles correspond to literature values, blue circles denote SED analysis results, and yellow plus signs show parameters determined via spectroscopic analysis.}
\label{fig:literature}
\end{figure}

To evaluate the accuracy and precision of the model atmospheric parameters derived for the four metal-poor main-sequence stars analyzed in this study, previous spectroscopic studies were examined. Most studies have focused on chemical abundance, Galactic evolution, and stellar populations. However, no prior work, except for \citet{Sahin2020} and \citet{Marismak2024}, has combined multiple techniques to determine the fundamental astrophysical parameters of these stars, thus underscoring the originality of this study. Literature surveys revealed that HD\,22879, HD\,144579, HD\,188510, and HD\,201891 were identified in 139, 48, 57, and 83 spectroscopic studies, respectively, up to January 1, 2025. The ranges of atmospheric parameters reported in the literature are $5100 < T_\text{eff}\, (\text{K}) < 6250$, $3.3 < \log g \, (\text{cgs}) < 5.0$, and $-2 < \text{[Fe/H]} \, (\text{dex}) < -0.5$ (Figure~\ref{fig:literature}). However, several gaps and inconsistencies remain in the literature regarding atmospheric parameters and chemical abundances. For HD\,22879, previous studies have reported effective temperatures ($T_{\rm eff}$) ranging from 5486 to 6058 K and surface gravities ($\log g$) between 3.3 and 4.6 cgs, with metallicity ([Fe/H]) estimates varying from $-0.51$ to $-0.97$ dex. Additionally, the ionization balance between the Fe\,{\sc i} and Fe\,{\sc ii} lines has not been consistently achieved across studies, suggesting potential systematic errors in the adopted atmospheric models or line parameters. HD\,144579 has been characterized with $T_{\rm eff}$ values between 5139 K and 5446 K, $\log g$ from 4.1 to 4.8 cgs, and [Fe/H] from $-0.55$ to $-0.76$ dex. Despite its frequent use as a calibration star, there is a lack of consensus on its microturbulence, with values ranging from 0.6 to 1.0 km~s$^{-1}$. HD\,188510, with reported $T_{\rm eff}$ values of 5379–5618 K, $\log g$ of 4.3–4.7 cgs, and [Fe/H] of $-0.77$ to $-1.88$ dex, have been studied less extensively than other stars. HD\,201891 exhibits $T_{\rm eff}$ values between 5730 K and 6000 K, $\log g$ from 4.0 to 4.3 cgs, and [Fe/H] from $-1.10$ to $-0.90$ dex. This study derived model atmospheric parameters and fundamental properties (mass, radius, and age) of four stars using three independent methods. A comparison with the medians and standard deviations in the literature (Table~\ref{tab:main_results}) shows overall consistency. However, discrepancies were identified, such as the photometric and astrometric temperatures of HD\,22879 and the metallicity of HD\,144579, which fall outside the reported ranges. 

In the following subsections, the results of the spectroscopic studies of stars in the literature are compared with the results obtained in this study.

\subsection{Individual Stars in the Literature}

A comprehensive comparison with previous spectroscopic studies is essential to contextualize our findings within the broader framework of stellar abundance analyses. In this section, we study the consistency of our derived atmospheric parameters and chemical abundances with the values reported in the literature for each star in our sample. 

\subsubsection{HD 22879}

The first comprehensive abundance analysis of HD 22879 was conducted by \cite{edvardsson1993}, who derived the atmospheric parameters $T_{\rm eff} =$ 5826 K and $\log g =$ 4.27 cgs, reporting a metallicity of [Fe/H] = -0.84 dex and an $\alpha$-enhancement of [$\alpha$/Fe] = +0.25 dex. Our study derived atmospheric parameters consistent with these literature values, and the $\alpha$-element abundances presented here align with those reported by \cite{edvardsson1993}. Subsequent work by \cite{jofre2015} refined the metallicity to [Fe/H] = -0.86 $\pm$ 0.05 dex using high-resolution HARPS spectra, along with the updated parameters of $T_{\rm eff} =$ 5868 $\pm$ 89 K and $\log g =$ 4.27 $\pm$ 0.03 cgs.  

The $\alpha$-rich composition of HD 22879 was corroborated by GALAH DR4, which reported enhanced [X/Fe] ratios of 0.33 $\pm$ 0.01 dex for Mg and 0.21 $\pm$ 0.01 dex for Si.  

\cite{Nissen2024} recently analyzed HD 22879 through a differential spectroscopic approach relative to the Sun, focusing on Sc\,{\sc ii}, V\,{\sc i}, and Co\,{\sc i} abundances. Their adoption of revised photometric temperatures yielded $T_{\rm eff}$ values approximately 100 K higher than those in earlier studies, although their final spectroscopically derived $T_{\rm eff}$ and $\log g$ agreed closely with our determinations. \cite{Nissen2024} reported 3D-corrected metallicities of [Fe/H] = -0.86 dex and -0.85 dex, consistent with our result of [Fe/H] = -0.86 $\pm$ 0.08 dex. The authors applied 3D non-LTE corrections to the Mg\,{\sc i} 5711.1 \AA\, line-the sole Mg line analyzed, yielding [Mg\,{\sc i}/Fe] = 0.34 dex. Although uncertainties for individual abundances were not provided, their [X/Fe] ratios for Na, Si, Ca, Sc, Ti, and Ni align closely (within 0.03 dex) with those derived in this work. Notably, our Mg abundances were determined using three Mg\,{\sc i} lines that were evaluated for blending effects using the Solar Flux Atlas.  

\subsubsection{HD 144579}

The study by \cite{daSilva2015} analyzed the high-resolution spectra of FGK stars, encompassing both planet-hosting and non-planet-hosting stars, to determine photospheric parameters, masses, ages, and elemental abundances. A comparison between the elemental abundances derived in this study and those reported by \cite{daSilva2015} reveals both consistencies and notable discrepancies. Stellar atmospheric parameters show strong agreement between the two studies, suggesting that differences in derived abundances are not driven by systematic parameter uncertainties, but rather by atomic data variations, line selection, or methodological approaches.  

The [Mg/Fe] and [Mn/Fe] ratios align with uncertainties: this study reports [Mg/Fe] = 0.17 $\pm$ 0.10 dex and [Mn/Fe] = -0.05 $\pm$ 0.06 dex, consistent with literature values of [Mg/Fe] = 0.20 $\pm$ 0.08 dex and [Mn/Fe] = -0.09 $\pm$ 0.06 dex. However, significant discrepancies are observed for [Si/Fe], [Ca/Fe], [Ti/Fe], and [V/Fe]. For [Si/Fe], a shared silicon line at 5793.08 \AA\ exhibits a 0.06 dex offset in $\log gf$ values. In the case of [Ca/Fe], although both studies employed five common calcium lines, our result (0.32 $\pm$ 0.04 dex) exceeds the literature value (0.16 $\pm$ 0.09 dex). This divergence is partially attributable to systematic offsets in oscillator strengths ($\log gf$), which differ by 0.13 $\pm$ 0.11 dex between line lists. Such discrepancies in the atomic data can directly propagate into abundance offsets of comparable magnitudes.  

For [Ti/Fe], the 12 shared Ti\,{\sc i} lines exhibited $\log gf$ differences of 0.11 $\pm$ 0.08 dex (this study minus the literature), introducing a systematic bias that likely contributes to the abundance discrepancy. Conversely, the two common Ti\,{\sc ii} lines show excellent agreement in $\log gf$ values. The most pronounced disparity occurred in [V/Fe], where our measurement (0.52 $\pm$ 0.06 dex) substantially exceeded the value reported by \cite{daSilva2015} (0.21 $\pm$ 0.12 dex). This discrepancy is particularly notable because the studies employed entirely distinct vanadium lines; \cite{daSilva2015} relied on weak V lines sensitive to continuum placement or blending, whereas our analysis utilized V lines that are relatively stronger and validated for blending effects via the solar flux spectrum. The line lists and atomic data adopted in this study were calibrated to reproduce solar abundances in the solar flux spectrum.  

\cite{Tautvaivsien2021} analyzed high-resolution spectra of 506 solar-neighborhood F, G, and K stars, including HD 144579, to investigate trends in n-capture element-to-iron ratios. The derived atmospheric parameters for HD 144579 ($T_{\rm eff} =$ 5296 $\pm$ 48 K, $\log g =$ 4.59 $\pm$ 0.30 cgs, and [Fe/H] = -0.65 $\pm$ 0.11 dex) are consistent with those reported in Table \ref{tab:atmosphere_parameters}. Among the shared elements, strontium abundances agree within 0.1 dex, while yttrium abundances differ by only +0.01 dex. The barium abundance reported by \cite{Tautvaivsien2021} also fell within our reported uncertainty range.  

\cite{John2023} conducted a comprehensive search for planetary systems around HD 144579, a G8V-type star with high proper motion, using HARPS-N spectroscopic data. Kinematic analysis classified the star as a member of the thick-disk population, with no detected planetary companions. The derived atmospheric parameters ($T_{\rm eff} =$ 5296 $\pm$ 37 K, $\log g =$ 4.67 $\pm$ 0.08 cgs, [Fe/H] = -0.65 $\pm$ 0.04 dex), and $\alpha$-enhancement ([$\alpha$/Fe] = 0.26 dex) are consistent with the values reported in this study. 

\subsubsection{HD 188510}

The study by \cite{Fulbright2000} provides a comprehensive analysis of the abundances and kinematics of 168 metal-poor stars, including HD 188510, using high-resolution spectroscopy. The derived atmospheric parameters, $T_{\rm eff} = 5370 \pm 60$ K, $\log g = 4.57 \pm 0.13$ cgs, and [Fe/H] = $-1.60 \pm 0.07$ dex, align closely with those reported by \cite{Fulbright2000}, who performed the first detailed spectroscopic analysis of HD\,188510. Their results ($T_{\rm eff} = 5325 \pm 150$ K, $\log g = 4.6 \pm 0.2$ cgs, [Fe/H] = $-1.6 \pm 0.3$ dex) and $\alpha$-element abundances ([Mg/Fe] = 0.26, [Ca/Fe] = 0.27, [Ti/Fe] = 0.28, [Cr/Fe] = 0.04 dex) are consistent with our findings. However, the [Ni/Fe] and [Ba/Fe] ratios exhibit a notable discrepancy of $\approx$0.2 dex.  

\cite{Roederer2014} conducted detailed chemical analyses of 313 metal-poor stars, including HD\,188510, and reported no significant radial velocity variations for this star. Their derived surface gravity ($\log g = 4.10 \pm 0.27$ cgs) exhibits notable inconsistency with our results. As discussed in their Section 9.4.1, \cite{Roederer2014} explicitly noted that their [X/Fe] abundance ratios were systematically higher than those reported in previous surveys. To investigate this discrepancy, we employed the Stellar Parameters and Chemical Abundance Estimator (SP\_Ace) code \citep{Boeche2016}, which generates real-time spectral models using a Generalized Curve of Growth (GCOG) library. Our SP\_Ace analysis yielded $T_{\rm eff} = 5128 \pm 18$ K, $\log g = 4.48 \pm 0.07$ cgs, and [Fe/H] = $-1.65 \pm 0.02$ dex. These values confirm the robustness of our $\log g$ determination, but suggest an 80 K lower $T_{\rm eff}$ compared to \cite{Roederer2014}. Despite the disagreement in $\log g$ between the two studies, key elemental ratios, such as [Mg/Fe], [Ca/Fe], [Ti/Fe], and [Zn/Fe], reported by \cite{Roederer2014} align closely with our findings (within 0.1 dex), while [Mn/Fe] differs by only 0.05 dex. In their recent study, \cite{Roederer2025} reanalyzed the stellar parameters of metal-poor stars, including HD\,188510, and compared their results with those of \cite{Roederer2014}. They reported systematically warmer effective temperatures across all evolutionary tracks, thus resolving this discrepancy. The derived parameters ($T_{\rm eff} = 5549 \pm 36$ K, $\log g = 4.67 \pm 0.15$ cgs, and [Fe/H] = $-1.50 \pm 0.3$ dex) showed a marked increase in $T_{\rm eff}$ compared to their 2014 values.

\cite{Santos-Peral2023} characterized magnesium and calcium abundances in the X-shooter Spectral Library (XSL), including HD\,188510 (designated X0697). Using the ULySS package \citep{Koleva2009} tied to the PASTEL catalog \citep{Soubiran2016}, they derived $T_{\rm eff} = 5531$ K and $\log g = 4.29$ cgs. Their automated code GAUGUIN \citep{BIJAOUI201255, Recio-Blanco2016}, part of the $Gaia$-RVS pipeline, reported [Mg/Fe] = $0.51 \pm 0.03$ and [Ca/Fe] = $0.33 \pm 0.02$ dex. Notably, GAUGUIN iteratively determines the continuum placement within local spectral windows, a process that is sensitive to normalization intervals. No magnesium lines are shared between the studies, and our [Mg/Fe] = $0.25 \pm 0.07$ dex derives from the spectral synthesis of the 4571 and 5711 \AA\, lines, while \cite{Santos-Peral2023} analyzed ionized calcium triplet lines in the NIR. Our [Ca/Fe] = $0.24 \pm 0.05$ dex, based on optical lines (5512 \AA\ and 6493 \AA), highlights differences in line selection and spectral regions.  

\subsubsection{HD 201891}

\cite{Soubiran2024} have introduced an updated set of {\it Gaia} FGK benchmark stars, including HD\,201891, with derived parameters $T_{\rm eff}$ and $\log g$ consistent with those reported in this study. The authors highlighted the persistent scarcity of metal-poor stars in the solar neighborhood within their catalogue, emphasizing that the current version does not fully represent the metallicity distribution. In this context, the four metal-poor stars analyzed in this study, including HD\,201891, serve as critical benchmarks. \cite{Hawkins2016} previously expanded the {\it Gaia} Benchmark Stars catalogue with five metal-poor candidates (-1.3 $<$ [Fe/H] (dex) $<$ -1.0), designating HD\,201891 as a robust candidate for calibrating and validating stellar atmospheric parameters. Their compilation from the PASTEL database \citep{Soubiran2016} reported $T_{\rm eff} = 5883 \pm 68$ K, $\log g = 4.33 \pm 0.15$ cgs, and [Fe/H] = -1.05 $\pm$ 0.08 dex \citep[see Table \ref{tab:spectroscopic_data};][]{Hawkins2016}, with dispersions in the stellar parameters reflecting variations across literature. These values closely align with the results of this study. In addition, their IRTF-derived $T_{\rm eff} = 5948 \pm 80$ K and $\log g = 4.30 \pm 0.04$ cgs (based on stellar evolutionary tracks and angular diameter measurements) agree with our parameters within uncertainties.  

For this $\alpha$-enhanced metal-poor halo star, \cite{edvardsson1993} reported $T_{\rm eff} = 5867$~K, $\log g = 4.46$~cgs, and [Fe/H] = -1.06~dex, which are consistent with our results. Their derived [$\alpha$/Fe] = 0.27 $\pm$ 0.09 dex (averaged over Mg, Si, Ca, and Ti abundances) further corroborates the $\alpha$-rich composition identified in this study.  

\subsection{Galactic Population Types} \label{sec:population}

To analyze the radial and vertical kinetic energies of the stars, a Toomre diagram was generated, marking four metal-poor stars, as shown in Figure \ref{fig:toomre}. The space velocity ($S$) of these four stars is below 230 km s$^{\rm -1}$, which corresponds to the velocity of the Sun around the Galactic center \citep{Bovy2012}. According to \citet{Nissen2004}, stars belonging to a thin disk typically exhibit space velocities of $S < 60$ km s$^{\rm -1}$, whereas thick-disk stars exhibit a broader range, with velocities in the range of $80 < S ~(\text{km~s}^{\rm -1}) < 180$. The space velocity of halo stars in the vicinity of the Sun exceeds 180 km s$^{\rm -1}$. Based on the kinematic criteria of \citet{Nissen2004}, HD\,188510 stars in the same area were classified as members of the halo population. Under the kinematic criteria delineated in \citep{Bensby2003}, three stars in the present sample have $TD/D$ values greater than 10 (see Table \ref{tab:pop_parameters}), indicating their membership in the halo population. Conversely, HD\,144579, with a $TD/D$ value of less than 10, is classified as a thin-disk star.

In this study, Galactic orbital parameters were employed to examine the population membership of stars. For this purpose, the stars are plotted in the $Z_{\rm max} \times R_{\rm gc}$ plane, where $Z_{\rm max}$ increases with the vertical eccentricity, as shown in Figure \ref{fig:orbit}. Additionally, based on studies of Galactic structures \citep{Karaali2003, Ak2007a, Ak2007b, Bilir2008, Coskunoglu2012, Bilir2012, Plevne2015, Guctekin2019}, stars located within 2 kpc from the Galactic plane are classified as members of the thin disk, those between 2 and 5 kpc belong to the thick disk, and stars situated beyond 5 kpc are considered halo members. When analyzing the spatial distribution of the stars in the sample, three stars were identified as thin disk members and one star was categorized as a member of the thick disk.

\begin{table}
\caption{Parameters and population types including their kinematic, dynamical orbital and chemical properties of the four studied stars.}
\label{tab:pop_parameters}
\centering
\tiny
\setlength{\tabcolsep}{3.5pt}
\begin{tabular}{ll|cccc}
\hline
& Parameter     & HD\,22879     & HD\,144579        & HD\,188510    & HD\,201891        \\
\hline \hline
&     &   \multicolumn{4}{c}{Kinematic Parameters}   \\
 \hline
&$U_{\rm LSR}$ (km s$^{-1}$) & -106.62$\pm$0.16 & -54.50$\pm$0.11& -156.13$\pm$0.33 & ~~~66.38$\pm$0.12 \\
&$V_{\rm LSR}$ (km s$^{-1}$ )& ~-71.76$\pm$0.09 & -17.31$\pm$0.02& ~-68.91$\pm$0.28 & -121.79$\pm$0.15 \\
&$W_{\rm LSR}$ (km s$^{-1}$) & ~-34.56$\pm$0.15 & ~-0.95$\pm$0.08& ~~83.36$\pm$0.11 & ~~-21.78$\pm$0.10  \\
&  $TD$/$D$    & 66        & 0.01   & 1.62$\times 10^5$ & 1.31$\times 10^4$    \\
&  Population Type: & Halo    & Thin Disk   & Halo    & Halo     \\
 \hline
&     &   \multicolumn{4}{c}{Galactic Orbit Parameters}   \\
     \hline
\multirow{5}{*}{\rotatebox{90}{\underline{MW}}}&  $R_{\rm a}$ (pc)    & 9514      & 8132      & 10745      & 12427     \\
&  $R_{\rm p}$ (pc)    & 3349      & 4958      & 2637       & 4022      \\
&  $R_{\rm m}$ (pc)    & 6431      & 6545      & 6691       & 8225      \\
&  $Z_{\rm max}$ (pc)  & 712       & 161       & 2289       & 258       \\
&  $e$                 & 0.48      & 0.24      & 0.61       & 0.51      \\
\hline
\multirow{5}{*}{\rotatebox{90}{\underline{MWBS}}}&   \textbf{$R_{\rm a}$ (pc)}  & \textbf{9419} & \textbf{11583}   &   \textbf{12726}   &  \textbf{11729}   \\
 &  \textbf{$R_{\rm p}$ (pc)}    &   \textbf{2209}    & \textbf{4082}   &   \textbf{2139}     & \textbf{2501}   \\
 &  \textbf{$R_{\rm m}$ (pc)}    &   \textbf{5814}  &   \textbf{7832}   &   \textbf{7432}    &  \textbf{7115}  \\
 &  \textbf{$Z_{\rm max}$ (pc)}  &   \textbf{705}    &   \textbf{222}   &   \textbf{3027}  &   \textbf{341}\\
&   \textbf{$e$}                 &   \textbf{0.62}   &  \textbf{0.48}   &   \textbf{0.71} &  \textbf{0.65} \\
&  $T_{\rm p}$ (Myr)   & 189       & 182       & 207        & 245       \\  
&  Population Type:     & Thin Disk & Thin Disk & Thick Disk & Thin Disk \\
 \hline
&     & \multicolumn{4}{c}{Abundances}   \\
\hline
& {[}Mg/Fe] (dex)  & 0.16$\pm$0.09  & 0.17$\pm$0.10   & 0.25$\pm$0.07   & 0.23$\pm$0.09   \\ 
& {[}Si/Fe] (dex)  & 0.27$\pm$0.03  & 0.27$\pm$0.05   & --              &  --  \\ 
& {[}Ca/Fe] (dex)  & 0.26$\pm$0.03  & 0.32$\pm$0.04   &  0.24$\pm$0.05  & 0.18$\pm$0.03     \\ 
& {[}Ti/Fe] (dex)  & 0.23$\pm$0.03  & 0.42$\pm$0.04   & 0.17$\pm$0.03   &  0.20$\pm$0.05    \\
& {[}Fe/H] (dex)   & -0.86$\pm$0.08 & -0.55$\pm$0.12  & -1.60$\pm$0.07  & -1.15$\pm$0.07  \\
& Population Type:  & Thick Disk     & Thick Disk      & Halo            & Halo            \\ 
\hline
\end{tabular}
\noindent
\underline{MW}: MWPotential2014 \citep{Bovy2015}, \\
\underline{MWBS}: DehnenBarPotential \citep{Dehnen2000, Monari2016}, SpiralArmsPotential \citep{CoxGomes2002}.

\end{table}

To chemically distinguish the stellar populations, the sample analyzed by \citet{Bensby2014} was considered because of the insufficient number of stars in this study's dataset. Their method provides a homogeneous analysis of main-sequence stars of intermediate spectral types. Using spectroscopic data obtained with ESO's 1.5m, 2.2m, and 3.6m telescopes (SOFIN, FIES, UVES, and HARPS spectrographs), \citet{Bensby2014} conducted detailed abundance analyses of 714 F- and G-type dwarf stars near the Sun. The derived parameters included elements (e.g., O, Na, Mg, Al, Si, Ca, Ti, Cr, Fe, Ni, Zn, Y, and Ba), stellar ages, kinematics, and Galactic orbital properties.

\begin{figure} 
\centering
\centerline{\includegraphics[width=0.45\textwidth]{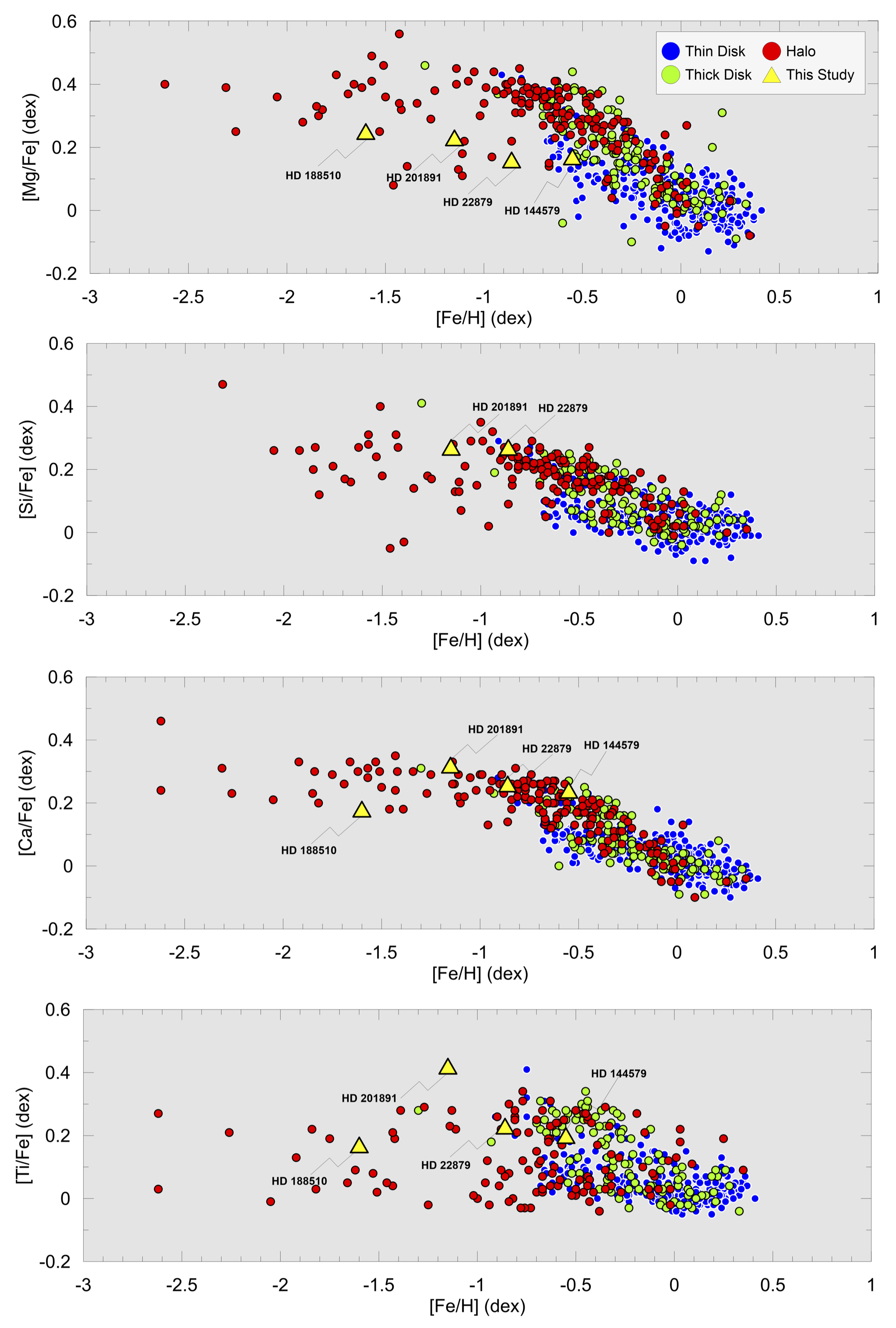}}
\caption{The chemical and kinematic classification of Galactic populations for the stars examined in this study, compared to 714 FGK stars from \citet{Bensby2014}. The red circle denotes a halo star, the green circle indicates a thick-disk star, the blue circle represents a thin-disk star, and the yellow triangle corresponds to the four stars analyzed in this study.}
\label{fig:bensby_pop}
\end{figure}

Among the stars examined in this study, HD\,22879, HD\,144579, HD\,188510, and HD\,201891 were identified in the catalogue by \citet{Bensby2014}. The chemical abundances [Mg/Fe], [Si/Fe], [Ca/Fe], [Ti/Fe], and [Fe/H], along with kinematic criteria were used to classify the star populations. Figure~\ref{fig:bensby_pop} shows stars mapped onto the [X/Fe] versus [Fe/H] planes, where X represents Mg, Si, Ca, and Ti. \citet{Bensby2014} classified stars with $TD/D \leq 1$ as thin disk, $1 < TD/D \leq 10$ as thick disk, and $TD/D > 10$ as halo stars. 

In the analyzed sample, stars with lower metallicity ([Fe/H]) tended to exhibit higher [Mg/Fe], [Si/Fe], [Ca/Fe], and [Ti/Fe] ratios, whereas metal-rich stars showed lower values of these abundances. The computed abundances for HD\,22879, HD\,144579, HD\,188510, and HD\,201891 were plotted for comparison. {In the study by \citet{Nieuwmunster2023}, it was shown that the [Mg/Fe] abundance of stars in the bulge region is comparatively lower than that of other Galactic components. The derived Mg abundance was reported to be, on mean, 0.14 dex lower than those in previous studies, and this decrease was suggested to be due to the increased sensitivity of the Mg lines used to non-LTE effects. It was further demonstrated that Mg abundance exhibits the steepest declining trend with metallicity and that the scatter observed in inner bulge stars at supersolar metallicities is more pronounced \citep{Matteucci1986, Lian2020}. This trend indicates that the contribution of Type Ia supernovae (SN Ia) to iron production became dominant, leading to relatively lower Mg abundances. Conversely, the contribution of $\alpha$ elements from the core-collapse supernovae (SN II) appears to be more limited than that of the outer disk and halo regions \citep{McWilliam2016}. Consequently, the relatively low [Mg/Fe] ratio observed in bulge stars implies that the star formation history in this region may have been influenced by the gas enriched in iron and that the long-term chemical evolution has been significantly shaped by SN Ia contributions \citep{Lian2020}. The results indicate that HD\,188510 and HD\,201891 are members of the halo population, whereas HD\,22879 and HD\,144579 occupy the transition region between the thick disk and halo.

\subsection{Dynamical Origins Beyond Tidal Escape: Motivation for Identifying Non-Coherently Ejected Stars}

While tidal stripping remains the canonical mechanism through which GCs lose stars---producing extended, coherent tidal tails---an increasing body of theoretical and observational evidence reveals that this process alone cannot account for all GC escapees \citep{Gnedin1999, galacticdynamics2, Kupper2012}. In particular, a population of non-coherently ejected stars---those not aligned with any identifiable stream---can arise from both internal and external dynamical mechanisms and may now be observed as chemically distinct field stars in the Galactic halo or thick disk.

The escape of stars from GCs occurs through a diverse set of mechanisms broadly categorized into gradual evaporation and dynamical ejection. Evaporation is primarily driven by two-body relaxation and tidal stripping, which act over long timescales to remove stars near the escape energy \citep{Weatherford2023}. In contrast, dynamical ejection encompasses rapid, high-velocity events, such as strong gravitational interactions (e.g., binary–single and binary–binary encounters), three-body binary formation (3BBF), and supernova-induced kicks. These interactions can impart escape velocities far exceeding those achieved through tidal evaporation \citep{Fragione2016, Weatherford2023, Weatherford2024}. This study highlights 3BBF, particularly in the presence of massive black holes, as a previously underappreciated but significant contributor to high-speed escapers. Additional channels include stellar evolution recoil, gravitational wave-driven mergers, and tidal disruption events.

These internal processes are especially effective in post-core-collapse clusters or those retaining a black hole subsystem, as the increased central densities and encounter rates enhance ejection activity \citep{Weatherford2023}. Additionally, asymmetric supernova kicks and gravitational-wave recoil from compact object mergers can further contribute to high-velocity ejections \citep{Fragione2016, Tomas2023}.

Equally significant are external dynamical perturbations experienced by GCs along their orbits in the Galactic potential. As clusters pass through the Galactic disk, interact with the bar or spiral arm resonances, or undergo pericentric passages near the bulge, time-varying tidal forces can rapidly heat the cluster and trigger an enhanced mass loss \citep{Gnedin1999}. Such processes can expel stars in directions that are not strictly aligned with Lagrange points, contributing to a diffuse and incoherent distribution of former cluster members. These stars may appear as spatial or kinematic outliers and are easily missed in stream-based searches.

Understanding the mechanisms responsible for stellar escape is therefore essential for interpreting the subsequent trajectories and present-day locations of these stars within the Galactic potential. The methodology employed in this study—integrating high-resolution chemical abundances, precise \textit{Gaia} DR3 astrometry, and full Galactic orbit modeling—provides a powerful tool for detecting such non-tidally ejected GC stars. By identifying outliers in orbital eccentricity, angular momentum, or vertical action, and contrasting them with their chemical signatures, it is possible to trace stars whose escape dynamics deviate from stream-like behavior.

Incorporating both coherent and stochastic escape mechanisms into our interpretation of stellar kinematics is essential for a complete understanding of the Milky Way’s assembly history. It reinforces the importance of dynamical context when identifying field stars of GC origin, especially in the absence of tidal structures.

\subsubsection{Galactic Possible Origins} \label{sec:origin}

The presence of high-proper-motion, metal-poor stars near the Sun is a phenomenon that \citet{Sahin2020} attributed to stars potentially ejected from globular clusters (GCs). To understand the Galactic origins of the two stars analyzed in this study, their orbital parameters as well as the orbits of GCs within the Milky Way were reconstructed using available observational data, including equatorial coordinates, distances, proper motions, and radial velocities sourced from the literature \citep{Baumgardt2019, Vasiliev2021}. Detailed kinematic and dynamic analyses were performed for 170 known GCs in our Galaxy. 

We utilized the Galactic potential models from \citet{Bovy2015}, implemented in the {\sc galpy} Python library, to calculate the orbital parameters of the GCs, along with symmetric ({\sc MWPotential2014}) and asymmetric ({\sc MWPotential2014 + DehnenBarPotential + SpiralArmsPotential}) Galactic potential models. The {\sc DehnenBarPotential} \citep{Dehnen2000} was implemented as a function, generalized to three dimensions following the approach outlined by \citet{Monari2016}, along with the implementation of the {\sc SpiralArmsPotential} from \citet{CoxGomes2002}. These calculations spanned from the present day to 13 billion years, considering orbital motions represented by 20 million data points. A time interval of 650 years between consecutive points was chosen after multiple trials to balance precision and computational efficiency. This approach provides sufficient resolution for capturing the interactions between stars and GCs, with multicore processors enabling faster computations. The estimated the Galactic orbital parameters of the analyzed stars are listed in Table \ref{tab:pop_parameters}.

The orbital parameters HD\,22879, HD\,144579, HD\,188510, and HD\,201891, including their equatorial coordinates, proper motion components, trigonometric parallaxes, and radial velocities, were derived from the \textit{Gaia} DR3 catalogue \citep{GaiaDR3}. The approach applied to these stars for orbital modelling follows the same methodology used for stars in GCs. The stellar trajectories were calculated to extend back 13 Gyr of alignment using the same procedure applied to the GCs orbital computations. However, the encounter probabilities were only evaluated for the periods after star formation, omitting any pre-formation epochs from the analysis.

\begin{figure*}   
\centerline{\includegraphics[width=0.95\textwidth,clip=]{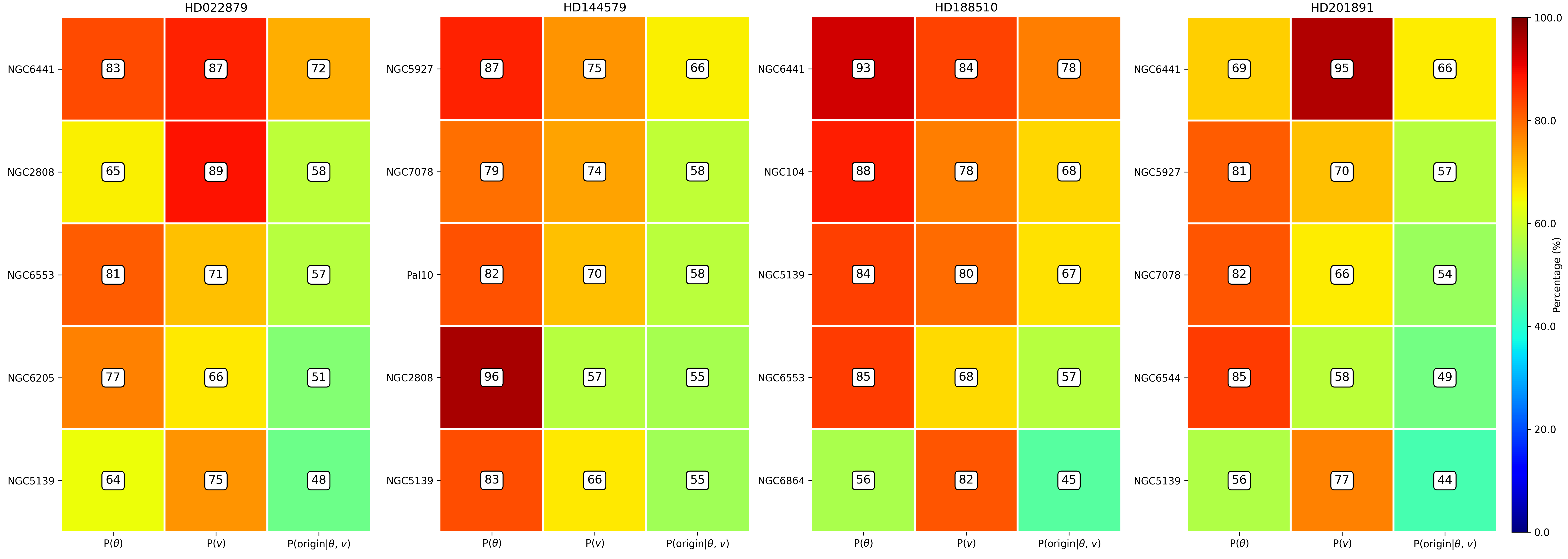}}
\caption{Likelihood of stars encountering various GCs, based on spatial and velocity parameters. The matrix displays three key probability values: spatial probability ($P(\theta)$), velocity probability ($P(\nu)$), and combined probability ($P({\rm origin | \theta, \nu})$) for each cluster. Colors in the matrix reflect the probability values, as indicated by the scale.}
\label{fig:origin}
\end{figure*}

In this study, the synchronous orbital paths of four stars were examined to determine their proximity to the centers of approximately 170 GCs \citep{Vasiliev2021}, specifically within the region defined by the five tidal radii. At each time step, the distance between the stars and GC centers was evaluated. If the calculated separation was found to be less than five tidal radii, an encounter was considered to have occurred. The encounter parameters, including the relative velocity ($\Delta \nu$) and position ($\Delta \theta$), were computed according to the following set of equations:
\begin{equation}
    \Delta \theta = \sqrt{(X_{\rm s}-X_{\rm GC})^2+(Y_{\rm s}-Y_{\rm GC})^2+(Z_{\rm s}-Z_{\rm GC})^2},
\end{equation}
\begin{equation}
    \Delta \nu = \sqrt{(U_{\rm s}-U_{\rm GC})^2+(V_{\rm s}-V_{\rm GC})^2+(W_{\rm s}-W_{\rm GC})^2}.
\end{equation}
where $\Delta \theta$ represents the spatial separation between the star and GC in the Cartesian coordinate system at a specific time $\tau$. Likewise, $\Delta \nu$ denotes the difference in the velocity components between the star and the GC. The position coordinates $X$, $Y$, and $Z$ represent the respective locations of the objects, while $U$, $V$, and $W$ indicate the corresponding space-velocity components. If multiple encounters occurred between the GC and star over several time steps, the probability for that time interval was derived by summing the probabilities of the individual encounters. In cases where encounters occurred at multiple time steps, the highest probability observed across these time intervals was selected for each GC and compared with the probabilities of encounters with other clusters. The probability of the position ($P(\theta)$) and velocity ($P(\nu)$) differences, assuming a Gaussian distribution over an extended time frame, was computed using the following equations \citep[see also][]{Marismak2024}:
\begin{equation}
    P({\rm \theta}) = \frac{1}{\sqrt{2\pi R_{\rm tidal}}}\times \exp\left(-\frac{(\Delta \theta)^2}{2\times R_{\rm tidal}}\right),
\end{equation}
\begin{equation}
    P({\rm \nu})= \frac{1}{\sqrt{2\pi V_{\rm escape}}}\times \exp\left(-\frac{(\Delta \nu)^2}{2\times V_{\rm escape}}\right).
\end{equation}
In this context, $R_{\rm tidal}$ represents five times the tidal radius of the cluster and $V_{\rm escape}$ denotes the escape velocity of the GC. These values are sourced from the 4th version of the Milky Way GC Database\footnote{\href{https://people.smp.uq.edu.au/HolgerBaumgardt/globular/}{https://people.smp.uq.edu.au/HolgerBaumgardt/globular/}}. For the joint evaluation of the spatial and velocity probabilities for each GC, the product of the two probability values, denoted by $P({\rm origin | \theta, \nu})$, was employed, and the following equation was applied:
\begin{equation}
    P({\rm origin | \theta, \nu})= P({\rm \theta}) \times P({\rm \nu}),
\end{equation}

The probabilities derived from the spatial and velocity comparisons of HD\,22879, HD\,144579, HD\,188510, and HD\,201891 with 170 GCs in the Milky Way \citep{Vasiliev2021}, along with their combined probability values, are illustrated in Figure \ref{fig:origin} as probability matrices that highlight the five most probable GCs. Table \ref{tab:origin} summarizes the encounter probabilities and provides an additional diagnostic framework for exploring the dynamic origins of four stars. These probabilities were determined through detailed kinematic analyses of stars and GCs. 

\begin{table*}
\setlength{\tabcolsep}{4pt}
\centering
\caption{The Galactic coordinates ($l$, $b$), position ($P(\Theta$)), velocity ($P(\nu)$) and their combined probability values ($P$(origin\textbar{}$\Theta$,$\nu$)) maximum distance away from the Galactic plane ($Z_{\rm max}$), iron ($\rm [Fe/H]$) abundances and ages ($\tau$), and population types (Pop) of the five most probable clusters under the scenario of the four stars leaving the GCs. The last column of the table contains the reference for [Fe/H] abundances and ages.}
\label{tab:origin}
\begin{tabular}{lcccccccccc} 
\hline
Cluster           & $l$      & $b$      & $P(\Theta$) & $P(\nu$) & $P$(origin$\mid$$\Theta$,$\nu$) & $Z_{\rm max}$ & $[\rm Fe/H]$       & $\tau$            & Pop  & Ref.   \\
                  & (°)      & ~(°)     & ~(\%)       & ~(\%)    & ~(\%)                             & (kpc)       & (dex)          & (Gyr)          &       &        \\ 
\hline
\textbf{NGC\,6441} & 353.532  & -5.005   & 83          & 87       & 72                                & 1.1         & -0.65          & 10.44          & Bulge & 1   \\
NGC\,2808          & 282.193  & -11.253  & 65          & 89       & 58                                & 2.2         & -1.32          & 10.93          & Disk  & 2   \\
NGC\,6553          & 005.250  & -3.023   & 81          & 71       & 57                                & 0.7         & -0.10          & 13.00          & Bulge & 3    \\
NGC\,6205          & 059.009  & 40.912   & 77          & 66       & 51                                & 6.2         & -1.48          & 13.49          & Halo  & 2    \\
NGC\,5139          & 309.102  & 14.968   & 64          & 75       & 48                                & 2.8         & -1.53          & 12.75          & Halo  & 4   \\ 
\hline
HD22879            & –        & –        & –           & –        & –                                 & –           & -0.86$\pm$0.08 & $10.02\pm1.15$ & –     & –      \\ 
\hline\hline
\textbf{NGC\,5927} & 326.604  & 4.860    & 87          & 75       & 66                                & 0.8         & -0.49          & 12.25          & Bulge & 5    \\
PAL\,10            & 052.436  & 2.725    & 82          & 70       & 58                                & 1.1         & -0.10          & –              & Disk  & 6    \\
NGC\,4372          & 300.993  & -9.884   & 72          & 59       & 43                                & 2.0         & -2.34          & 12.50          & Bulge & 7    \\
NGC\,104           & 305.895  & -44.889  & 45          & 91       & 41                                & 3.2         & -0.81          & 13.54          & Halo  & 2    \\
NGC\,1851          & 244.513  & -35.036  & 64          & 44       & 28                                & 11.5        & -1.14          & 12.27          & Halo  & 2    \\ 
\hline
HD144579           & –        & –        & –           & –        & –                                 & –           & -0.55$\pm$0.12 & $10.99\pm1.15$ & –     & –      \\ 
\hline\hline
NGC\,6441          & 353.532  & -5.006   & 93          & 84       & 78                                & 1.1         & -0.65          & 10.44          & Bulge & 2   \\
NGC\,104           & 305.895  & -44.889  & 88          & 78       & 68                                & 3.2         & -0.81          & 13.54          & Halo  & 2    \\
\textbf{NGC\,5139} & 309.102  & 14.968   & 84          & 80       & 67                                & 2.8         & -1.53          & 12.75          & Halo  & 5   \\
NGC\,6864          & 020.303  & -25.748  & 56          & 82       & 45                                & 12.0        & -1.16          & 9.98           & Halo  & 8   \\
NGC\,6569          & 000.481  & -6.681   & 69          & 56       & 39                                & 2.2         & -0.79          & 10.90          & Bulge & 9    \\ 
\hline
HD188510           & –        & –        & –           & –        & –                                 & –           & -1.60$\pm$0.07 & $11.42\pm1.09$ & –     & –      \\ 
\hline\hline
NGC\,6441          & 353.532  & -5.0060  & 69          & 95       & 66                                & 1.1         & -0.65          & 10.44          & Bulge & 2   \\
NGC\,5927          & 326.604  & 4.8600   & 81          & 70       & 57                                & 0.8         & -0.49          & 12.25          & Bulge & 5    \\
NGC\,7078          & 065.012  & -27.312  & 82          & 66       & 54                                & 4.5         & -2.36          & 13.28          & Halo  & 2    \\
\textbf{NGC\,6544} & 005.838  & -2.204   & 85          & 58       & 49                                & 1.1         & -1.44          & 12.00          & Bulge & 10  \\
NGC\,5139          & 309.102  & 14.968   & 56          & 77       & 44                                & 2.8         & -1.53          & 12.75          & Halo  & 5   \\ 
\hline
HD201891           & –        & –        & –           & –        & –                                 & –           & -1.15$\pm$0.07 & $11.41\pm1.15$ & –     & –      \\
\hline
\end{tabular}
\\
\noindent
(1) \citet{Zhang2012}
(2) \citet{Valcin2020}, (3) \citet{Montecinos2021}, (4) \citet{Baldwin2016}, (5) \citet{Usher2019}, (6) \citet{Bica2006}, (7) \citet{Kovalev2019}, (8) \citet{Kirby2016}, (9) \citet{Marsakov2019}, (10) \citet{Gran2021}.
\end{table*}

\begin{figure}
\centering
\includegraphics[width=0.95\linewidth]{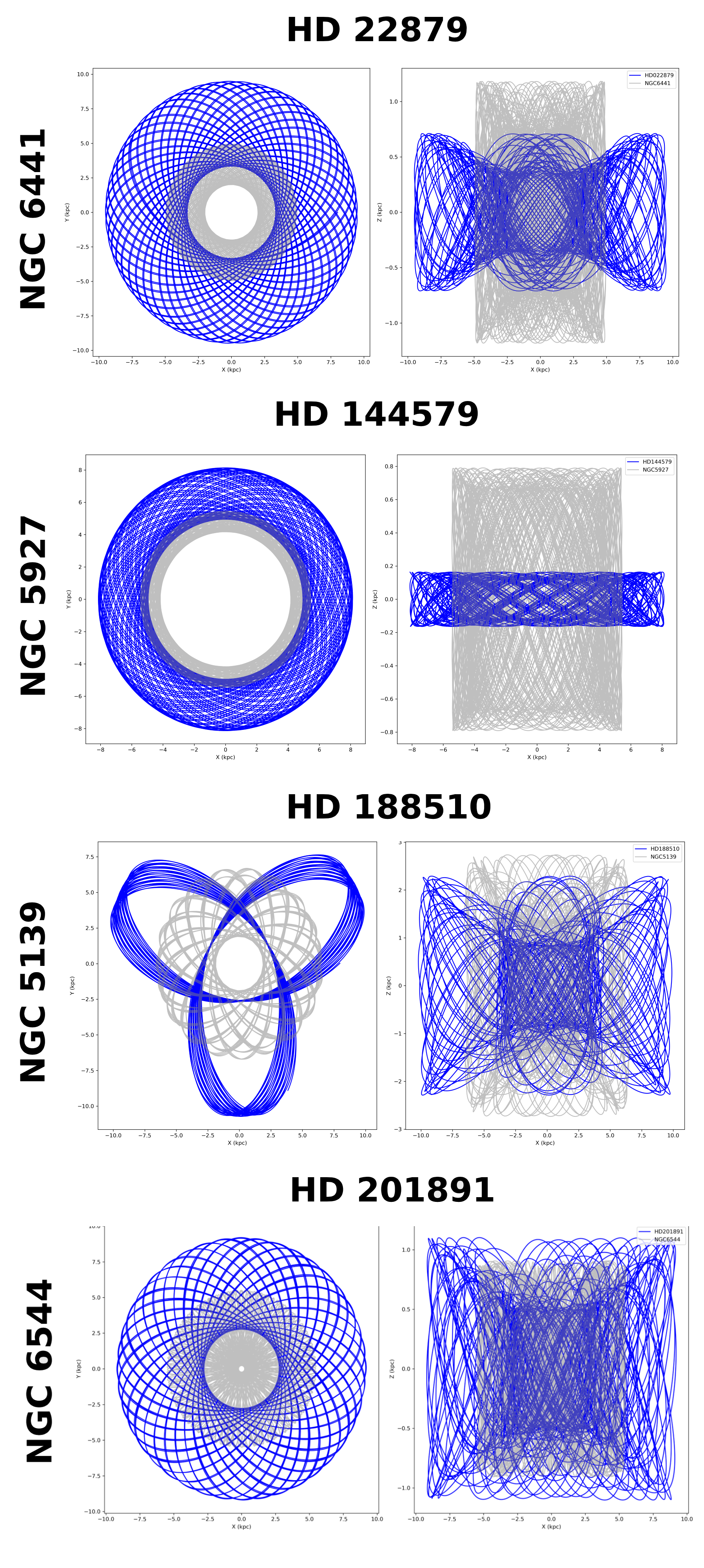}
\caption{The Galactic orbits on the Galactic $X-Y$ and $X-Z$ planes have been calculated for the GC of which the four stars studied are members with the highest probability. The blue line indicates the orbits of the stars, while the gray lines represent the orbits of the GCs.}
\label{fig:GC_Orbit}
\end{figure}

\begin{figure*}   
\centerline{\includegraphics[width=0.95\textwidth,clip=]{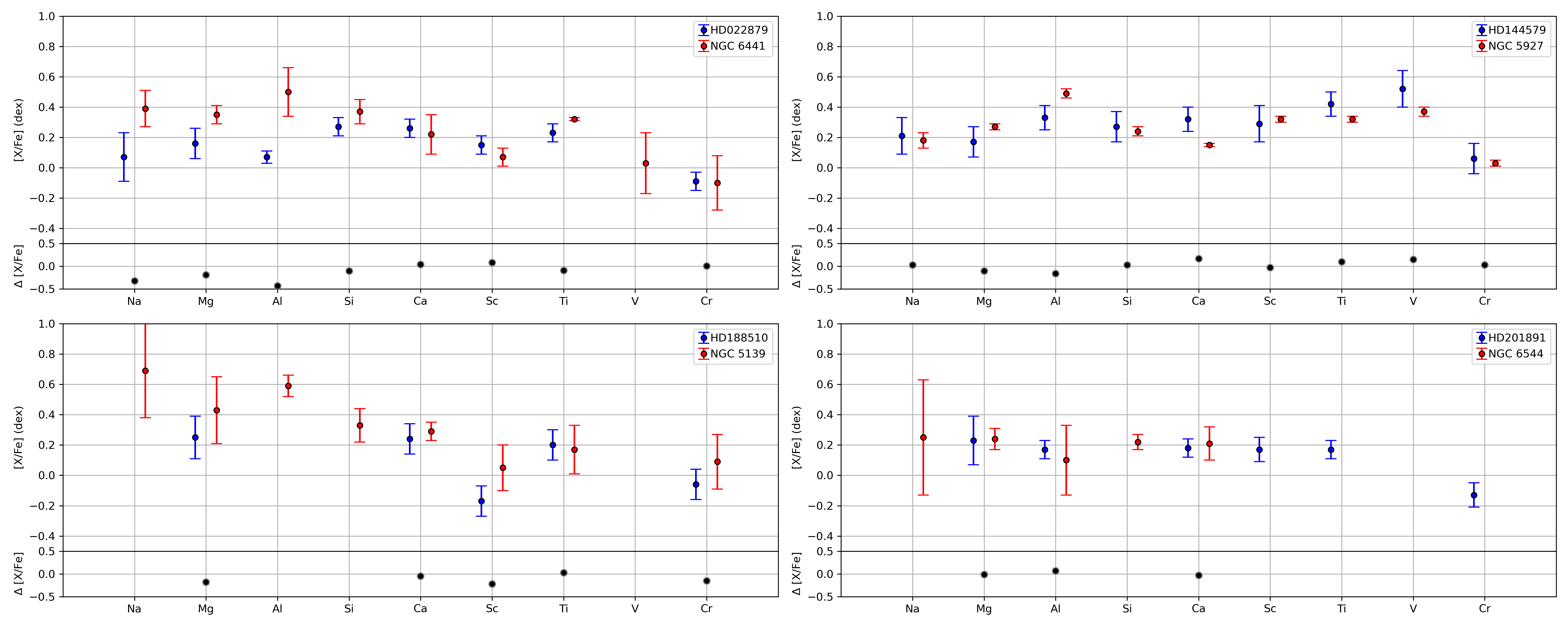}}
\caption{Element abundance patterns ([X/Fe]) for four stars (blue) and GCs (red). Each panel compares the chemical abundances of a star with those of a globular cluster, showing variations across different elements (Na, Mg, Al, Si, Ca, Sc, Ti, V, and Cr). Error bars represent the measurement uncertainties. The lower section of each panel illustrates the abundance differences ($\Delta$[X/Fe]) between the star and the GC.}
\label{fig:GCs_Abundances}
\end{figure*}

To evaluate the potential dynamical escape scenarios from GCs, we analyzed the spatial and kinematic alignment of four target stars with known GCs. While encounter probabilities provide initial clues, they are insufficient to conclusively trace the stellar kinematic origins. To augment this analysis, we compared the stars’ ages ($\tau$) and iron abundances ([Fe/H]) with those of candidate GCs, compiling the literature-derived values for these parameters in Table \ref{tab:origin}. A critical methodological consideration involves reconstructing Galactic orbits under the influence of gravitational perturbations. For stars hypothesized to have escaped from GCs, we backward-integrated their current velocities, positions, and ages to model their orbital trajectories around the Galactic center. This approach accounts for Galactic potential perturbations and yields closed Galactic orbits, thereby enabling the derivation of their orbital parameters. Importantly, these reconstructions represent backward extrapolations from present-day conditions rather than direct recordings of past motions. The resulting orbits, calculated in tandem with the trajectories of the candidate GCs, are shown in Figure \ref{fig:GC_Orbit}. This methodology assumes that escaped stars may retain similar orbital paths to their parent clusters if perturbations (e.g., tidal interactions or dynamical heating) do not drastically alter their trajectories. However, inherent uncertainties in backward integration, such as incomplete knowledge of the time evolution of galactic potential, limit the precision of these reconstructions.

For HD\,22879, it was determined that the metallicity and age parameters of NGC\,6441 were closely aligned with the values obtained for this star, which was one-quarter of the value derived in this study. This comparison provides significant evidence that HD\,22879 may have escaped the NGC\,6441, a bulge GC ($Z_{\rm max} \approx 1.1$ kpc).

The kinematic behavior of star HD\,144579 indicates compatibility with the escape scenario of NGC\,5927 in terms of both position and velocity. Based on a literature review, the age ($\tau$) and iron ([Fe/H]) abundances reported for NGC\,5927 aligned well with the parameters calculated for HD\,144579 in this study, which were determined to be one-quarter lower for the cluster. These findings offer robust evidence to support the possibility that HD\,144579 escapes from NGC\,5927, a bulge GC ($Z_{\rm max} \approx 0.8$) kpc

For HD\,188510, it was revealed that its position and velocity were compatible with those of NGC\,6441, with a probability of $P({\rm origin | \theta, \nu}) = 78\%$. However, the iron ([Fe/H]) abundance and age of the clusters derived from the literature differed significantly from those of the star. This discrepancy limits the escape scenario of HD\,188510 from NGC\,6441. Further analysis considered four other candidate clusters, revealing that NGC\,5189, with an escape probability of $P({\rm origin | \theta, \nu}) = 67\%$, exhibited better consistency in terms of [Fe/H] abundance and age. These findings support the escape scenario of HD\,188510 from NGC\,5189, which is a halo member GC ($Z_{\rm max} \approx 2.8$ kpc) with a metallicity [Fe/H] = $-1.53$ dex. As demonstrated in Appendix \ref{A1}, an animation of the HD 188510 encounter with NGC 6441 is presented.

The final star analyzed, HD\,201891, demonstrated compatibility with the escape scenario from the bulge GC NGC\,6441, with a probability of $P({\rm origin | \theta, \nu}) = 66\%$. In contrast, among the five candidate GCs examined, NGC\,6544, a bulge Gc ($Z_{\rm max} \approx 1.1$ kpc), exhibited a lower escape probability of $P({\rm origin | \theta, \nu}) = 49\%$ but showed greater consistency in terms of iron ([Fe/H]) abundance and age parameters when compared to the values derived for HD\,201891.

The chemical composition of stars originating from GCs serves as a key diagnostic tool for tracing their origins and evolutionary history. As shown in Table~\ref{tab:origin}, the iron abundances of the stars were in general agreement with the metallicities of their associated globular clusters, thereby supporting the hypothesis that these stars originated from these systems. However, subtle differences in the [Fe/H] values may reflect the effects of chemical self-enrichment within the clusters or possible contamination from field populations.

For instance, HD\,22879 is linked to NGC\,6441, a metal-rich bulge cluster with $\rm [Fe/H] = -0.65$ dex, while the star exhibits a slightly lower iron abundance of $\rm [Fe/H] = -0.86\pm0.08$ dex. A similar trend is observed for HD\,144579, which is associated with NGC\,5927 ($\rm [Fe/H] = -0.49$ dex) but has a slightly lower metallicity of $\rm [Fe/H] = -0.55\pm0.12$ dex. These minor discrepancies may be attributed to variations in enrichment histories or differences in measurement techniques.  HD\,188510, likely originating from NGC\,5139 ($\omega$ Centauri), showed excellent agreement in iron abundance ($\rm [Fe/H] = -1.60\pm0.07$ dex) compared to the metallicity of the cluster ($\rm [Fe/H] = -1.53$ dex). This consistency strengthens the case for NGC\,5139, which is the progenitor of dynamically unbound stars in the Galactic halo. In contrast, HD\,201891 exhibited a slightly higher metallicity ($\rm [Fe/H] = -1.15\pm0.07$ dex) than its associated bulge cluster, NGC\,6544 ([Fe/H] = -1.44 dex), suggesting either a different star formation history within the cluster or enrichment from external sources.

These abundance comparisons, along with dynamic evidence, provide strong support for the GC origin hypothesis of these stars. Future studies should focus on detailed chemical abundance patterns, including $\alpha$ and neutron-capture elements of GCs, to further refine the connection between these stars and their host GCs. The elemental abundance patterns of stars and their associated GCs provide additional evidence supporting their possible origins, as shown in Figure~\ref{fig:GCs_Abundances}. The chemical compositions of stars largely align with those of their respective GC, particularly in elements such as magnesium (Mg), aluminum (Al), and calcium (Ca), which serve as key tracers of nucleosynthesis in massive stars. However, certain elements, including sodium (Na) and vanadium (V), exhibit notable deviations, with stars displaying abundance differences relative to their associated clusters. These discrepancies may result from intrinsic chemical inhomogeneities within clusters, evolutionary effects, or contamination from field populations. Specifically, HD\,22879 and HD\,144579 exhibit slightly lower Mg and Al abundances than NGC\,6441 and NGC\,5927, respectively, while HD\,188510 and HD\,201891 show $\alpha$-element abundance patterns that are broadly consistent with those of NGC\,5139 and NGC\,6544. Such detailed chemical abundance comparisons, in conjunction with kinematic evidence, provide important constraints on the mechanisms by which stars are dynamically ejected from GCs into the Galactic field.

In conclusion, considering the positions, velocities, iron and individual abundances, and ages of the four stars analyzed in this study, it was determined that HD\,22879 likely escaped from NGC\,6441, HD\,144579 from NGC\,5927, HD\,188510 from NGC\,5139, and HD\,201891 from NGC\,6544. These findings indicate that, except for HD\,188510, the escape scenarios of the other three stars originated from the bulge GCs. \citet{Lucey2022} classified Galactic GCs based on their orbital parameters, adopting criteria to distinguish bulge, disk, and halo populations. Bulge GCs generally consist of objects that satisfy the conditions $R_{\rm a} \leq 5$ kpc and $Z_{\rm max} \leq 2.5$ kpc \citep{Lucey2022}. This implies that bulge GCs are unlikely to reach the solar neighborhood in terms of their Galactic orbits. However, the ability of the four metal-poor stars analyzed in this study to leave the bulge region for the solar neighborhood remains an unresolved mystery.

Globular clusters are capable of dynamically ejecting individual stars through internal gravitational interactions, such as three-body encounters or close binary scattering \citep[\textbf{cf.}][]{Tomas2023, Weatherford2023, Weatherford2024}. These ejected stars, as they traverse through the Galactic potential, can interact with large-scale structural features such as spiral arms and the central bar, particularly within the resonance regions. Such interactions induce significant changes in the orbits of the stars, enabling them to travel considerable distances from their birthplaces, even reaching regions near the Solar neighborhood.

Among these Galactic resonances, the corotation resonance (CR) and the outer Lindblad resonance (OLR) are particularly influential in altering stellar angular momenta, which consequently leads to notable modifications in orbital radii and geometries. In this context, \citet{Minchev2010} demonstrated that the overlap of bar and spiral resonances allows stars to effectively disperse throughout the galactic disk. Similarly, \citet{Roskar2008}, through high-resolution simulations, showed that stars can migrate far from their formation radii because of resonant interactions with spiral arms while maintaining nearly circular orbits during the process. This orbital migration is not confined solely to dynamical effects; when combined with stellar chemical compositions and ages, it offers crucial insights into the structural and evolutionary characteristics of the Galaxy. The study by \citet{Onal-tas2018}, based on RAVE data, provided statistical evidence of how stars in the solar neighborhood are influenced by Galactic resonances and highlighted the impact of these resonances on stellar metallicity gradients and kinematic structures.

Tracking the orbits of stars that have escaped from GCs is of paramount importance for a comprehensive understanding of Galactic structure and evolution. The analysis of these orbits, particularly through numerical modeling supported by high-precision astrometric observations, such as those from the {\it Gaia} mission \citep{Gaia_Mission}, has proven to be essential for elucidating the interactions between migrating stars and Galactic resonant structures. In conclusion, the orbital modifications of GC escapees through interactions with Galactic resonances play a key role in understanding radial migration processes extending to the solar neighborhood and present a rich field for future research endeavors.

\begin{figure*} 
\centerline{\includegraphics[width=0.85\textwidth]{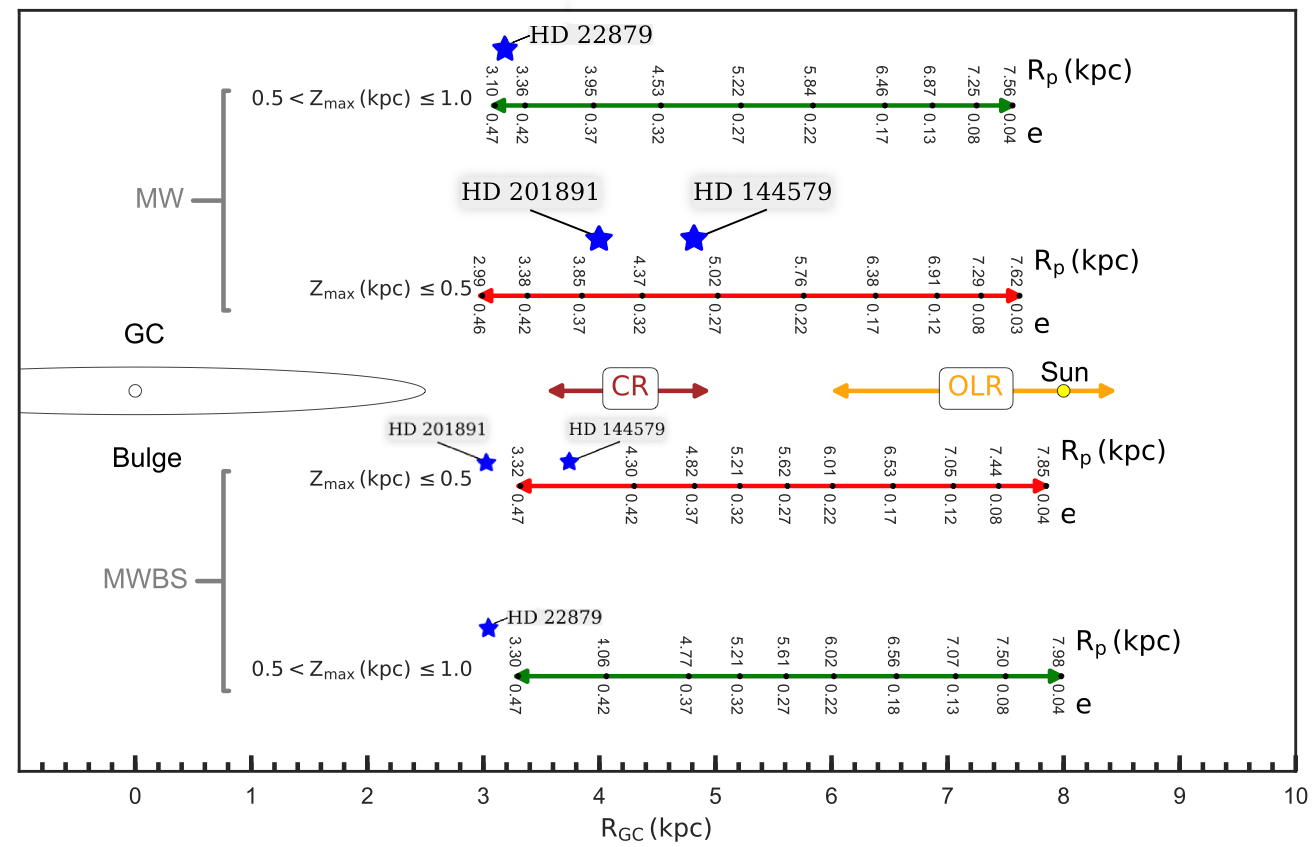}}
\caption{\citet{Onal-tas2018} showing the Galactic center, bulge, corotation resonance (CR), outer Lindblad resonance (OLR) \citep{Dehnen2000}, and the Sun's position ($R_0 = 8$ kpc). It also illustrates the metallicity variation within the two $Z_{\rm max}$ ranges ($0 < Z_{\rm max}~{\rm (kpc)} \leq$ 0.5, and $0.5 < Z_{\rm max}~{\rm (kpc)} \leq$ 1) based on the MW and {\bf MWBS models} represented by different colors. The blue star symbol represents the three stars studied in this study.}
\label{fig:onal-tas}
\end{figure*}

To investigate the effects of Galactic perturbations, \citet{Onal-tas2018} analyzed radial metallicity gradients and dynamical structures in the solar neighborhood using orbital parameters derived from symmetric and asymmetric Galactic potential models. Their study used red clump (RC) stars from the fourth data release of the Radial Velocity Experiment \citep[RAVE DR4,][]{RaveDR4}, focusing on metallicity gradients as functions of two parameters: (1) the maximum vertical distance from the Galactic plane (\(Z_{\rm max}\)) and (2) orbital eccentricity (\(e\)). The steepest gradient of \(-0.065\pm0.005\) dex kpc\(^{-1}\) was identified for stars with \(0 < Z_{\rm max}~{\rm (kpc)} \leq 0.5\) and low eccentricity (\(e \leq 0.1\)), while the shallowest gradient (\(-0.014\pm0.006\) dex kpc\(^{-1}\)) corresponded to the same \(Z_{\rm max}\) range but higher eccentricity (\(e \leq 0.5\)). For stars with \(Z_{\rm max} > 1\) kpc, the gradients flattened or became positive, exhibiting no dependence on orbital eccentricity. The authors suggested that the observed gradient variations within \(Z_{\rm max} = 1\) kpc result from dynamic interactions between the stellar orbits and the resonance points of the Galactic bar, as depicted schematically in Figure \ref{fig:onal-tas}.

\citet{Onal-tas2018} analyzed the mean perigalactic radius ($R_{\rm p}$) and orbital eccentricity ($e$) of RC stars within different maximum the Galactic height ($Z_{\rm max}$) ranges. Among the four stars considered in this study, their vertical orbital motion excursions were examined under both symmetric and asymmetric Galactic potential models. Within these models, HD\,201891 and HD\,144579 are found to remain within the $Z_{\rm max}< 0.5$ kpc interval, while HD\,22879 falls within the $0.5 < Z_{\rm max}~{\rm (kpc)} < 1$ range (see Table \ref{tab:pop_parameters}). However, HD\,188510, with a $Z_{\rm max}$ value of 2.2 kpc, is outside the scale established by \citet{Onal-tas2018}.  When the $R_{\rm p}$ values of both stars are considered, HD\,144579 and HD\,22879 are found to be consistent with the proposed symmetric and asymmetric Galactic potential model scales in both $R_{\rm p}$ and $e$. In contrast, HD\,201891 exhibits an eccentricity value approximately 0.15 higher than the expected range. Overall, the results largely agree with the scale suggested by \citet{Onal-tas2018}. Additionally, this study investigated the escape scenarios of stars from GCs based on kinematic and dynamical considerations. 

Specifically, stars HD\,22879, HD\,144579, and HD\,201891 are hypothesized to have escaped from NGC\,6441, NGC\,5927, and NGC\,6544 GCs. To assess this scenario, the perigalactic and apogalactic radii ($R_{\rm p}$ and $R_{\rm a}$) of these clusters were examined. The globular cluster parameters were obtained from \citet{Baumgardt2019}, with the values for the NGC\,6441, NGC\,5927, and NGC\,6544 clusters given by ($R_{\rm p}$, $R_{\rm a}$) = (1.00, 3.91) kpc, (3.99, 5.42) kpc, and (0.62, 5.49) kpc, respectively. Considering the orbital paths of these GCs around the Galactic center, all three stars appear to be kinematically associated with their respective GCs. The escape mechanism from GCs can occur at any point along the orbit, and may be influenced by perturbative forces within the Galactic disk, potentially leading to the orbit of these stars in the solar neighborhood \citep{Baumgardt2018}. In this context, orbital analyses and dynamical modeling are crucial for understanding possible escape scenarios and the role of these stars in the broader context of Galactic evolution. 

This analysis suggests that three of the stars (HD 22879, HD 144579, and HD 201891) may have reached the solar neighborhood after evaporating from GCs located in the bulge region of the Galaxy, driven by resonances associated with the Galactic bar and spiral arms. In contrast, HD 188510 is likely to have been separated from halo NGC 5139. As noted above, the standard Milky Way model was adopted to compute the Galactic orbits of these stars. The orbital analyses indicate a potential association between the studied stars and bulge GCs, highlighting the importance of considering additional dynamical influences—such as those arising from the Galactic bar and spiral arms—in future orbit modeling.

The results obtained from orbit analyses under the MWBS potential for the four stars are summarized in Table \ref{tab:pop_parameters}. When these results are placed within the schematic diagram obtained for the MWBS model (Figure \ref{fig:onal-tas}), it becomes evident that relative to the standard MW model, the stars may follow orbits that can penetrate deeper into the inner regions of the Galaxy. More complex potential calculations further support that three of the studied stars could have traveled from their parent GCs to the solar vicinity by escaping from regions where the co-rotation effect of the Galactic bar is dominant.

Additionally, the fact that all four stars are G-type main-sequence stars with ages comparable to those of GCs suggests that they are old enough to have plausibly moved from their parent clusters to the solar neighborhood. These results predict that a fraction of metal-poor stars in the solar neighborhood may originate from disrupted or evaporated GCs, especially those associated with the Galactic bar. The distinct chemical signatures observed in these stars further emphasize the role of detailed abundance analyses in tracing their Galactic origins. Despite the use of high-resolution spectral data and space-based photometric and astrometric observations, precise population classification of these stars remains challenging. This study underscores the necessity of high-resolution spectroscopy and space-based observations of GCs to refine our understanding of the Galactic origins of metal-poor stars.

\subsection{Future Works and Sights}

As new data from various sky survey programmes is obtained on a daily basis, our understanding of the structural and evolutionary processes of our Galaxy is being significantly expanded. In particular, $Gaia$ DR4 data, which are expected to be released in the near future, will enable the kinematic and dynamical orbital parameters of stars to be determined more accurately and precisely. This development holds immense potential for unravelling the enigmatic origins and intrinsic nature of metal-poor stars.

In parallel to these upcoming efforts, recent high-resolution spectroscopic studies have already made significant strides in characterizing metal-poor stellar populations. Surveys such as COMBS I/II/III \citep{Lucey2019, Lucey2021, Lucey2022} SDSS/SEGUE \citep{Caffau2011} and the Pristine Survey \citep{Venn2020, Martin2024,Viswanathan2025} have delivered high-quality abundance measurements for large samples of metal-poor stars, probing nucleosynthetic signatures and Galactic chemical evolution. Individual high-resolution studies \citep[e.g.,][]{Norris2013, Garcia2013, Keller2014, Anna2015} have further resolved the detailed chemistry of ultra metal-poor stars, offering critical insights into early stellar generations. These efforts complement large-scale projects like GALAH survey \citep{Buder2024} and LAMOST \citep{Li2015}, which combine moderate- and high-resolution spectroscopy to map metallicity gradients and substructures in the halo. 

Beyond the data provided by $Gaia$, large-scale observing projects such as 4MOST \citep{4MOST}, H3 \citep{H3}, and Milky Way Mapper \citep{SDSSV} aim to study the chemical evolution of our Galaxy through high-resolution and high signal-to-noise ratio spectra obtained at different wavelengths, while also providing radial velocity measurements, which are essential data for kinematic analyses. These programs will synergize with existing high-resolution campaigns \citep[e.g., TOPoS][]{Caffau2013} to bridge gaps between kinematic and chemical diagnostics for metal-poor populations.

In the near future, these high-accuracy data will facilitate the detection and deeper understanding of the origin of the structures of uncertain origin that are the subject of our study, as well as structural anomalies caused by intrinsic gravitational interactions beyond the chemical and dynamical evolution of our Galaxy. Consequently, it is anticipated that the solution approaches developed in this study will provide a framework for further analyses with larger and more sensitive data sets.

\section*{Acknowledgments}
We would like to express our sincere gratitude to the anonymous referee for their invaluable comments and suggestions, which were instrumental in enhancing the quality of our manuscript. This study was partially supported by the Scientific and Technological Research Council (TÜBİTAK) MFAG-121F265. This study was funded by the Scientific Research Projects Coordination Unit of the Istanbul University. Project number: 40044. This study is part of the MSc thesis of Deniz Cennet \c{C}ınar. This research used NASA's (National Aeronautics and Space Administration) Astrophysics Data System and the SIMBAD Astronomical Database, operated at CDS, Strasbourg, France, and the NASA/IPAC Infrared Science Archive, which is operated by the Jet Propulsion Laboratory, California Institute of Technology, under contract with the National Aeronautics and Space Administration. This study used data from the European Space Agency (ESA) mission {\it Gaia} (\mbox{https://www.cosmos.esa.int/gaia}) and, processed by the {\it Gaia} Data Processing and Analysis Consortium (DPAC, \mbox{https://www.cosmos.esa.int/web/gaia/dpac/consortium}). Funding for the DPAC was provided by national institutions, particularly those participating in the {\it Gaia} Multilateral Agreement.

\software{LIME \citep{Sahin2017}, MOOG \citep{Sneden1974}, {\sc galpy} \citep{Bovy2015}, MWPotential2014 \citep{Bovy2015}, DehnenBarPotential \citep{Dehnen2000, Monari2016}, SpiralArmsPotential \citep{CoxGomes2002}, ARIADNE \citep{Vines2022}}. 

%---------------------------------------------------------------------------------------------
\appendix
\input{A1}

\bibliography{Cinar_et_al}{}
\bibliographystyle{aasjournalv7}



\end{document}

%% file: A1.tex
\section{Orbital Evolution of HD 188510 and NGC 6441}\label{A1}

\begin{figure*}
\centering
\label{Animated_gif}
\includegraphics[width=0.95\textwidth]{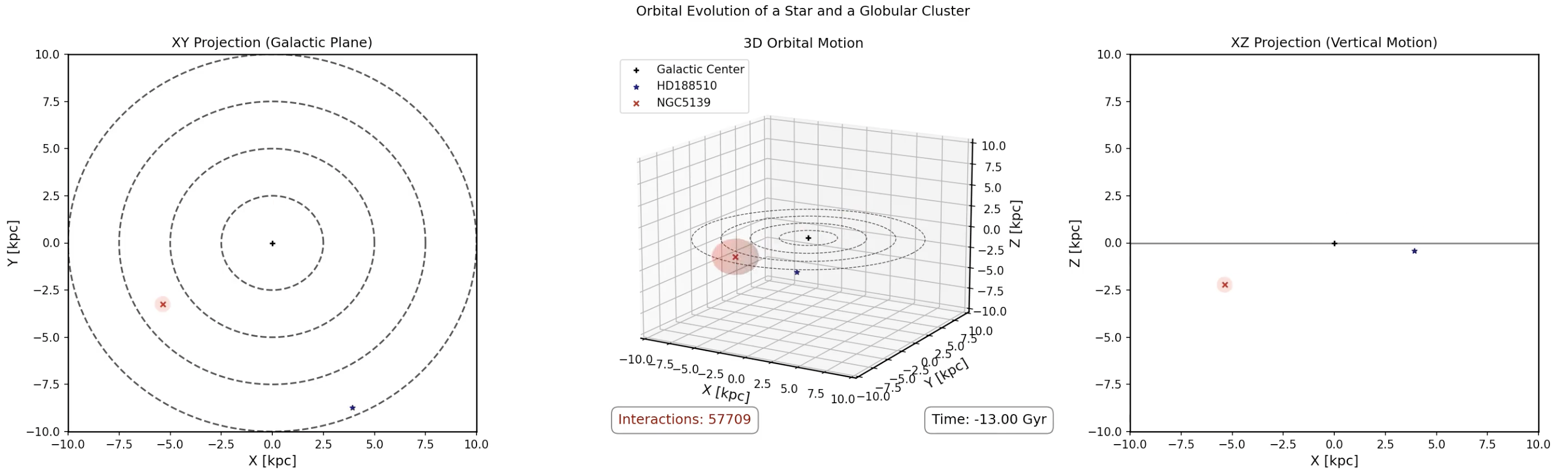}
\caption{Orbital evolution of HD\,188510 and NGC\,5139 ($\omega$ Cen) in the {\sc MWPotential2014}. The left panel shows the top-down ($X-Y$) projection in the Galactic plane, with dashed circles indicating constant Galactocentric radii. The central panel illustrates the full 3D orbital motion, while the right panel presents the vertical ($X-Z$) projection, highlighting motion perpendicular to the Galactic plane. The Galactic center is marked with a black cross, HD 188510 with a blue star, and NGC 5139 with a red cross, which corresponds to 5 tidal radii. The figure represents to a backward to 13 Gyr, with a total of 57,709 interaction calculations performed. An animated version showing the orbital evolution over 13 Gyr back integration is available as a supplementary material. The animation captures the full 3D motion, including projection onto the Galactic plane and vertical oscillations relative to it. The total duration of the animation is 1 min 20 s.}
\end{figure*}

%% file: Cinar_et_al.bbl
\begin{thebibliography}{}
\expandafter\ifx\csname natexlab\endcsname\relax\def\natexlab#1{#1}\fi
\providecommand{\url}[1]{\href{#1}{#1}}
\providecommand{\dodoi}[1]{doi:~\href{http://doi.org/#1}{\nolinkurl{#1}}}
\providecommand{\doeprint}[1]{\href{http://ascl.net/#1}{\nolinkurl{http://ascl.net/#1}}}
\providecommand{\doarXiv}[1]{\href{https://arxiv.org/abs/#1}{\nolinkurl{https://arxiv.org/abs/#1}}}

\bibitem[{S. {Ak} {et~al.}(2007{\natexlab{a}}){Ak}, {Bilir}, {Karaali}, \& {Buser}}]{Ak2007a}
{Ak}, S., {Bilir}, S., {Karaali}, S., \& {Buser}, R. 2007{\natexlab{a}}, \bibinfo{title}{{Estimation of galactic model parameters with the Sloan Digital Sky Survey and the metallicity distribution in two fields in the anti-centre direction of the Galaxy},} Astronomische Nachrichten, 328, 169, \dodoi{10.1002/asna.200610709}

\bibitem[{S. {Ak} {et~al.}(2007{\natexlab{b}}){Ak}, {Bilir}, {Karaali}, {Buser}, \& {Cabrera-Lavers}}]{Ak2007b}
{Ak}, S., {Bilir}, S., {Karaali}, S., {Buser}, R., \& {Cabrera-Lavers}, A. 2007{\natexlab{b}}, \bibinfo{title}{{The metallicity distributions in high-latitudes with SDSS},} \na, 12, 605, \dodoi{10.1016/j.newast.2007.04.005}

\bibitem[{M. {Asplund} {et~al.}(2009){Asplund}, {Grevesse}, {Sauval}, \& {Scott}}]{Asplund2009}
{Asplund}, M., {Grevesse}, N., {Sauval}, A.~J., \& {Scott}, P. 2009, \bibinfo{title}{{The Chemical Composition of the Sun},} \araa, 47, 481, \dodoi{10.1146/annurev.astro.46.060407.145222}

\bibitem[{A.~T. {Baldwin} {et~al.}(2016){Baldwin}, {Watkins}, {van der Marel}, {Bianchini}, {Bellini}, \& {Anderson}}]{Baldwin2016}
{Baldwin}, A.~T., {Watkins}, L.~L., {van der Marel}, R.~P., {et~al.} 2016, \bibinfo{title}{{Hubble Space Telescope Proper Motion (HSTPROMO) Catalogs of Galactic Globular Clusters. IV. Kinematic Profiles and Average Masses of Blue Straggler Stars},} \apj, 827, 12, \dodoi{10.3847/0004-637X/827/1/12}

\bibitem[{C. {Battistini} \& T. {Bensby}(2016){Battistini} \& {Bensby}}]{Battistini2016}
{Battistini}, C., \& {Bensby}, T. 2016, \bibinfo{title}{{The origin and evolution of r- and s-process elements in the Milky Way stellar disk},} \aap, 586, A49, \dodoi{10.1051/0004-6361/201527385}

\bibitem[{H. {Baumgardt} \& M. {Hilker}(2018){Baumgardt} \& {Hilker}}]{Baumgardt2018}
{Baumgardt}, H., \& {Hilker}, M. 2018, \bibinfo{title}{{A catalogue of masses, structural parameters, and velocity dispersion profiles of 112 Milky Way globular clusters},} \mnras, 478, 1520, \dodoi{10.1093/mnras/sty1057}

\bibitem[{H. {Baumgardt} {et~al.}(2019){Baumgardt}, {Hilker}, {Sollima}, \& {Bellini}}]{Baumgardt2019}
{Baumgardt}, H., {Hilker}, M., {Sollima}, A., \& {Bellini}, A. 2019, \bibinfo{title}{{Mean proper motions, space orbits, and velocity dispersion profiles of Galactic globular clusters derived from Gaia DR2 data},} \mnras, 482, 5138, \dodoi{10.1093/mnras/sty2997}

\bibitem[{T.~C. {Beers} \& N. {Christlieb}(2005){Beers} \& {Christlieb}}]{Beers2005}
{Beers}, T.~C., \& {Christlieb}, N. 2005, \bibinfo{title}{{The Discovery and Analysis of Very Metal-Poor Stars in the Galaxy},} \araa, 43, 531, \dodoi{10.1146/annurev.astro.42.053102.134057}

\bibitem[{K. {Bekki} \& M. {Chiba}(2001){Bekki} \& {Chiba}}]{Bekki2001}
{Bekki}, K., \& {Chiba}, M. 2001, \bibinfo{title}{{Formation of the Galactic Stellar Halo. I. Structure and Kinematics},} \apj, 558, 666, \dodoi{10.1086/322300}

\bibitem[{T. {Bensby} {et~al.}(2003){Bensby}, {Feltzing}, \& {Lundstr{\"o}m}}]{Bensby2003}
{Bensby}, T., {Feltzing}, S., \& {Lundstr{\"o}m}, I. 2003, \bibinfo{title}{{Elemental abundance trends in the Galactic thin and thick disks as traced by nearby F and G dwarf stars},} \aap, 410, 527, \dodoi{10.1051/0004-6361:20031213}

\bibitem[{T. {Bensby} {et~al.}(2005){Bensby}, {Feltzing}, {Lundstr{\"o}m}, \& {Ilyin}}]{Bensby2005}
{Bensby}, T., {Feltzing}, S., {Lundstr{\"o}m}, I., \& {Ilyin}, I. 2005, \bibinfo{title}{{{\ensuremath{\alpha}}-, r-, and s-process element trends in the Galactic thin and thick disks},} \aap, 433, 185, \dodoi{10.1051/0004-6361:20040332}

\bibitem[{T. {Bensby} {et~al.}(2014){Bensby}, {Feltzing}, \& {Oey}}]{Bensby2014}
{Bensby}, T., {Feltzing}, S., \& {Oey}, M.~S. 2014, \bibinfo{title}{{Exploring the Milky Way stellar disk. A detailed elemental abundance study of 714 F and G dwarf stars in the solar neighbourhood},} \aap, 562, A71, \dodoi{10.1051/0004-6361/201322631}

\bibitem[{M. {Bergemann} {et~al.}(2012){Bergemann}, {Lind}, {Collet}, {Magic}, \& {Asplund}}]{Bergemann2012}
{Bergemann}, M., {Lind}, K., {Collet}, R., {Magic}, Z., \& {Asplund}, M. 2012, \bibinfo{title}{{Non-LTE line formation of Fe in late-type stars - I. Standard stars with 1D and <3D> model atmospheres},} \mnras, 427, 27, \dodoi{10.1111/j.1365-2966.2012.21687.x}

\bibitem[{E. {Bica} {et~al.}(2006){Bica}, {Bonatto}, {Barbuy}, \& {Ortolani}}]{Bica2006}
{Bica}, E., {Bonatto}, C., {Barbuy}, B., \& {Ortolani}, S. 2006, \bibinfo{title}{{Globular cluster system and Milky Way properties revisited},} \aap, 450, 105, \dodoi{10.1051/0004-6361:20054351}

\bibitem[{A. Bijaoui {et~al.}(2012)Bijaoui, Recio-Blanco, {de Laverny}, \& Ordenovic}]{BIJAOUI201255}
Bijaoui, A., Recio-Blanco, A., {de Laverny}, P., \& Ordenovic, C. 2012, \bibinfo{title}{Parameter estimation from a model grid application to the Gaia RVS spectra,} Statistical Methodology, 9, 55, \dodoi{https://doi.org/10.1016/j.stamet.2011.07.004}

\bibitem[{S. {Bilir} {et~al.}(2008){Bilir}, {Cabrera-Lavers}, {Karaali}, {Ak}, {Yaz}, \& {L{\'o}pez-Corredoira}}]{Bilir2008}
{Bilir}, S., {Cabrera-Lavers}, A., {Karaali}, S., {et~al.} 2008, \bibinfo{title}{{Estimation of Galactic Model Parameters in High Latitudes with SDSS},} \pasa, 25, 69, \dodoi{10.1071/AS07026}

\bibitem[{S. {Bilir} {et~al.}(2012){Bilir}, {Karaali}, {Ak}, {{\"O}nal}, {Da{\v{g}}tekin}, {Yontan}, {Gilmore}, \& {Seabroke}}]{Bilir2012}
{Bilir}, S., {Karaali}, S., {Ak}, S., {et~al.} 2012, \bibinfo{title}{{Local stellar kinematics from RAVE data - III. Radial and vertical metallicity gradients based on red clump stars},} \mnras, 421, 3362, \dodoi{10.1111/j.1365-2966.2012.20561.x}

\bibitem[{S. {Bilir} {et~al.}(2006{\natexlab{a}}){Bilir}, {Karaali}, {Ak}, {Yaz}, \& {Hamzao{\u{g}}lu}}]{Bilir2006a}
{Bilir}, S., {Karaali}, S., {Ak}, S., {Yaz}, E., \& {Hamzao{\u{g}}lu}, E. 2006{\natexlab{a}}, \bibinfo{title}{{Galactic longitude dependent galactic model parameters},} \na, 12, 234, \dodoi{10.1016/j.newast.2006.10.001}

\bibitem[{S. {Bilir} {et~al.}(2006{\natexlab{b}}){Bilir}, {Karaali}, \& {Gilmore}}]{Bilir2006b}
{Bilir}, S., {Karaali}, S., \& {Gilmore}, G. 2006{\natexlab{b}}, \bibinfo{title}{{Investigation of the ELAIS field by Vega photometry: absolute magnitude-dependent Galactic model parameters},} \mnras, 366, 1295, \dodoi{10.1111/j.1365-2966.2006.09891.x}

\bibitem[{S. {Bilir} {et~al.}(2006{\natexlab{c}}){Bilir}, {Karaali}, {G{\"u}ver}, {Karata{\c{s}}}, \& {Ak}}]{Bilir2006c}
{Bilir}, S., {Karaali}, S., {G{\"u}ver}, T., {Karata{\c{s}}}, Y., \& {Ak}, S.~G. 2006{\natexlab{c}}, \bibinfo{title}{{Galactic model parameters for field giants separated from field dwarfs by their 2MASS and V apparent magnitudes},} Astronomische Nachrichten, 327, 72, \dodoi{10.1002/asna.200510480}

\bibitem[{J. {Binney} \& S. {Tremaine}(2008){Binney} \& {Tremaine}}]{galacticdynamics2}
{Binney}, J., \& {Tremaine}, S. 2008, {Galactic Dynamics: Second Edition}

\bibitem[{C. {Boeche} \& E.~K. {Grebel}(2016){Boeche} \& {Grebel}}]{Boeche2016}
{Boeche}, C., \& {Grebel}, E.~K. 2016, \bibinfo{title}{{SP\_Ace: a new code to derive stellar parameters and elemental abundances},} \aap, 587, A2, \dodoi{10.1051/0004-6361/201526758}

\bibitem[{A. Bonaca {et~al.}(2019)Bonaca, Conroy, Price-Whelan, \& Hogg}]{Bonaca2019}
Bonaca, A., Conroy, C., Price-Whelan, A.~M., \& Hogg, D.~W. 2019, \bibinfo{title}{Multiple Components of the Jhelum Stellar Stream,} The Astrophysical Journal Letters, 881, L37, \dodoi{10.3847/2041-8213/ab36ba}

\bibitem[{A. {Bonaca} {et~al.}(2017){Bonaca}, {Conroy}, {Wetzel}, {Hopkins}, \& {Kere{\v{s}}}}]{Bonaca2017}
{Bonaca}, A., {Conroy}, C., {Wetzel}, A., {Hopkins}, P.~F., \& {Kere{\v{s}}}, D. 2017, \bibinfo{title}{{Gaia Reveals a Metal-rich, in situ Component of the Local Stellar Halo},} \apj, 845, 101, \dodoi{10.3847/1538-4357/aa7d0c}

\bibitem[{J. {Bovy}(2015){Bovy}}]{Bovy2015}
{Bovy}, J. 2015, \bibinfo{title}{{galpy: A python Library for Galactic Dynamics},} \apjs, 216, 29, \dodoi{10.1088/0067-0049/216/2/29}

\bibitem[{J. Bovy \& S. Tremaine(2012)Bovy \& Tremaine}]{Bovy2012}
Bovy, J., \& Tremaine, S. 2012, \bibinfo{title}{On the Local Dark Matter Density,} The Astrophysical Journal, 756, 89, \dodoi{10.1088/0004-637X/756/1/89}

\bibitem[{A. {Bressan} {et~al.}(2012){Bressan}, {Marigo}, {Girardi}, {Salasnich}, {Dal Cero}, {Rubele}, \& {Nanni}}]{Bressan2012}
{Bressan}, A., {Marigo}, P., {Girardi}, L., {et~al.} 2012, \bibinfo{title}{{PARSEC: stellar tracks and isochrones with the PAdova and TRieste Stellar Evolution Code},} \mnras, 427, 127, \dodoi{10.1111/j.1365-2966.2012.21948.x}

\bibitem[{S. {Buder} {et~al.}(2018){Buder}, {Asplund}, {Duong}, {Kos}, {Lind}, {Ness}, {Sharma}, {Bland-Hawthorn}, {Casey}, {de Silva}, {D'Orazi}, {Freeman}, {Lewis}, {Lin}, {Martell}, {Schlesinger}, {Simpson}, {Zucker}, {Zwitter}, {Amarsi}, {Anguiano}, {Carollo}, {Casagrande}, {{\v{C}}otar}, {Cottrell}, {da Costa}, {Gao}, {Hayden}, {Horner}, {Ireland}, {Kafle}, {Munari}, {Nataf}, {Nordlander}, {Stello}, {Ting}, {Traven}, {Watson}, {Wittenmyer}, {Wyse}, {Yong}, {Zinn}, {{\v{Z}}erjal}, \& {Galah Collaboration}}]{Buder2018}
{Buder}, S., {Asplund}, M., {Duong}, L., {et~al.} 2018, \bibinfo{title}{{The GALAH Survey: second data release},} \mnras, 478, 4513, \dodoi{10.1093/mnras/sty1281}

\bibitem[{S. {Buder} {et~al.}(2024){Buder}, {Kos}, {Wang}, {McKenzie}, {Howell}, {Martell}, {Hayden}, {Zucker}, {Nordlander}, {Montet}, {Traven}, {Bland-Hawthorn}, {De Silva}, {Freeman}, {Lewis}, {Lind}, {Sharma}, {Simpson}, {Stello}, {Zwitter}, {Amarsi}, {Armstrong}, {Banks}, {Beavis}, {Beeson}, {Chen}, {Ciuc{\u{a}}}, {Da Costa}, {de Grijs}, {Martin}, {Nataf}, {Ness}, {Rains}, {Scarr}, {Vogrin{\v{c}}i{\v{c}}}, {Wang}, {Wittenmyer}, {Xie}, \& {The GALAH Collaboration}}]{Buder2024}
{Buder}, S., {Kos}, J., {Wang}, E.~X., {et~al.} 2024, \bibinfo{title}{{The GALAH Survey: Data Release 4},} arXiv e-prints, arXiv:2409.19858, \dodoi{10.48550/arXiv.2409.19858}

\bibitem[{T. {Cabrera} \& C.~L. {Rodriguez}(2023){Cabrera} \& {Rodriguez}}]{Tomas2023}
{Cabrera}, T., \& {Rodriguez}, C.~L. 2023, \bibinfo{title}{{Runaway and Hypervelocity Stars from Compact Object Encounters in Globular Clusters},} \apj, 953, 19, \dodoi{10.3847/1538-4357/acdc22}

\bibitem[{A. {Cabrera-Lavers} {et~al.}(2007){Cabrera-Lavers}, {Bilir}, {Ak}, {Yaz}, \& {L{\'o}pez-Corredoira}}]{Cabrera-Lavers2007}
{Cabrera-Lavers}, A., {Bilir}, S., {Ak}, S., {Yaz}, E., \& {L{\'o}pez-Corredoira}, M. 2007, \bibinfo{title}{{Estimation of Galactic model parameters in high latitudes with 2MASS},} \aap, 464, 565, \dodoi{10.1051/0004-6361:20066475}

\bibitem[{E. {Caffau} {et~al.}(2011){Caffau}, {Bonifacio}, {Fran{\c{c}}ois}, {Spite}, {Spite}, {Zaggia}, {Ludwig}, {Monaco}, {Sbordone}, {Cayrel}, {Hammer}, {Randich}, {Hill}, \& {Molaro}}]{Caffau2011}
{Caffau}, E., {Bonifacio}, P., {Fran{\c{c}}ois}, P., {et~al.} 2011, \bibinfo{title}{{X-Shooter GTO: chemical analysis of a sample of EMP candidates},} \aap, 534, A4, \dodoi{10.1051/0004-6361/201117530}

\bibitem[{E. {Caffau} {et~al.}(2013){Caffau}, {Bonifacio}, {Sbordone}, {Fran{\c{c}}ois}, {Monaco}, {Spite}, {Plez}, {Cayrel}, {Christlieb}, {Clark}, {Glover}, {Klessen}, {Koch}, {Ludwig}, {Spite}, {Steffen}, \& {Zaggia}}]{Caffau2013}
{Caffau}, E., {Bonifacio}, P., {Sbordone}, L., {et~al.} 2013, \bibinfo{title}{{TOPoS. I. Survey design and analysis of the first sample},} \aap, 560, A71, \dodoi{10.1051/0004-6361/201322488}

\bibitem[{R. {Canbay} {et~al.}(2023){Canbay}, {Bilir}, {{\"O}zd{\"o}nmez}, \& {Ak}}]{Canbay2023}
{Canbay}, R., {Bilir}, S., {{\"O}zd{\"o}nmez}, A., \& {Ak}, T. 2023, \bibinfo{title}{{Galactic Model Parameters and Spatial Density of Cataclysmic Variables in the Gaia Era: New Constraints on Population Models},} \aj, 165, 163, \dodoi{10.3847/1538-3881/acbead}

\bibitem[{J.~A. {Cardelli} {et~al.}(1989){Cardelli}, {Clayton}, \& {Mathis}}]{Cardelli1989}
{Cardelli}, J.~A., {Clayton}, G.~C., \& {Mathis}, J.~S. 1989, \bibinfo{title}{{The Relationship between Infrared, Optical, and Ultraviolet Extinction},} \apj, 345, 245, \dodoi{10.1086/167900}

\bibitem[{F. {Castelli} \& R.~L. {Kurucz}(2003){Castelli} \& {Kurucz}}]{Castelli2003}
{Castelli}, F., \& {Kurucz}, R.~L. 2003, in Modelling of Stellar Atmospheres, ed. N.~{Piskunov}, W.~W. {Weiss}, \& D.~F. {Gray}, Vol. 210, A20, \dodoi{10.48550/arXiv.astro-ph/0405087}

\bibitem[{F. {Castelli} \& R.~L. {Kurucz}(2004){Castelli} \& {Kurucz}}]{Castelli2004}
{Castelli}, F., \& {Kurucz}, R.~L. 2004, \bibinfo{title}{{Is missing Fe I opacity in stellar atmospheres a significant problem?},} \aap, 419, 725, \dodoi{10.1051/0004-6361:20040079}

\bibitem[{J.~W. {Chamberlain} \& L.~H. {Aller}(1951){Chamberlain} \& {Aller}}]{Chamberlain1951}
{Chamberlain}, J.~W., \& {Aller}, L.~H. 1951, \bibinfo{title}{{The Atmospheres of A-Type Subdwarfs and 95 Leonis.},} \apj, 114, 52, \dodoi{10.1086/145451}

\bibitem[{H.-W. {Chen} {et~al.}(2001){Chen}, {Lanzetta}, {Webb}, \& {Barcons}}]{Chen2001}
{Chen}, H.-W., {Lanzetta}, K.~M., {Webb}, J.~K., \& {Barcons}, X. 2001, \bibinfo{title}{{The Gaseous Extent of Galaxies and the Origin of Ly{\ensuremath{\alpha}} Absorption Systems. V. Optical and Near-Infrared Photometry of Ly{\ensuremath{\alpha}}-absorbing Galaxies at z<1},} \apj, 559, 654, \dodoi{10.1086/322414}

\bibitem[{B. {Co{\c{s}}kuno{\v{g}}lu} {et~al.}(2012){Co{\c{s}}kuno{\v{g}}lu}, {Ak}, {Bilir}, {Karaali}, {{\"O}nal}, {Yaz}, {Gilmore}, \& {Seabroke}}]{Coskunoglu2012}
{Co{\c{s}}kuno{\v{g}}lu}, B., {Ak}, S., {Bilir}, S., {et~al.} 2012, \bibinfo{title}{{Local stellar kinematics from RAVE data - II. Radial metallicity gradient},} \mnras, 419, 2844, \dodoi{10.1111/j.1365-2966.2011.19925.x}

\bibitem[{C. {Conroy} {et~al.}(2019){Conroy}, {Bonaca}, {Cargile}, {Johnson}, {Caldwell}, {Naidu}, {Zaritsky}, {Fabricant}, {Moran}, {Rhee}, {Szentgyorgyi}, {Berlind}, {Calkins}, {Kattner}, \& {Ly}}]{H3}
{Conroy}, C., {Bonaca}, A., {Cargile}, P., {et~al.} 2019, \bibinfo{title}{{Mapping the Stellar Halo with the H3 Spectroscopic Survey},} \apj, 883, 107, \dodoi{10.3847/1538-4357/ab38b8}

\bibitem[{B. Co{\c{s}}kuno{\v{g}}lu {et~al.}(2011)Co{\c{s}}kuno{\v{g}}lu, Ak, Bilir, Karaali, Yaz, Gilmore, Seabroke, Bienaym{\'e}, Bland-Hawthorn, Campbell, Freeman, Gibson, Grebel, Munari, Navarro, Parker, Siebert, Siviero, Steinmetz, Watson, Wyse, \& Zwitter}]{Coskunoglu2011}
Co{\c{s}}kuno{\v{g}}lu, B., Ak, S., Bilir, S., {et~al.} 2011, \bibinfo{title}{{Local stellar kinematics from {RAVE} data - {I.} Local standard of rest},} \mnras, 412, 1237, \dodoi{10.1111/j.1365-2966.2010.17983.x}

\bibitem[{D.~P. {Cox} \& G.~C. {G{\'o}mez}(2002){Cox} \& {G{\'o}mez}}]{CoxGomes2002}
{Cox}, D.~P., \& {G{\'o}mez}, G.~C. 2002, \bibinfo{title}{{Analytical Expressions for Spiral Arm Gravitational Potential and Density},} \apjs, 142, 261, \dodoi{10.1086/341946}

\bibitem[{T. \c{S}ahin(2017)\c{S}ahin}]{Sahin2017}
\c{S}ahin, T. 2017, \bibinfo{title}{Turkish Journal of Physics LIME: Semiautomated line measurement and identification from stellar spectra,} Turkish Journal of Physics, 41, 367, \dodoi{10.3906/fiz-1704-13}

\bibitem[{T. {{\c{S}}ahin} \& S. {Bilir}(2020){{\c{S}}ahin} \& {Bilir}}]{Sahin2020}
{{\c{S}}ahin}, T., \& {Bilir}, S. 2020, \bibinfo{title}{{On the Origin of Metal-poor Stars in the Solar Neighborhood},} \apj, 899, 41, \dodoi{10.3847/1538-4357/aba2d2}

\bibitem[{T. {{\c{S}ahin}} {et~al.}(2024){{\c{S}ahin}}, {Guney}, {Aleyna Senturk}, {{\c{C}}{\i}nar}, \& {Marismak}}]{Sahin2024}
{{\c{S}ahin}}, T., {Guney}, F., {Aleyna Senturk}, S., {{\c{C}}{\i}nar}, N., \& {Marismak}, M. 2024, \bibinfo{title}{{An Updated Line List for Spectroscopic Investigation of G Stars II: Refined Solar Abundances via Extended Wavelength Coverage to 10 000 A},} Physics and Astronomy Reports, 2, 65, \dodoi{10.26650/PAR.2024.00007}

\bibitem[{T. {{\c{S}}ahin} \& D.~L. {Lambert}(2009){{\c{S}}ahin} \& {Lambert}}]{Sahin2009}
{{\c{S}}ahin}, T., \& {Lambert}, D.~L. 2009, \bibinfo{title}{{High-resolution optical spectroscopy of a newly discovered post-AGB star with a surprising metallicity in the globular cluster M79},} \mnras, 398, 1730, \dodoi{10.1111/j.1365-2966.2009.15251.x}

\bibitem[{T. {{\c{S}}ahin} {et~al.}(2016){{\c{S}}ahin}, {Lambert}, {Klochkova}, \& {Panchuk}}]{Sahin2016}
{{\c{S}}ahin}, T., {Lambert}, D.~L., {Klochkova}, V.~G., \& {Panchuk}, V.~E. 2016, \bibinfo{title}{{HD 179821 (V1427 Aql, IRAS 19114+0002) - a massive post-red supergiant star?},} \mnras, 461, 4071, \dodoi{10.1093/mnras/stw1586}

\bibitem[{T. {{\c{S}}ahin} {et~al.}(2011){{\c{S}}ahin}, {Lambert}, {Klochkova}, \& {Tavolganskaya}}]{Sahin2011}
{{\c{S}}ahin}, T., {Lambert}, D.~L., {Klochkova}, V.~G., \& {Tavolganskaya}, N.~S. 2011, \bibinfo{title}{{High-resolution optical spectroscopy of the F supergiant protoplanetary nebula IRAS 18095+2704},} \mnras, 410, 612, \dodoi{10.1111/j.1365-2966.2010.17467.x}

\bibitem[{S.~A. {{\c{S}}ent{\"u}rk} {et~al.}(2024){{\c{S}}ent{\"u}rk}, {{\c{S}}ahin}, {G{\"u}ney}, {Bilir}, \& {Mar{\i}{\c{s}}mak}}]{Senturk2024}
{{\c{S}}ent{\"u}rk}, S.~A., {{\c{S}}ahin}, T., {G{\"u}ney}, F., {Bilir}, S., \& {Mar{\i}{\c{s}}mak}, M. 2024, \bibinfo{title}{{Near-infrared Spectroscopy of the Sun and Solar Analog Star HD 76151: Compiling an Extensive Line List in the Y, J, H, and K Bands},} \apj, 976, 175, \dodoi{10.3847/1538-4357/ad85e4}

\bibitem[{R. {da Silva} {et~al.}(2015){da Silva}, {Milone}, \& {Rocha-Pinto}}]{daSilva2015}
{da Silva}, R., {Milone}, A. d.~C., \& {Rocha-Pinto}, H.~J. 2015, \bibinfo{title}{{Homogeneous abundance analysis of FGK dwarf, subgiant, and giant stars with and without giant planets},} \aap, 580, A24, \dodoi{10.1051/0004-6361/201525770}

\bibitem[{R.~S. {de Jong} {et~al.}(2012){de Jong}, {Bellido-Tirado}, {Chiappini}, {Depagne}, {Haynes}, {Johl}, {Schnurr}, {Schwope}, {Walcher}, {Dionies}, {Haynes}, {Kelz}, {Kitaura}, {Lamer}, {Minchev}, {M{\"u}ller}, {Nuza}, {Olaya}, {Piffl}, {Popow}, {Steinmetz}, {Ural}, {Williams}, {Winkler}, {Wisotzki}, {Ansorge}, {Banerji}, {Gonzalez Solares}, {Irwin}, {Kennicutt}, {King}, {McMahon}, {Koposov}, {Parry}, {Sun}, {Walton}, {Finger}, {Iwert}, {Krumpe}, {Lizon}, {Vincenzo}, {Amans}, {Bonifacio}, {Cohen}, {Francois}, {Jagourel}, {Mignot}, {Royer}, {Sartoretti}, {Bender}, {Grupp}, {Hess}, {Lang-Bardl}, {Muschielok}, {B{\"o}hringer}, {Boller}, {Bongiorno}, {Brusa}, {Dwelly}, {Merloni}, {Nandra}, {Salvato}, {Pragt}, {Navarro}, {Gerlofsma}, {Roelfsema}, {Dalton}, {Middleton}, {Tosh}, {Boeche}, {Caffau}, {Christlieb}, {Grebel}, {Hansen}, {Koch}, {Ludwig}, {Quirrenbach}, {Sbordone}, {Seifert}, {Thimm}, {Trifonov}, {Helmi}, {Trager}, {Feltzing}, {Korn}, \& {Boland}}]{4MOST}
{de Jong}, R.~S., {Bellido-Tirado}, O., {Chiappini}, C., {et~al.} 2012, in Society of Photo-Optical Instrumentation Engineers (SPIE) Conference Series, Vol. 8446, Ground-based and Airborne Instrumentation for Astronomy IV, ed. I.~S. {McLean}, S.~K. {Ramsay}, \& H.~{Takami}, 84460T, \dodoi{10.1117/12.926239}

\bibitem[{W. {Dehnen}(2000){Dehnen}}]{Dehnen2000}
{Dehnen}, W. 2000, \bibinfo{title}{{The Effect of the Outer Lindblad Resonance of the Galactic Bar on the Local Stellar Velocity Distribution},} \aj, 119, 800, \dodoi{10.1086/301226}

\bibitem[{P. {Di Matteo} {et~al.}(2019){Di Matteo}, {Haywood}, {Lehnert}, {Katz}, {Khoperskov}, {Snaith}, {G{\'o}mez}, \& {Robichon}}]{DiMatteo2019}
{Di Matteo}, P., {Haywood}, M., {Lehnert}, M.~D., {et~al.} 2019, \bibinfo{title}{{The Milky Way has no in-situ halo other than the heated thick disc. Composition of the stellar halo and age-dating the last significant merger with Gaia DR2 and APOGEE},} \aap, 632, A4, \dodoi{10.1051/0004-6361/201834929}

\bibitem[{A. {Dotter}(2016){Dotter}}]{Dotter2016}
{Dotter}, A. 2016, \bibinfo{title}{{MESA Isochrones and Stellar Tracks (MIST) 0: Methods for the Construction of Stellar Isochrones},} \apjs, 222, 8, \dodoi{10.3847/0067-0049/222/1/8}

\bibitem[{D.~C. {Dursun} {et~al.}(2024){Dursun}, {Tasdemir}, {Koc}, \& {Iyer}}]{Dursun2024}
{Dursun}, D.~C., {Tasdemir}, S., {Koc}, S., \& {Iyer}, S. 2024, \bibinfo{title}{{SED Analysis of the Old Open Cluster NGC 188},} Physics and Astronomy Reports, 2, 1, \dodoi{10.26650/PAR.2024.00002}

\bibitem[{B. {Edvardsson} {et~al.}(1993){Edvardsson}, {Andersen}, {Gustafsson}, {Lambert}, {Nissen}, \& {Tomkin}}]{edvardsson1993}
{Edvardsson}, B., {Andersen}, J., {Gustafsson}, B., {et~al.} 1993, \bibinfo{title}{{The Chemical Evolution of the Galactic Disk - Part One - Analysis and Results},} \aap, 275, 101

\bibitem[{O.~J. {Eggen} {et~al.}(1962){Eggen}, {Lynden-Bell}, \& {Sandage}}]{Eggen1962}
{Eggen}, O.~J., {Lynden-Bell}, D., \& {Sandage}, A.~R. 1962, \bibinfo{title}{{Evidence from the motions of old stars that the Galaxy collapsed.},} \apj, 136, 748, \dodoi{10.1086/147433}

\bibitem[{W.~H. {Elsanhoury} {et~al.}(2025){Elsanhoury}, {Haroon}, {Elkholy}, \& {{\c{C}}{\i}nar}}]{Elsanhoury2024}
{Elsanhoury}, W.~H., {Haroon}, A.~A., {Elkholy}, E.~A., \& {{\c{C}}{\i}nar}, D.~C. 2025, \bibinfo{title}{{Deeply comprehensive astrometric, photometric, and kinematic studies of the three OCSN open clusters with Gaia DR3: Deeply comprehensive astrometric, photometric, and kinematic studies},} Journal of Astrophysics and Astronomy, 46, 21, \dodoi{10.1007/s12036-025-10044-0}

\bibitem[{ {ESA}(1997){ESA}}]{ESA1997}
{ESA}, ed. 1997, ESA Special Publication, Vol. 1200, {The HIPPARCOS and TYCHO catalogues. Astrometric and photometric star catalogues derived from the ESA HIPPARCOS Space Astrometry Mission}

\bibitem[{E.~L. {Fitzpatrick}(1999){Fitzpatrick}}]{Fitzpatrick1999}
{Fitzpatrick}, E.~L. 1999, \bibinfo{title}{{Correcting for the Effects of Interstellar Extinction},} \pasp, 111, 63, \dodoi{10.1086/316293}

\bibitem[{G. {Fragione} \& R. {Capuzzo-Dolcetta}(2016){Fragione} \& {Capuzzo-Dolcetta}}]{Fragione2016}
{Fragione}, G., \& {Capuzzo-Dolcetta}, R. 2016, \bibinfo{title}{{High-velocity stars from the interaction of a globular cluster and a massive black hole binary},} \mnras, 458, 2596, \dodoi{10.1093/mnras/stw531}

\bibitem[{A. {Frebel}(2010){Frebel}}]{Frebel2010}
{Frebel}, A. 2010, \bibinfo{title}{{Stellar archaeology: Exploring the Universe with metal-poor stars},} Astronomische Nachrichten, 331, 474, \dodoi{10.1002/asna.201011362}

\bibitem[{A. {Frebel} \& J.~E. {Norris}(2015{\natexlab{a}}){Frebel} \& {Norris}}]{Frebel2015}
{Frebel}, A., \& {Norris}, J.~E. 2015{\natexlab{a}}, \bibinfo{title}{{Near-Field Cosmology with Extremely Metal-Poor Stars},} \araa, 53, 631, \dodoi{10.1146/annurev-astro-082214-122423}

\bibitem[{A. {Frebel} \& J.~E. {Norris}(2015{\natexlab{b}}){Frebel} \& {Norris}}]{Anna2015}
{Frebel}, A., \& {Norris}, J.~E. 2015{\natexlab{b}}, \bibinfo{title}{{Near-Field Cosmology with Extremely Metal-Poor Stars},} \araa, 53, 631, \dodoi{10.1146/annurev-astro-082214-122423}

\bibitem[{K. {Freeman} \& J. {Bland-Hawthorn}(2002){Freeman} \& {Bland-Hawthorn}}]{Freeman2002}
{Freeman}, K., \& {Bland-Hawthorn}, J. 2002, \bibinfo{title}{{The New Galaxy: Signatures of Its Formation},} \araa, 40, 487, \dodoi{10.1146/annurev.astro.40.060401.093840}

\bibitem[{J.~P. {Fulbright}(2000){Fulbright}}]{Fulbright2000}
{Fulbright}, J.~P. 2000, \bibinfo{title}{{Abundances and Kinematics of Field Halo and Disk Stars. I. Observational Data and Abundance Analysis},} \aj, 120, 1841, \dodoi{10.1086/301548}

\bibitem[{ {Gaia Collaboration} {et~al.}(2016{\natexlab{a}}){Gaia Collaboration}, {Brown}, {Vallenari}, {Prusti}, {de Bruijne}, {Mignard}, {Drimmel}, {Babusiaux}, {Bailer-Jones}, {Bastian}, {Biermann}, {Evans}, {Eyer}, {Jansen}, {Jordi}, {Katz}, {Klioner}, {Lammers}, {Lindegren}, {Luri}, {O'Mullane}, {Panem}, {Pourbaix}, {Randich}, {Sartoretti}, {Siddiqui}, {Soubiran}, {Valette}, {van Leeuwen}, {Walton}, {Aerts}, {Arenou}, {Cropper}, {H{\o}g}, {Lattanzi}, {Grebel}, {Holland}, {Huc}, {Passot}, {Perryman}, {Bramante}, {Cacciari}, {Casta{\~n}eda}, {Chaoul}, {Cheek}, {De Angeli}, {Fabricius}, {Guerra}, {Hern{\'a}ndez}, {Jean-Antoine-Piccolo}, {Masana}, {Messineo}, {Mowlavi}, {Nienartowicz}, {Ord{\'o}{\~n}ez-Blanco}, {Panuzzo}, {Portell}, {Richards}, {Riello}, {Seabroke}, {Tanga}, {Th{\'e}venin}, {Torra}, {Els}, {Gracia-Abril}, {Comoretto}, {Garcia-Reinaldos}, {Lock}, {Mercier}, {Altmann}, {Andrae}, {Astraatmadja}, {Bellas-Velidis}, {Benson}, {Berthier}, {Blomme}, {Busso}, {Carry}, {Cellino}, {Clementini},
  {Cowell}, {Creevey}, {Cuypers}, {Davidson}, {De Ridder}, {de Torres}, {Delchambre}, {Dell'Oro}, {Ducourant}, {Fr{\'e}mat}, {Garc{\'\i}a-Torres}, {Gosset}, {Halbwachs}, {Hambly}, {Harrison}, {Hauser}, {Hestroffer}, {Hodgkin}, {Huckle}, {Hutton}, {Jasniewicz}, {Jordan}, {Kontizas}, {Korn}, {Lanzafame}, {Manteiga}, {Moitinho}, {Muinonen}, {Osinde}, {Pancino}, {Pauwels}, {Petit}, {Recio-Blanco}, {Robin}, {Sarro}, {Siopis}, {Smith}, {Smith}, {Sozzetti}, {Thuillot}, {van Reeven}, {Viala}, {Abbas}, {Abreu Aramburu}, {Accart}, {Aguado}, {Allan}, {Allasia}, {Altavilla}, {{\'A}lvarez}, {Alves}, {Anderson}, {Andrei}, {Anglada Varela}, {Antiche}, {Antoja}, {Ant{\'o}n}, {Arcay}, {Bach}, {Baker}, {Balaguer-N{\'u}{\~n}ez}, {Barache}, {Barata}, {Barbier}, {Barblan}, {Barrado y Navascu{\'e}s}, {Barros}, {Barstow}, {Becciani}, {Bellazzini}, {Bello Garc{\'\i}a}, {Belokurov}, {Bendjoya}, {Berihuete}, {Bianchi}, {Bienaym{\'e}}, {Billebaud}, {Blagorodnova}, {Blanco-Cuaresma}, {Boch}, {Bombrun}, {Borrachero}, {Bouquillon},
  {Bourda}, {Bouy}, {Bragaglia}, {Breddels}, {Brouillet}, {Br{\"u}semeister}, {Bucciarelli}, {Burgess}, {Burgon}, {Burlacu}, {Busonero}, {Buzzi}, {Caffau}, {Cambras}, {Campbell}, {Cancelliere}, {Cantat-Gaudin}, {Carlucci}, {Carrasco}, {Castellani}, {Charlot}, {Charnas}, {Chiavassa}, {Clotet}, {Cocozza}, {Collins}, {Costigan}, {Crifo}, {Cross}, {Crosta}, {Crowley}, {Dafonte}, {Damerdji}, {Dapergolas}, {David}, {David}, \& {De Cat}}]{GaiaDR1}
{Gaia Collaboration}, {Brown}, A.~G.~A., {Vallenari}, A., {et~al.} 2016{\natexlab{a}}, \bibinfo{title}{{Gaia Data Release 1. Summary of the astrometric, photometric, and survey properties},} \aap, 595, A2, \dodoi{10.1051/0004-6361/201629512}

\bibitem[{ {Gaia Collaboration} {et~al.}(2016{\natexlab{b}}){Gaia Collaboration}, {Prusti}, {de Bruijne}, {Brown}, {Vallenari}, {Babusiaux}, {Bailer-Jones}, {Bastian}, {Biermann}, {Evans}, {Eyer}, {Jansen}, {Jordi}, {Klioner}, {Lammers}, {Lindegren}, {Luri}, {Mignard}, {Milligan}, {Panem}, {Poinsignon}, {Pourbaix}, {Randich}, {Sarri}, {Sartoretti}, {Siddiqui}, {Soubiran}, {Valette}, {van Leeuwen}, {Walton}, {Aerts}, {Arenou}, {Cropper}, {Drimmel}, {H{\o}g}, {Katz}, {Lattanzi}, {O'Mullane}, {Grebel}, {Holland}, {Huc}, {Passot}, {Bramante}, {Cacciari}, {Casta{\~n}eda}, {Chaoul}, {Cheek}, {De Angeli}, {Fabricius}, {Guerra}, {Hern{\'a}ndez}, {Jean-Antoine-Piccolo}, {Masana}, {Messineo}, {Mowlavi}, {Nienartowicz}, {Ord{\'o}{\~n}ez-Blanco}, {Panuzzo}, {Portell}, {Richards}, {Riello}, {Seabroke}, {Tanga}, {Th{\'e}venin}, {Torra}, {Els}, {Gracia-Abril}, {Comoretto}, {Garcia-Reinaldos}, {Lock}, {Mercier}, {Altmann}, {Andrae}, {Astraatmadja}, {Bellas-Velidis}, {Benson}, {Berthier}, {Blomme}, {Busso}, {Carry},
  {Cellino}, {Clementini}, {Cowell}, {Creevey}, {Cuypers}, {Davidson}, {De Ridder}, {de Torres}, {Delchambre}, {Dell'Oro}, {Ducourant}, {Fr{\'e}mat}, {Garc{\'\i}a-Torres}, {Gosset}, {Halbwachs}, {Hambly}, {Harrison}, {Hauser}, {Hestroffer}, {Hodgkin}, {Huckle}, {Hutton}, {Jasniewicz}, {Jordan}, {Kontizas}, {Korn}, {Lanzafame}, {Manteiga}, {Moitinho}, {Muinonen}, {Osinde}, {Pancino}, {Pauwels}, {Petit}, {Recio-Blanco}, {Robin}, {Sarro}, {Siopis}, {Smith}, {Smith}, {Sozzetti}, {Thuillot}, {van Reeven}, {Viala}, {Abbas}, {Abreu Aramburu}, {Accart}, {Aguado}, {Allan}, {Allasia}, {Altavilla}, {{\'A}lvarez}, {Alves}, {Anderson}, {Andrei}, {Anglada Varela}, {Antiche}, {Antoja}, {Ant{\'o}n}, {Arcay}, {Atzei}, {Ayache}, {Bach}, {Baker}, {Balaguer-N{\'u}{\~n}ez}, {Barache}, {Barata}, {Barbier}, {Barblan}, {Baroni}, {Barrado y Navascu{\'e}s}, {Barros}, {Barstow}, {Becciani}, {Bellazzini}, {Bellei}, {Bello Garc{\'\i}a}, {Belokurov}, {Bendjoya}, {Berihuete}, {Bianchi}, {Bienaym{\'e}}, {Billebaud}, {Blagorodnova},
  {Blanco-Cuaresma}, {Boch}, {Bombrun}, {Borrachero}, {Bouquillon}, {Bourda}, {Bouy}, {Bragaglia}, {Breddels}, {Brouillet}, {Br{\"u}semeister}, {Bucciarelli}, {Budnik}, {Burgess}, {Burgon}, {Burlacu}, {Busonero}, {Buzzi}, {Caffau}, {Cambras}, {Campbell}, {Cancelliere}, {Cantat-Gaudin}, {Carlucci}, {Carrasco}, {Castellani}, {Charlot}, {Charnas}, {Charvet}, {Chassat}, {Chiavassa}, {Clotet}, {Cocozza}, {Collins}, {Collins}, \& {Costigan}}]{Gaia_Mission}
{Gaia Collaboration}, {Prusti}, T., {de Bruijne}, J.~H.~J., {et~al.} 2016{\natexlab{b}}, \bibinfo{title}{{The Gaia mission},} \aap, 595, A1, \dodoi{10.1051/0004-6361/201629272}

\bibitem[{ {Gaia Collaboration} {et~al.}(2018){Gaia Collaboration}, {Brown}, {Vallenari}, {Prusti}, {de Bruijne}, {Babusiaux}, {Bailer-Jones}, {Biermann}, {Evans}, {Eyer}, {Jansen}, {Jordi}, {Klioner}, {Lammers}, {Lindegren}, {Luri}, {Mignard}, {Panem}, {Pourbaix}, {Randich}, {Sartoretti}, {Siddiqui}, {Soubiran}, {van Leeuwen}, {Walton}, {Arenou}, {Bastian}, {Cropper}, {Drimmel}, {Katz}, {Lattanzi}, {Bakker}, {Cacciari}, {Casta{\~n}eda}, {Chaoul}, {Cheek}, {De Angeli}, {Fabricius}, {Guerra}, {Holl}, {Masana}, {Messineo}, {Mowlavi}, {Nienartowicz}, {Panuzzo}, {Portell}, {Riello}, {Seabroke}, {Tanga}, {Th{\'e}venin}, {Gracia-Abril}, {Comoretto}, {Garcia-Reinaldos}, {Teyssier}, {Altmann}, {Andrae}, {Audard}, {Bellas-Velidis}, {Benson}, {Berthier}, {Blomme}, {Burgess}, {Busso}, {Carry}, {Cellino}, {Clementini}, {Clotet}, {Creevey}, {Davidson}, {De Ridder}, {Delchambre}, {Dell'Oro}, {Ducourant}, {Fern{\'a}ndez-Hern{\'a}ndez}, {Fouesneau}, {Fr{\'e}mat}, {Galluccio}, {Garc{\'\i}a-Torres},
  {Gonz{\'a}lez-N{\'u}{\~n}ez}, {Gonz{\'a}lez-Vidal}, {Gosset}, {Guy}, {Halbwachs}, {Hambly}, {Harrison}, {Hern{\'a}ndez}, {Hestroffer}, {Hodgkin}, {Hutton}, {Jasniewicz}, {Jean-Antoine-Piccolo}, {Jordan}, {Korn}, {Krone-Martins}, {Lanzafame}, {Lebzelter}, {L{\"o}ffler}, {Manteiga}, {Marrese}, {Mart{\'\i}n-Fleitas}, {Moitinho}, {Mora}, {Muinonen}, {Osinde}, {Pancino}, {Pauwels}, {Petit}, {Recio-Blanco}, {Richards}, {Rimoldini}, {Robin}, {Sarro}, {Siopis}, {Smith}, {Sozzetti}, {S{\"u}veges}, {Torra}, {van Reeven}, {Abbas}, {Abreu Aramburu}, {Accart}, {Aerts}, {Altavilla}, {{\'A}lvarez}, {Alvarez}, {Alves}, {Anderson}, {Andrei}, {Anglada Varela}, {Antiche}, {Antoja}, {Arcay}, {Astraatmadja}, {Bach}, {Baker}, {Balaguer-N{\'u}{\~n}ez}, {Balm}, {Barache}, {Barata}, {Barbato}, {Barblan}, {Barklem}, {Barrado}, {Barros}, {Barstow}, {Bartholom{\'e} Mu{\~n}oz}, {Bassilana}, {Becciani}, {Bellazzini}, {Berihuete}, {Bertone}, {Bianchi}, {Bienaym{\'e}}, {Blanco-Cuaresma}, {Boch}, {Boeche}, {Bombrun}, {Borrachero},
  {Bossini}, {Bouquillon}, {Bourda}, {Bragaglia}, {Bramante}, {Breddels}, {Bressan}, {Brouillet}, {Br{\"u}semeister}, {Brugaletta}, {Bucciarelli}, {Burlacu}, {Busonero}, {Butkevich}, {Buzzi}, {Caffau}, {Cancelliere}, {Cannizzaro}, {Cantat-Gaudin}, {Carballo}, {Carlucci}, {Carrasco}, {Casamiquela}, {Castellani}, {Castro-Ginard}, {Charlot}, {Chemin}, {Chiavassa}, {Cocozza}, {Costigan}, {Cowell}, {Crifo}, {Crosta}, {Crowley}, {Cuypers}, {Dafonte}, {Damerdji}, {Dapergolas}, {David}, {David}, {de Laverny}, {De Luise}, {De March}, {de Martino}, {de Souza}, {de Torres}, {Debosscher}, {del Pozo}, {Delbo}, {Delgado}, {Delgado}, {Di Matteo}, {Diakite}, {Diener}, {Distefano}, {Dolding}, {Drazinos}, {Dur{\'a}n}, {Edvardsson}, {Enke}, {Eriksson}, {Esquej}, {Eynard Bontemps}, {Fabre}, {Fabrizio}, {Faigler}, {Falc{\~a}o}, {Farr{\`a}s Casas}, {Federici}, {Fedorets}, {Fernique}, {Figueras}, {Filippi}, {Findeisen}, {Fonti}, {Fraile}, {Fraser}, {Fr{\'e}zouls}, {Gai}, {Galleti}, {Garabato}, {Garc{\'\i}a-Sedano}, {Garofalo},
  {Garralda}, {Gavel}, {Gavras}, {Gerssen}, {Geyer}, {Giacobbe}, {Gilmore}, {Girona}, {Giuffrida}, {Glass}, {Gomes}, {Granvik}, {Gueguen}, {Guerrier}, {Guiraud}, {Guti{\'e}rrez-S{\'a}nchez}, {Haigron}, {Hatzidimitriou}, {Hauser}, {Haywood}, {Heiter}, {Helmi}, {Heu}, {Hilger}, {Hobbs}, {Hofmann}, {Holland}, {Huckle}, {Hypki}, {Icardi}, {Jan{\ss}en}, {Jevardat de Fombelle}, {Jonker}, {Juh{\'a}sz}, {Julbe}, {Karampelas}, {Kewley}, {Klar}, {Kochoska}, {Kohley}, {Kolenberg}, {Kontizas}, {Kontizas}, {Koposov}, {Kordopatis}, {Kostrzewa-Rutkowska}, {Koubsky}, {Lambert}, {Lanza}, {Lasne}, {Lavigne}, {Le Fustec}, {Le Poncin-Lafitte}, {Lebreton}, {Leccia}, {Leclerc}, {Lecoeur-Taibi}, {Lenhardt}, {Leroux}, {Liao}, {Licata}, {Lindstr{\o}m}, {Lister}, {Livanou}, {Lobel}, {L{\'o}pez}, {Managau}, {Mann}, {Mantelet}, {Marchal}, {Marchant}, {Marconi}, {Marinoni}, {Marschalk{\'o}}, {Marshall}, {Martino}, {Marton}, {Mary}, {Massari}, {Matijevi{\v{c}}}, {Mazeh}, {McMillan}, {Messina}, {Michalik}, {Millar}, {Molina}, {Molinaro},
  {Moln{\'a}r}, {Montegriffo}, {Mor}, {Morbidelli}, {Morel}, {Morris}, {Mulone}, {Muraveva}, {Musella}, {Nelemans}, {Nicastro}, {Noval}, {O'Mullane}, {Ord{\'e}novic}, {Ord{\'o}{\~n}ez-Blanco}, {Osborne}, {Pagani}, {Pagano}, {Pailler}, {Palacin}, {Palaversa}, {Panahi}, {Pawlak}, {Piersimoni}, {Pineau}, {Plachy}, {Plum}, {Poggio}, {Poujoulet}, {Pr{\v{s}}a}, {Pulone}, {Racero}, {Ragaini}, {Rambaux}, {Ramos-Lerate}, {Regibo}, {Reyl{\'e}}, {Riclet}, {Ripepi}, {Riva}, {Rivard}, {Rixon}, {Roegiers}, {Roelens}, {Romero-G{\'o}mez}, {Rowell}, {Royer}, {Ruiz-Dern}, {Sadowski}, {Sagrist{\`a} Sell{\'e}s}, {Sahlmann}, {Salgado}, {Salguero}, {Sanna}, {Santana-Ros}, {Sarasso}, {Savietto}, {Schultheis}, {Sciacca}, {Segol}, {Segovia}, {S{\'e}gransan}, {Shih}, {Siltala}, {Silva}, {Smart}, {Smith}, {Solano}, {Solitro}, {Sordo}, {Soria Nieto}, {Souchay}, {Spagna}, {Spoto}, {Stampa}, {Steele}, {Steidelm{\"u}ller}, {Stephenson}, {Stoev}, {Suess}, {Surdej}, {Szabados}, {Szegedi-Elek}, {Tapiador}, {Taris}, {Tauran}, {Taylor},
  {Teixeira}, {Terrett}, {Teyssandier}, {Thuillot}, {Titarenko}, {Torra Clotet}, {Turon}, {Ulla}, {Utrilla}, {Uzzi}, {Vaillant}, {Valentini}, {Valette}, {van Elteren}, {Van Hemelryck}, {van Leeuwen}, {Vaschetto}, {Vecchiato}, {Veljanoski}, {Viala}, {Vicente}, {Vogt}, {von Essen}, {Voss}, {Votruba}, {Voutsinas}, {Walmsley}, {Weiler}, {Wertz}, {Wevers}, {Wyrzykowski}, {Yoldas}, {{\v{Z}}erjal}, {Ziaeepour}, {Zorec}, {Zschocke}, {Zucker}, {Zurbach}, \& {Zwitter}}]{GaiaDR2}
{Gaia Collaboration}, {Brown}, A.~G.~A., {Vallenari}, A., {et~al.} 2018, \bibinfo{title}{{Gaia Data Release 2. Summary of the contents and survey properties},} \aap, 616, A1, \dodoi{10.1051/0004-6361/201833051}

\bibitem[{ {Gaia Collaboration} {et~al.}(2021){Gaia Collaboration}, {Brown}, {Vallenari}, {Prusti}, {de Bruijne}, {Babusiaux}, {Biermann}, {Creevey}, {Evans}, {Eyer}, {Hutton}, {Jansen}, {Jordi}, {Klioner}, {Lammers}, {Lindegren}, {Luri}, {Mignard}, {Panem}, {Pourbaix}, {Randich}, {Sartoretti}, {Soubiran}, {Walton}, {Arenou}, {Bailer-Jones}, {Bastian}, {Cropper}, {Drimmel}, {Katz}, {Lattanzi}, {van Leeuwen}, {Bakker}, {Cacciari}, {Casta{\~n}eda}, {De Angeli}, {Ducourant}, {Fabricius}, {Fouesneau}, {Fr{\'e}mat}, {Guerra}, {Guerrier}, {Guiraud}, {Jean-Antoine Piccolo}, {Masana}, {Messineo}, {Mowlavi}, {Nicolas}, {Nienartowicz}, {Pailler}, {Panuzzo}, {Riclet}, {Roux}, {Seabroke}, {Sordo}, {Tanga}, {Th{\'e}venin}, {Gracia-Abril}, {Portell}, {Teyssier}, {Altmann}, {Andrae}, {Bellas-Velidis}, {Benson}, {Berthier}, {Blomme}, {Brugaletta}, {Burgess}, {Busso}, {Carry}, {Cellino}, {Cheek}, {Clementini}, {Damerdji}, {Davidson}, {Delchambre}, {Dell'Oro}, {Fern{\'a}ndez-Hern{\'a}ndez}, {Galluccio}, {Garc{\'\i}a-Lario},
  {Garcia-Reinaldos}, {Gonz{\'a}lez-N{\'u}{\~n}ez}, {Gosset}, {Haigron}, {Halbwachs}, {Hambly}, {Harrison}, {Hatzidimitriou}, {Heiter}, {Hern{\'a}ndez}, {Hestroffer}, {Hodgkin}, {Holl}, {Jan{\ss}en}, {Jevardat de Fombelle}, {Jordan}, {Krone-Martins}, {Lanzafame}, {L{\"o}ffler}, {Lorca}, {Manteiga}, {Marchal}, {Marrese}, {Moitinho}, {Mora}, {Muinonen}, {Osborne}, {Pancino}, {Pauwels}, {Petit}, {Recio-Blanco}, {Richards}, {Riello}, {Rimoldini}, {Robin}, {Roegiers}, {Rybizki}, {Sarro}, {Siopis}, {Smith}, {Sozzetti}, {Ulla}, {Utrilla}, {van Leeuwen}, {van Reeven}, {Abbas}, {Abreu Aramburu}, {Accart}, {Aerts}, {Aguado}, {Ajaj}, {Altavilla}, {{\'A}lvarez}, {{\'A}lvarez Cid-Fuentes}, {Alves}, {Anderson}, {Anglada Varela}, {Antoja}, {Audard}, {Baines}, {Baker}, {Balaguer-N{\'u}{\~n}ez}, {Balbinot}, {Balog}, {Barache}, {Barbato}, {Barros}, {Barstow}, {Bartolom{\'e}}, {Bassilana}, {Bauchet}, {Baudesson-Stella}, {Becciani}, {Bellazzini}, {Bernet}, {Bertone}, {Bianchi}, {Blanco-Cuaresma}, {Boch}, {Bombrun}, {Bossini},
  {Bouquillon}, {Bragaglia}, {Bramante}, {Breedt}, {Bressan}, {Brouillet}, {Bucciarelli}, {Burlacu}, {Busonero}, {Butkevich}, {Buzzi}, {Caffau}, {Cancelliere}, {C{\'a}novas}, {Cantat-Gaudin}, {Carballo}, {Carlucci}, {Carnerero}, {Carrasco}, {Casamiquela}, {Castellani}, {Castro-Ginard}, {Castro Sampol}, {Chaoul}, {Charlot}, {Chemin}, {Chiavassa}, {Cioni}, {Comoretto}, {Cooper}, {Cornez}, {Cowell}, {Crifo}, {Crosta}, {Crowley}, {Dafonte}, {Dapergolas}, {David}, {David}, {de Laverny}, {De Luise}, {De March}, {De Ridder}, {de Souza}, {de Teodoro}, {de Torres}, {del Peloso}, {del Pozo}, {Delbo}, {Delgado}, {Delgado}, {Delisle}, {Di Matteo}, {Diakite}, {Diener}, {Distefano}, {Dolding}, {Eappachen}, {Edvardsson}, {Enke}, {Esquej}, {Fabre}, {Fabrizio}, {Faigler}, {Fedorets}, {Fernique}, {Fienga}, {Figueras}, {Fouron}, {Fragkoudi}, {Fraile}, {Franke}, {Gai}, {Garabato}, {Garcia-Gutierrez}, {Garc{\'\i}a-Torres}, {Garofalo}, {Gavras}, {Gerlach}, {Geyer}, {Giacobbe}, {Gilmore}, {Girona}, {Giuffrida}, {Gomel}, {Gomez},
  {Gonzalez-Santamaria}, {Gonz{\'a}lez-Vidal}, {Granvik}, {Guti{\'e}rrez-S{\'a}nchez}, {Guy}, {Hauser}, {Haywood}, {Helmi}, {Hidalgo}, {Hilger}, {H{\l}adczuk}, {Hobbs}, {Holland}, {Huckle}, {Jasniewicz}, {Jonker}, {Juaristi Campillo}, {Julbe}, {Karbevska}, {Kervella}, {Khanna}, {Kochoska}, {Kontizas}, {Kordopatis}, {Korn}, {Kostrzewa-Rutkowska}, {Kruszy{\'n}ska}, {Lambert}, {Lanza}, {Lasne}, {Le Campion}, {Le Fustec}, {Lebreton}, {Lebzelter}, {Leccia}, {Leclerc}, {Lecoeur-Taibi}, {Liao}, {Licata}, {Lindstr{\o}m}, {Lister}, {Livanou}, {Lobel}, {Madrero Pardo}, {Managau}, {Mann}, {Marchant}, {Marconi}, {Marcos Santos}, {Marinoni}, {Marocco}, {Marshall}, {Martin Polo}, {Mart{\'\i}n-Fleitas}, {Masip}, {Massari}, {Mastrobuono-Battisti}, {Mazeh}, {McMillan}, {Messina}, {Michalik}, {Millar}, {Mints}, {Molina}, {Molinaro}, {Moln{\'a}r}, {Montegriffo}, {Mor}, {Morbidelli}, {Morel}, {Morris}, {Mulone}, {Munoz}, {Muraveva}, {Murphy}, {Musella}, {Noval}, {Ord{\'e}novic}, {Orr{\`u}}, {Osinde}, {Pagani}, {Pagano},
  {Palaversa}, {Palicio}, {Panahi}, {Pawlak}, {Pe{\~n}alosa Esteller}, {Penttil{\"a}}, {Piersimoni}, {Pineau}, {Plachy}, {Plum}, {Poggio}, {Poretti}, {Poujoulet}, {Pr{\v{s}}a}, {Pulone}, {Racero}, {Ragaini}, {Rainer}, {Raiteri}, {Rambaux}, {Ramos}, {Ramos-Lerate}, {Re Fiorentin}, {Regibo}, {Reyl{\'e}}, {Ripepi}, {Riva}, {Rixon}, {Robichon}, {Robin}, {Roelens}, {Rohrbasser}, {Romero-G{\'o}mez}, {Rowell}, {Royer}, {Rybicki}, {Sadowski}, {Sagrist{\`a} Sell{\'e}s}, {Sahlmann}, {Salgado}, {Salguero}, {Samaras}, {Sanchez Gimenez}, {Sanna}, {Santove{\~n}a}, {Sarasso}, {Schultheis}, {Sciacca}, {Segol}, {Segovia}, {S{\'e}gransan}, {Semeux}, {Shahaf}, {Siddiqui}, {Siebert}, {Siltala}, {Slezak}, {Smart}, {Solano}, {Solitro}, {Souami}, {Souchay}, {Spagna}, {Spoto}, {Steele}, {Steidelm{\"u}ller}, {Stephenson}, {S{\"u}veges}, {Szabados}, {Szegedi-Elek}, {Taris}, {Tauran}, {Taylor}, {Teixeira}, {Thuillot}, {Tonello}, {Torra}, {Torra}, {Turon}, {Unger}, {Vaillant}, {van Dillen}, {Vanel}, {Vecchiato}, {Viala}, {Vicente},
  {Voutsinas}, {Weiler}, {Wevers}, {Wyrzykowski}, {Yoldas}, {Yvard}, {Zhao}, {Zorec}, {Zucker}, {Zurbach}, \& {Zwitter}}]{GaiaEDR3}
{Gaia Collaboration}, {Brown}, A.~G.~A., {Vallenari}, A., {et~al.} 2021, \bibinfo{title}{{Gaia Early Data Release 3. Summary of the contents and survey properties},} \aap, 649, A1, \dodoi{10.1051/0004-6361/202039657}

\bibitem[{ {Gaia Collaboration} {et~al.}(2023){Gaia Collaboration}, {Vallenari}, {Brown}, {Prusti}, {de Bruijne}, {Arenou}, {Babusiaux}, {Biermann}, {Creevey}, {Ducourant}, {Evans}, {Eyer}, {Guerra}, {Hutton}, {Jordi}, {Klioner}, {Lammers}, {Lindegren}, {Luri}, {Mignard}, {Panem}, {Pourbaix}, {Randich}, {Sartoretti}, {Soubiran}, {Tanga}, {Walton}, {Bailer-Jones}, {Bastian}, {Drimmel}, {Jansen}, {Katz}, {Lattanzi}, {van Leeuwen}, {Bakker}, {Cacciari}, {Casta{\~n}eda}, {De Angeli}, {Fabricius}, {Fouesneau}, {Fr{\'e}mat}, {Galluccio}, {Guerrier}, {Heiter}, {Masana}, {Messineo}, {Mowlavi}, {Nicolas}, {Nienartowicz}, {Pailler}, {Panuzzo}, {Riclet}, {Roux}, {Seabroke}, {Sordo}, {Th{\'e}venin}, {Gracia-Abril}, {Portell}, {Teyssier}, {Altmann}, {Andrae}, {Audard}, {Bellas-Velidis}, {Benson}, {Berthier}, {Blomme}, {Burgess}, {Busonero}, {Busso}, {C{\'a}novas}, {Carry}, {Cellino}, {Cheek}, {Clementini}, {Damerdji}, {Davidson}, {de Teodoro}, {Nu{\~n}ez Campos}, {Delchambre}, {Dell'Oro}, {Esquej},
  {Fern{\'a}ndez-Hern{\'a}ndez}, {Fraile}, {Garabato}, {Garc{\'\i}a-Lario}, {Gosset}, {Haigron}, {Halbwachs}, {Hambly}, {Harrison}, {Hern{\'a}ndez}, {Hestroffer}, {Hodgkin}, {Holl}, {Jan{\ss}en}, {Jevardat de Fombelle}, {Jordan}, {Krone-Martins}, {Lanzafame}, {L{\"o}ffler}, {Marchal}, {Marrese}, {Moitinho}, {Muinonen}, {Osborne}, {Pancino}, {Pauwels}, {Recio-Blanco}, {Reyl{\'e}}, {Riello}, {Rimoldini}, {Roegiers}, {Rybizki}, {Sarro}, {Siopis}, {Smith}, {Sozzetti}, {Utrilla}, {van Leeuwen}, {Abbas}, {{\'A}brah{\'a}m}, {Abreu Aramburu}, {Aerts}, {Aguado}, {Ajaj}, {Aldea-Montero}, {Altavilla}, {{\'A}lvarez}, {Alves}, {Anders}, {Anderson}, {Anglada Varela}, {Antoja}, {Baines}, {Baker}, {Balaguer-N{\'u}{\~n}ez}, {Balbinot}, {Balog}, {Barache}, {Barbato}, {Barros}, {Barstow}, {Bartolom{\'e}}, {Bassilana}, {Bauchet}, {Becciani}, {Bellazzini}, {Berihuete}, {Bernet}, {Bertone}, {Bianchi}, {Binnenfeld}, {Blanco-Cuaresma}, {Blazere}, {Boch}, {Bombrun}, {Bossini}, {Bouquillon}, {Bragaglia}, {Bramante}, {Breedt},
  {Bressan}, {Brouillet}, {Brugaletta}, {Bucciarelli}, {Burlacu}, {Butkevich}, {Buzzi}, {Caffau}, {Cancelliere}, {Cantat-Gaudin}, {Carballo}, {Carlucci}, {Carnerero}, {Carrasco}, {Casamiquela}, {Castellani}, {Castro-Ginard}, {Chaoul}, {Charlot}, {Chemin}, {Chiaramida}, {Chiavassa}, {Chornay}, {Comoretto}, {Contursi}, {Cooper}, {Cornez}, {Cowell}, {Crifo}, {Cropper}, {Crosta}, {Crowley}, {Dafonte}, {Dapergolas}, {David}, {David}, {de Laverny}, {De Luise}, {De March}, {De Ridder}, {de Souza}, {de Torres}, {del Peloso}, {del Pozo}, {Delbo}, {Delgado}, {Delisle}, {Demouchy}, {Dharmawardena}, {Di Matteo}, {Diakite}, {Diener}, {Distefano}, {Dolding}, {Edvardsson}, {Enke}, {Fabre}, {Fabrizio}, {Faigler}, {Fedorets}, {Fernique}, {Fienga}, {Figueras}, {Fournier}, {Fouron}, {Fragkoudi}, {Gai}, {Garcia-Gutierrez}, {Garcia-Reinaldos}, {Garc{\'\i}a-Torres}, {Garofalo}, {Gavel}, {Gavras}, {Gerlach}, {Geyer}, {Giacobbe}, {Gilmore}, {Girona}, {Giuffrida}, {Gomel}, {Gomez}, {Gonz{\'a}lez-N{\'u}{\~n}ez},
  {Gonz{\'a}lez-Santamar{\'\i}a}, {Gonz{\'a}lez-Vidal}, {Granvik}, {Guillout}, {Guiraud}, {Guti{\'e}rrez-S{\'a}nchez}, {Guy}, {Hatzidimitriou}, {Hauser}, {Haywood}, {Helmer}, {Helmi}, {Sarmiento}, {Hidalgo}, {Hilger}, {H{\l}adczuk}, {Hobbs}, {Holland}, {Huckle}, {Jardine}, {Jasniewicz}, {Jean-Antoine Piccolo}, {Jim{\'e}nez-Arranz}, {Jorissen}, {Juaristi Campillo}, {Julbe}, {Karbevska}, {Kervella}, {Khanna}, {Kontizas}, {Kordopatis}, {Korn}, {K{\'o}sp{\'a}l}, {Kostrzewa-Rutkowska}, {Kruszy{\'n}ska}, {Kun}, {Laizeau}, {Lambert}, {Lanza}, {Lasne}, {Le Campion}, {Lebreton}, {Lebzelter}, {Leccia}, {Leclerc}, {Lecoeur-Taibi}, {Liao}, {Licata}, {Lindstr{\o}m}, {Lister}, {Livanou}, {Lobel}, {Lorca}, {Loup}, {Madrero Pardo}, {Magdaleno Romeo}, {Managau}, {Mann}, {Manteiga}, {Marchant}, {Marconi}, {Marcos}, {Marcos Santos}, {Mar{\'\i}n Pina}, {Marinoni}, {Marocco}, {Marshall}, {Martin Polo}, {Mart{\'\i}n-Fleitas}, {Marton}, {Mary}, {Masip}, {Massari}, {Mastrobuono-Battisti}, {Mazeh}, {McMillan}, {Messina}, {Michalik},
  {Millar}, {Mints}, {Molina}, {Molinaro}, {Moln{\'a}r}, {Monari}, {Mongui{\'o}}, {Montegriffo}, {Montero}, {Mor}, {Mora}, {Morbidelli}, {Morel}, {Morris}, {Muraveva}, {Murphy}, {Musella}, {Nagy}, {Noval}, {Oca{\~n}a}, {Ogden}, {Ordenovic}, {Osinde}, {Pagani}, {Pagano}, {Palaversa}, {Palicio}, {Pallas-Quintela}, {Panahi}, {Payne-Wardenaar}, {Pe{\~n}alosa Esteller}, {Penttil{\"a}}, {Pichon}, {Piersimoni}, {Pineau}, {Plachy}, {Plum}, {Poggio}, {Pr{\v{s}}a}, {Pulone}, {Racero}, {Ragaini}, {Rainer}, {Raiteri}, {Rambaux}, {Ramos}, {Ramos-Lerate}, {Re Fiorentin}, {Regibo}, {Richards}, {Rios Diaz}, {Ripepi}, {Riva}, {Rix}, {Rixon}, {Robichon}, {Robin}, {Robin}, {Roelens}, {Rogues}, {Rohrbasser}, {Romero-G{\'o}mez}, {Rowell}, {Royer}, {Ruz Mieres}, {Rybicki}, {Sadowski}, {S{\'a}ez N{\'u}{\~n}ez}, {Sagrist{\`a} Sell{\'e}s}, {Sahlmann}, {Salguero}, {Samaras}, {Sanchez Gimenez}, {Sanna}, {Santove{\~n}a}, {Sarasso}, {Schultheis}, {Sciacca}, {Segol}, {Segovia}, {S{\'e}gransan}, {Semeux}, {Shahaf}, {Siddiqui}, {Siebert},
  {Siltala}, {Silvelo}, {Slezak}, {Slezak}, {Smart}, {Snaith}, {Solano}, {Solitro}, {Souami}, {Souchay}, {Spagna}, {Spina}, {Spoto}, {Steele}, {Steidelm{\"u}ller}, {Stephenson}, {S{\"u}veges}, {Surdej}, {Szabados}, {Szegedi-Elek}, {Taris}, {Taylor}, {Teixeira}, {Tolomei}, {Tonello}, {Torra}, {Torra}, {Torralba Elipe}, {Trabucchi}, {Tsounis}, {Turon}, {Ulla}, {Unger}, {Vaillant}, {van Dillen}, {van Reeven}, {Vanel}, {Vecchiato}, {Viala}, {Vicente}, {Voutsinas}, {Weiler}, {Wevers}, {Wyrzykowski}, {Yoldas}, {Yvard}, {Zhao}, {Zorec}, {Zucker}, \& {Zwitter}}]{GaiaDR3}
{Gaia Collaboration}, {Vallenari}, A., {Brown}, A.~G.~A., {et~al.} 2023, \bibinfo{title}{{Gaia Data Release 3. Summary of the content and survey properties},} \aap, 674, A1, \dodoi{10.1051/0004-6361/202243940}

\bibitem[{A.~E. {Garc{\'\i}a P{\'e}rez} {et~al.}(2013){Garc{\'\i}a P{\'e}rez}, {Cunha}, {Shetrone}, {Majewski}, {Johnson}, {Smith}, {Schiavon}, {Holtzman}, {Nidever}, {Zasowski}, {Allende Prieto}, {Beers}, {Bizyaev}, {Ebelke}, {Eisenstein}, {Frinchaboy}, {Girardi}, {Hearty}, {Malanushenko}, {Malanushenko}, {Meszaros}, {O'Connell}, {Oravetz}, {Pan}, {Robin}, {Schneider}, {Schultheis}, {Skrutskie}, {Simmonsand}, \& {Wilson}}]{Garcia2013}
{Garc{\'\i}a P{\'e}rez}, A.~E., {Cunha}, K., {Shetrone}, M., {et~al.} 2013, \bibinfo{title}{{Very Metal-poor Stars in the Outer Galactic Bulge Found by the APOGEE Survey},} \apjl, 767, L9, \dodoi{10.1088/2041-8205/767/1/L9}

\bibitem[{G. {Gilmore} \& N. {Reid}(1983){Gilmore} \& {Reid}}]{Gilmore1983}
{Gilmore}, G., \& {Reid}, N. 1983, \bibinfo{title}{{New light on faint stars - III. Galactic structure towards the South Pole and the Galactic thick disc.},} \mnras, 202, 1025, \dodoi{10.1093/mnras/202.4.1025}

\bibitem[{O.~Y. {Gnedin} {et~al.}(1999){Gnedin}, {Hernquist}, \& {Ostriker}}]{Gnedin1999}
{Gnedin}, O.~Y., {Hernquist}, L., \& {Ostriker}, J.~P. 1999, \bibinfo{title}{{Tidal Shocking by Extended Mass Distributions},} \apj, 514, 109, \dodoi{10.1086/306910}

\bibitem[{F. {Gran} {et~al.}(2021){Gran}, {Zoccali}, {Rojas-Arriagada}, {Saviane}, {Contreras Ramos}, {Beaton}, {Bizyaev}, {Cohen}, {Fern{\'a}ndez-Trincado}, {Garc{\'\i}a-Hern{\'a}ndez}, {Geisler}, {Lane}, {Minniti}, {Moni Bidin}, {Nitschelm}, {Olivares Carvajal}, {Pan}, {Rojas}, \& {Villanova}}]{Gran2021}
{Gran}, F., {Zoccali}, M., {Rojas-Arriagada}, A., {et~al.} 2021, \bibinfo{title}{{APOGEE view of the globular cluster NGC 6544},} \mnras, 504, 3494, \dodoi{10.1093/mnras/stab1051}

\bibitem[{C.~K. {Harada} {et~al.}(2024){Harada}, {Dressing}, {Kane}, \& {Ardestani}}]{Harada2024}
{Harada}, C.~K., {Dressing}, C.~D., {Kane}, S.~R., \& {Ardestani}, B.~A. 2024, \bibinfo{title}{{Setting the Stage for the Search for Life with the Habitable Worlds Observatory: Properties of 164 Promising Planet-survey Targets},} \apjs, 272, 30, \dodoi{10.3847/1538-4365/ad3e81}

\bibitem[{A.~A. {Haroon} {et~al.}(2025){Haroon}, {Elsanhoury}, {Elkholy}, {Saad}, \& {{\c{C}}{\i}nar}}]{Haroon2025}
{Haroon}, A.~A., {Elsanhoury}, W.~H., {Elkholy}, E.~A., {Saad}, A.~S., \& {{\c{C}}{\i}nar}, D.~C. 2025, \bibinfo{title}{{Study of open star clusters using the gaia DR3: I-poorly studied king 2 and king 5},} \physscr, 100, 055006, \dodoi{10.1088/1402-4896/adbf71}

\bibitem[{K. {Hawkins} {et~al.}(2016){Hawkins}, {Jofr{\'e}}, {Heiter}, {Soubiran}, {Blanco-Cuaresma}, {Casagrande}, {Gilmore}, {Lind}, {Magrini}, {Masseron}, {Pancino}, {Randich}, \& {Worley}}]{Hawkins2016}
{Hawkins}, K., {Jofr{\'e}}, P., {Heiter}, U., {et~al.} 2016, \bibinfo{title}{{Gaia FGK benchmark stars: new candidates at low metallicities},} \aap, 592, A70, \dodoi{10.1051/0004-6361/201628268}

\bibitem[{M.~R. {Hayden} {et~al.}(2015){Hayden}, {Bovy}, {Holtzman}, {Nidever}, {Bird}, {Weinberg}, {Andrews}, {Majewski}, {Allende Prieto}, {Anders}, {Beers}, {Bizyaev}, {Chiappini}, {Cunha}, {Frinchaboy}, {Garc{\'\i}a-Her{\'n}andez}, {Garc{\'\i}a P{\'e}rez}, {Girardi}, {Harding}, {Hearty}, {Johnson}, {M{\'e}sz{\'a}ros}, {Minchev}, {O'Connell}, {Pan}, {Robin}, {Schiavon}, {Schneider}, {Schultheis}, {Shetrone}, {Skrutskie}, {Steinmetz}, {Smith}, {Wilson}, {Zamora}, \& {Zasowski}}]{Hayden2015}
{Hayden}, M.~R., {Bovy}, J., {Holtzman}, J.~A., {et~al.} 2015, \bibinfo{title}{{Chemical Cartography with APOGEE: Metallicity Distribution Functions and the Chemical Structure of the Milky Way Disk},} \apj, 808, 132, \dodoi{10.1088/0004-637X/808/2/132}

\bibitem[{S. {Hirano} {et~al.}(2014){Hirano}, {Hosokawa}, {Yoshida}, {Umeda}, {Omukai}, {Chiaki}, \& {Yorke}}]{Hirano2014}
{Hirano}, S., {Hosokawa}, T., {Yoshida}, N., {et~al.} 2014, \bibinfo{title}{{One Hundred First Stars: Protostellar Evolution and the Final Masses},} \apj, 781, 60, \dodoi{10.1088/0004-637X/781/2/60}

\bibitem[{N. {Holanda} {et~al.}(2024){Holanda}, {Flaulhabe}, {Quispe-Huaynasi}, {Sonally}, \& {Pereira}}]{Holanda2024}
{Holanda}, N., {Flaulhabe}, T., {Quispe-Huaynasi}, F., {Sonally}, A., \& {Pereira}, C.~B. 2024, \bibinfo{title}{{The Chemical Puzzle of Weak G-Band Stars: A Comprehensive Study of HD 54627, HD 105783, HD 198718, and HD 201557},} \apj, 971, 152, \dodoi{10.3847/1538-4357/ad58bf}

\bibitem[{T.~O. {Husser} {et~al.}(2013){Husser}, {Wende-von Berg}, {Dreizler}, {Homeier}, {Reiners}, {Barman}, \& {Hauschildt}}]{Husser2013}
{Husser}, T.~O., {Wende-von Berg}, S., {Dreizler}, S., {et~al.} 2013, \bibinfo{title}{{A new extensive library of PHOENIX stellar atmospheres and synthetic spectra},} \aap, 553, A6, \dodoi{10.1051/0004-6361/201219058}

\bibitem[{S. {{\.I}yisan} {et~al.}(2025){{\.I}yisan}, {Bilir}, {{\"O}nal Ta{\c{s}}}, \& {Plevne}}]{iyisan2025}
{{\.I}yisan}, S., {Bilir}, S., {{\"O}nal Ta{\c{s}}}, {\"O}., \& {Plevne}, O. 2025, \bibinfo{title}{{Structural Parameters of the Thin Disk Population from Evolved Stars in the Solar Neighborhood},} \aj, 169, 138, \dodoi{10.3847/1538-3881/ada952}

\bibitem[{P. {Jofr{\'e}} {et~al.}(2015){Jofr{\'e}}, {Heiter}, {Soubiran}, {Blanco-Cuaresma}, {Masseron}, {Nordlander}, {Chemin}, {Worley}, {Van Eck}, {Hourihane}, {Gilmore}, {Adibekyan}, {Bergemann}, {Cantat-Gaudin}, {Delgado-Mena}, {Gonz{\'a}lez Hern{\'a}ndez}, {Guiglion}, {Lardo}, {de Laverny}, {Lind}, {Magrini}, {Mikolaitis}, {Montes}, {Pancino}, {Recio-Blanco}, {Sordo}, {Sousa}, {Tabernero}, \& {Vallenari}}]{jofre2015}
{Jofr{\'e}}, P., {Heiter}, U., {Soubiran}, C., {et~al.} 2015, \bibinfo{title}{{Gaia FGK benchmark stars: abundances of {\ensuremath{\alpha}} and iron-peak elements},} \aap, 582, A81, \dodoi{10.1051/0004-6361/201526604}

\bibitem[{A.~A. {John} {et~al.}(2023){John}, {Collier Cameron}, {Faria}, {Mortier}, {Wilson}, {Malavolta}, {Buchhave}, {Dumusque}, {L{\'o}pez-Morales}, {Haywood}, {Rice}, {Sozzetti}, {Latham}, {Udry}, {Pepe}, {Pinamonti}, {Vanderburg}, {Ghedina}, {Cosentino}, {Stalport}, {Nicholson}, {Fiorenzano}, \& {Poretti}}]{John2023}
{John}, A.~A., {Collier Cameron}, A., {Faria}, J.~P., {et~al.} 2023, \bibinfo{title}{{Sub-m s$^{-1}$ upper limits from a deep HARPS-N radial-velocity search for planets orbiting HD 166620 and HD 144579},} \mnras, 525, 1687, \dodoi{10.1093/mnras/stad2381}

\bibitem[{D.~R.~H. {Johnson} \& D.~R. {Soderblom}(1987){Johnson} \& {Soderblom}}]{Johnson1987}
{Johnson}, D. R.~H., \& {Soderblom}, D.~R. 1987, \bibinfo{title}{{Calculating Galactic Space Velocities and Their Uncertainties, with an Application to the Ursa Major Group},} \aj, 93, 864, \dodoi{10.1086/114370}

\bibitem[{S. {Karaali} {et~al.}(2003){Karaali}, {Ak}, {Bilir}, {Karata{\c{s}}}, \& {Gilmore}}]{Karaali2003}
{Karaali}, S., {Ak}, S.~G., {Bilir}, S., {Karata{\c{s}}}, Y., \& {Gilmore}, G. 2003, \bibinfo{title}{{A charge-coupled device study of high-latitude Galactic structure: testing the model parameters},} \mnras, 343, 1013, \dodoi{10.1046/j.1365-8711.2003.06743.x}

\bibitem[{S. {Karaali} {et~al.}(2004){Karaali}, {Bilir}, \& {Hamzao{\v{g}}lu}}]{Karaali2004}
{Karaali}, S., {Bilir}, S., \& {Hamzao{\v{g}}lu}, E. 2004, \bibinfo{title}{{A different approach for the estimation of Galactic model parameters},} \mnras, 355, 307, \dodoi{10.1111/j.1365-2966.2004.08319.x}

\bibitem[{S.~C. {Keller} {et~al.}(2014){Keller}, {Bessell}, {Frebel}, {Casey}, {Asplund}, {Jacobson}, {Lind}, {Norris}, {Yong}, {Heger}, {Magic}, {da Costa}, {Schmidt}, \& {Tisserand}}]{Keller2014}
{Keller}, S.~C., {Bessell}, M.~S., {Frebel}, A., {et~al.} 2014, \bibinfo{title}{{A single low-energy, iron-poor supernova as the source of metals in the star SMSS J031300.36-670839.3},} \nat, 506, 463, \dodoi{10.1038/nature12990}

\bibitem[{E.~N. {Kirby} {et~al.}(2016){Kirby}, {Guhathakurta}, {Zhang}, {Hong}, {Guo}, {Guo}, {Cohen}, \& {Cunha}}]{Kirby2016}
{Kirby}, E.~N., {Guhathakurta}, P., {Zhang}, A.~J., {et~al.} 2016, \bibinfo{title}{{Lithium-rich Giants in Globular Clusters},} \apj, 819, 135, \dodoi{10.3847/0004-637X/819/2/135}

\bibitem[{M. {Koleva} {et~al.}(2009){Koleva}, {Prugniel}, {Bouchard}, \& {Wu}}]{Koleva2009}
{Koleva}, M., {Prugniel}, P., {Bouchard}, A., \& {Wu}, Y. 2009, \bibinfo{title}{{ULySS: a full spectrum fitting package},} \aap, 501, 1269, \dodoi{10.1051/0004-6361/200811467}

\bibitem[{J.~A. {Kollmeier} {et~al.}(2017){Kollmeier}, {Zasowski}, {Rix}, {Johns}, {Anderson}, {Drory}, {Johnson}, {Pogge}, {Bird}, {Blanc}, {Brownstein}, {Crane}, {De Lee}, {Klaene}, {Kreckel}, {MacDonald}, {Merloni}, {Ness}, {O'Brien}, {Sanchez-Gallego}, {Sayres}, {Shen}, {Thakar}, {Tkachenko}, {Aerts}, {Blanton}, {Eisenstein}, {Holtzman}, {Maoz}, {Nandra}, {Rockosi}, {Weinberg}, {Bovy}, {Casey}, {Chaname}, {Clerc}, {Conroy}, {Eracleous}, {G{\"a}nsicke}, {Hekker}, {Horne}, {Kauffmann}, {McQuinn}, {Pellegrini}, {Schinnerer}, {Schlafly}, {Schwope}, {Seibert}, {Teske}, \& {van Saders}}]{SDSSV}
{Kollmeier}, J.~A., {Zasowski}, G., {Rix}, H.-W., {et~al.} 2017, \bibinfo{title}{{SDSS-V: Pioneering Panoptic Spectroscopy},} arXiv e-prints, arXiv:1711.03234, \dodoi{10.48550/arXiv.1711.03234}

\bibitem[{G. {Kordopatis} {et~al.}(2013){Kordopatis}, {Gilmore}, {Steinmetz}, {Boeche}, {Seabroke}, {Siebert}, {Zwitter}, {Binney}, {de Laverny}, {Recio-Blanco}, {Williams}, {Piffl}, {Enke}, {Roeser}, {Bijaoui}, {Wyse}, {Freeman}, {Munari}, {Carrillo}, {Anguiano}, {Burton}, {Campbell}, {Cass}, {Fiegert}, {Hartley}, {Parker}, {Reid}, {Ritter}, {Russell}, {Stupar}, {Watson}, {Bienaym{\'e}}, {Bland-Hawthorn}, {Gerhard}, {Gibson}, {Grebel}, {Helmi}, {Navarro}, {Conrad}, {Famaey}, {Faure}, {Just}, {Kos}, {Matijevi{\v{c}}}, {McMillan}, {Minchev}, {Scholz}, {Sharma}, {Siviero}, {de Boer}, \& {{\v{Z}}erjal}}]{RaveDR4}
{Kordopatis}, G., {Gilmore}, G., {Steinmetz}, M., {et~al.} 2013, \bibinfo{title}{{The Radial Velocity Experiment (RAVE): Fourth Data Release},} \aj, 146, 134, \dodoi{10.1088/0004-6256/146/5/134}

\bibitem[{M. {Kovalev} {et~al.}(2019){Kovalev}, {Bergemann}, {Ting}, \& {Rix}}]{Kovalev2019}
{Kovalev}, M., {Bergemann}, M., {Ting}, Y.-S., \& {Rix}, H.-W. 2019, \bibinfo{title}{{Non-LTE chemical abundances in Galactic open and globular clusters},} \aap, 628, A54, \dodoi{10.1051/0004-6361/201935861}

\bibitem[{A.~H.~W. {K{\"u}pper} {et~al.}(2012){K{\"u}pper}, {Lane}, \& {Heggie}}]{Kupper2012}
{K{\"u}pper}, A. H.~W., {Lane}, R.~R., \& {Heggie}, D.~C. 2012, \bibinfo{title}{{More on the structure of tidal tails},} \mnras, 420, 2700, \dodoi{10.1111/j.1365-2966.2011.20242.x}

\bibitem[{R.~L. {Kurucz} {et~al.}(1984){Kurucz}, {Furenlid}, {Brault}, \& {Testerman}}]{Kurucz1984}
{Kurucz}, R.~L., {Furenlid}, I., {Brault}, J., \& {Testerman}, L. 1984, {Solar flux atlas from 296 to 1300 nm}

\bibitem[{H.-N. {Li} {et~al.}(2015){Li}, {Zhao}, {Christlieb}, {Wang}, {Wang}, {Zhang}, {Hou}, \& {Yuan}}]{Li2015}
{Li}, H.-N., {Zhao}, G., {Christlieb}, N., {et~al.} 2015, \bibinfo{title}{{Spectroscopic Analysis of Metal-poor Stars from LAMOST: Early Results},} \apj, 798, 110, \dodoi{10.1088/0004-637X/798/2/110}

\bibitem[{J. {Lian} {et~al.}(2020){Lian}, {Zasowski}, {Hasselquist}, {Nataf}, {Thomas}, {Moni Bidin}, {Fern{\'a}ndez-Trincado}, {Garcia-Hernandez}, {Lane}, {Majewski}, {Roman-Lopes}, \& {Schultheis}}]{Lian2020}
{Lian}, J., {Zasowski}, G., {Hasselquist}, S., {et~al.} 2020, \bibinfo{title}{{The Milky Way's bulge star formation history as constrained from its bimodal chemical abundance distribution},} \mnras, 497, 3557, \dodoi{10.1093/mnras/staa2205}

\bibitem[{K. {Lind} {et~al.}(2012){Lind}, {Bergemann}, \& {Asplund}}]{Lind2012}
{Lind}, K., {Bergemann}, M., \& {Asplund}, M. 2012, \bibinfo{title}{{Non-LTE line formation of Fe in late-type stars - II. 1D spectroscopic stellar parameters},} \mnras, 427, 50, \dodoi{10.1111/j.1365-2966.2012.21686.x}

\bibitem[{L. {Lindegren} {et~al.}(2021){Lindegren}, {Klioner}, {Hern{\'a}ndez}, {Bombrun}, {Ramos-Lerate}, {Steidelm{\"u}ller}, {Bastian}, {Biermann}, {de Torres}, {Gerlach}, {Geyer}, {Hilger}, {Hobbs}, {Lammers}, {McMillan}, {Stephenson}, {Casta{\~n}eda}, {Davidson}, {Fabricius}, {Gracia-Abril}, {Portell}, {Rowell}, {Teyssier}, {Torra}, {Bartolom{\'e}}, {Clotet}, {Garralda}, {Gonz{\'a}lez-Vidal}, {Torra}, {Abbas}, {Altmann}, {Anglada Varela}, {Balaguer-N{\'u}{\~n}ez}, {Balog}, {Barache}, {Becciani}, {Bernet}, {Bertone}, {Bianchi}, {Bouquillon}, {Brown}, {Bucciarelli}, {Busonero}, {Butkevich}, {Buzzi}, {Cancelliere}, {Carlucci}, {Charlot}, {Cioni}, {Crosta}, {Crowley}, {del Peloso}, {del Pozo}, {Drimmel}, {Esquej}, {Fienga}, {Fraile}, {Gai}, {Garcia-Reinaldos}, {Guerra}, {Hambly}, {Hauser}, {Jan{\ss}en}, {Jordan}, {Kostrzewa-Rutkowska}, {Lattanzi}, {Liao}, {Licata}, {Lister}, {L{\"o}ffler}, {Marchant}, {Masip}, {Mignard}, {Mints}, {Molina}, {Mora}, {Morbidelli}, {Murphy}, {Pagani}, {Panuzzo}, {Pe{\~n}alosa
  Esteller}, {Poggio}, {Re Fiorentin}, {Riva}, {Sagrist{\`a} Sell{\'e}s}, {Sanchez Gimenez}, {Sarasso}, {Sciacca}, {Siddiqui}, {Smart}, {Souami}, {Spagna}, {Steele}, {Taris}, {Utrilla}, {van Reeven}, \& {Vecchiato}}]{Lindegren2021}
{Lindegren}, L., {Klioner}, S.~A., {Hern{\'a}ndez}, J., {et~al.} 2021, \bibinfo{title}{{Gaia Early Data Release 3. The astrometric solution},} \aap, 649, A2, \dodoi{10.1051/0004-6361/202039709}

\bibitem[{M. Lucey {et~al.}(2019)Lucey, Hawkins, Ness, Asplund, Bensby, Casagrande, Feltzing, Freeman, Kobayashi, \& Marino}]{Lucey2019}
Lucey, M., Hawkins, K., Ness, M., {et~al.} 2019, \bibinfo{title}{The COMBS survey – I. Chemical origins of metal-poor stars in the Galactic bulge,} Monthly Notices of the Royal Astronomical Society, 488, 2283, \dodoi{10.1093/mnras/stz1847}

\bibitem[{M. Lucey {et~al.}(2021)Lucey, Hawkins, Ness, Debattista, Luna, Asplund, Bensby, Casagrande, Feltzing, Freeman, Kobayashi, \& Marino}]{Lucey2021}
Lucey, M., Hawkins, K., Ness, M., {et~al.} 2021, \bibinfo{title}{The COMBS Survey - II. Distinguishing the metal-poor bulge from the halo interlopers,} Monthly Notices of the Royal Astronomical Society, 501, 5981, \dodoi{10.1093/mnras/stab003}

\bibitem[{M. {Lucey} {et~al.}(2022){Lucey}, {Hawkins}, {Ness}, {Nelson}, {Debattista}, {Luna}, {Bensby}, {Freeman}, \& {Kobayashi}}]{Lucey2022}
{Lucey}, M., {Hawkins}, K., {Ness}, M., {et~al.} 2022, \bibinfo{title}{{The COMBS Survey - III. The chemodynamical origins of metal-poor bulge stars},} \mnras, 509, 122, \dodoi{10.1093/mnras/stab2878}

\bibitem[{S.~R. {Majewski}(1993){Majewski}}]{Majewski1993}
{Majewski}, S.~R. 1993, \bibinfo{title}{{Galactic structure surveys and the evolution of the Milky Way.},} \araa, 31, 575, \dodoi{10.1146/annurev.aa.31.090193.003043}

\bibitem[{S.~R. {Majewski} {et~al.}(2017){Majewski}, {Schiavon}, {Frinchaboy}, {Allende Prieto}, {Barkhouser}, {Bizyaev}, {Blank}, {Brunner}, {Burton}, {Carrera}, {Chojnowski}, {Cunha}, {Epstein}, {Fitzgerald}, {Garc{\'\i}a P{\'e}rez}, {Hearty}, {Henderson}, {Holtzman}, {Johnson}, {Lam}, {Lawler}, {Maseman}, {M{\'e}sz{\'a}ros}, {Nelson}, {Nguyen}, {Nidever}, {Pinsonneault}, {Shetrone}, {Smee}, {Smith}, {Stolberg}, {Skrutskie}, {Walker}, {Wilson}, {Zasowski}, {Anders}, {Basu}, {Beland}, {Blanton}, {Bovy}, {Brownstein}, {Carlberg}, {Chaplin}, {Chiappini}, {Eisenstein}, {Elsworth}, {Feuillet}, {Fleming}, {Galbraith-Frew}, {Garc{\'\i}a}, {Garc{\'\i}a-Hern{\'a}ndez}, {Gillespie}, {Girardi}, {Gunn}, {Hasselquist}, {Hayden}, {Hekker}, {Ivans}, {Kinemuchi}, {Klaene}, {Mahadevan}, {Mathur}, {Mosser}, {Muna}, {Munn}, {Nichol}, {O'Connell}, {Parejko}, {Robin}, {Rocha-Pinto}, {Schultheis}, {Serenelli}, {Shane}, {Silva Aguirre}, {Sobeck}, {Thompson}, {Troup}, {Weinberg}, \& {Zamora}}]{Majewski2017}
{Majewski}, S.~R., {Schiavon}, R.~P., {Frinchaboy}, P.~M., {et~al.} 2017, \bibinfo{title}{{The Apache Point Observatory Galactic Evolution Experiment (APOGEE)},} \aj, 154, 94, \dodoi{10.3847/1538-3881/aa784d}

\bibitem[{M. {Mar{\i}{\c{s}}mak} {et~al.}(2024){Mar{\i}{\c{s}}mak}, {{\c{S}}ahin}, {G{\"u}ney}, {Plevne}, \& {Bilir}}]{Marismak2024}
{Mar{\i}{\c{s}}mak}, M., {{\c{S}}ahin}, T., {G{\"u}ney}, F., {Plevne}, O., \& {Bilir}, S. 2024, \bibinfo{title}{{Spectroscopic and dynamic orbital analyses of metal-poor and high proper-motion stars: I. HD 8724 and HD 195633},} Astronomische Nachrichten, 345, e20240047, \dodoi{10.1002/asna.20240047}

\bibitem[{V.~A. {Marsakov} {et~al.}(2019){Marsakov}, {Koval'}, \& {Gozha}}]{Marsakov2019}
{Marsakov}, V.~A., {Koval'}, V.~V., \& {Gozha}, M.~L. 2019, \bibinfo{title}{{The Chemical Composition of Globular Clusters of Different Nature in Our Galaxy},} Astronomy Reports, 63, 274, \dodoi{10.1134/S1063772919040048}

\bibitem[{N.~F. {Martin} {et~al.}(2024){Martin}, {Starkenburg}, {Yuan}, {Fouesneau}, {Ardern-Arentsen}, {De Angeli}, {Gran}, {Montelius}, {Rusterucci}, {Andrae}, {Bellazzini}, {Montegriffo}, {Esselink}, {Zhang}, {Venn}, {Viswanathan}, {Aguado}, {Battaglia}, {Bayer}, {Bonifacio}, {Caffau}, {C{\^o}t{\'e}}, {Carlberg}, {Fabbro}, {Fern{\'a}ndez-Alvar}, {Gonz{\'a}lez Hern{\'a}ndez}, {Gonz{\'a}lez Rivera de La Vernhe}, {Hill}, {Ibata}, {Jablonka}, {Kordopatis}, {Lardo}, {McConnachie}, {Navarrete}, {Navarro}, {Recio-Blanco}, {S{\'a}nchez-Janssen}, {Sestito}, {Thomas}, {Vitali}, \& {Youakim}}]{Martin2024}
{Martin}, N.~F., {Starkenburg}, E., {Yuan}, Z., {et~al.} 2024, \bibinfo{title}{{The Pristine survey: XXIII. Data Release 1 and an all-sky metallicity catalogue based on Gaia DR3 BP/RP spectro-photometry},} \aap, 692, A115, \dodoi{10.1051/0004-6361/202347633}

\bibitem[{F. {Matteucci} \& L. {Greggio}(1986){Matteucci} \& {Greggio}}]{Matteucci1986}
{Matteucci}, F., \& {Greggio}, L. 1986, \bibinfo{title}{{Relative roles of type I and II supernovae in the chemical enrichment of the interstellar gas},} \aap, 154, 279

\bibitem[{A. {McWilliam}(2016){McWilliam}}]{McWilliam2016}
{McWilliam}, A. 2016, \bibinfo{title}{{The Chemical Composition of the Galactic Bulge and Implications for its Evolution},} \pasa, 33, e040, \dodoi{10.1017/pasa.2016.32}

\bibitem[{D. {Mihalas} \& J. {Binney}(1981){Mihalas} \& {Binney}}]{Mihalas1981}
{Mihalas}, D., \& {Binney}, J. 1981, {Galactic astronomy. Structure and kinematics} (W.H.Freeman \& Co Ltd)

\bibitem[{I. {Minchev} \& B. {Famaey}(2010){Minchev} \& {Famaey}}]{Minchev2010}
{Minchev}, I., \& {Famaey}, B. 2010, \bibinfo{title}{{A New Mechanism for Radial Migration in Galactic Disks: Spiral-Bar Resonance Overlap},} \apj, 722, 112, \dodoi{10.1088/0004-637X/722/1/112}

\bibitem[{S. {Mittal} \& I.~U. {Roederer}(2025){Mittal} \& {Roederer}}]{Roederer2025}
{Mittal}, S., \& {Roederer}, I.~U. 2025, \bibinfo{title}{{New Stellar Parameters, Metallicities, and Elemental Abundance Ratios for 311 Metal-poor Stars},} \aj, 169, 172, \dodoi{10.3847/1538-3881/adadf0}

\bibitem[{M. Miyamoto \& R. Nagai(1975)Miyamoto \& Nagai}]{Miyamoto1975}
Miyamoto, M., \& Nagai, R. 1975, \bibinfo{title}{Three-dimensional models for the distribution of mass in galaxies,} Publications of the Astronomical Society of Japan, 27, 533

\bibitem[{G. {Monari} {et~al.}(2016){Monari}, {Famaey}, \& {Siebert}}]{Monari2016}
{Monari}, G., {Famaey}, B., \& {Siebert}, A. 2016, \bibinfo{title}{{Modelling the Galactic disc: perturbed distribution functions in the presence of spiral arms},} \mnras, 457, 2569, \dodoi{10.1093/mnras/stw171}

\bibitem[{C. {Montecinos} {et~al.}(2021){Montecinos}, {Villanova}, {Mu{\~n}oz}, \& {Cort{\'e}s}}]{Montecinos2021}
{Montecinos}, C., {Villanova}, S., {Mu{\~n}oz}, C., \& {Cort{\'e}s}, C.~C. 2021, \bibinfo{title}{{Chemical analysis of the bulge globular cluster NGC 6553},} \mnras, 503, 4336, \dodoi{10.1093/mnras/stab712}

\bibitem[{T.~D. {Morton}(2015){Morton}}]{Morton2015}
{Morton}, T.~D. 2015, \bibinfo{title}{{isochrones: Stellar model grid package},}, Astrophysics Source Code Library, record ascl:1503.010

\bibitem[{J.~F. {Navarro} {et~al.}(1996){Navarro}, {Frenk}, \& {White}}]{Navarro1996}
{Navarro}, J.~F., {Frenk}, C.~S., \& {White}, S. D.~M. 1996, \bibinfo{title}{{The Structure of Cold Dark Matter Halos},} \apj, 462, 563, \dodoi{10.1086/177173}

\bibitem[{L. {Necib} \& T. {Lin}(2022){Necib} \& {Lin}}]{Necib2022}
{Necib}, L., \& {Lin}, T. 2022, \bibinfo{title}{{Substructure at High Speed. II. The Local Escape Velocity and Milky Way Mass with Gaia eDR3},} \apj, 926, 189, \dodoi{10.3847/1538-4357/ac4244}

\bibitem[{N. {Nieuwmunster} {et~al.}(2023){Nieuwmunster}, {Nandakumar}, {Spitoni}, {Ryde}, {Schultheis}, {Rich}, {Barklem}, {Agertz}, {Renaud}, \& {Matteucci}}]{Nieuwmunster2023}
{Nieuwmunster}, N., {Nandakumar}, G., {Spitoni}, E., {et~al.} 2023, \bibinfo{title}{{Detailed {\ensuremath{\alpha}} abundance trends in the inner Galactic bulge},} \aap, 671, A94, \dodoi{10.1051/0004-6361/202245374}

\bibitem[{P.~E. {Nissen}(2004){Nissen}}]{Nissen2004}
{Nissen}, P.~E. 2004, in Origin and Evolution of the Elements, ed. A.~{McWilliam} \& M.~{Rauch}, 154, \dodoi{10.48550/arXiv.astro-ph/0310326}

\bibitem[{P.~E. {Nissen} {et~al.}(2024){Nissen}, {Amarsi}, {Sk{\'u}lad{\'o}ttir}, \& {Schuster}}]{Nissen2024}
{Nissen}, P.~E., {Amarsi}, A.~M., {Sk{\'u}lad{\'o}ttir}, {\'A}., \& {Schuster}, W.~J. 2024, \bibinfo{title}{{Abundances of iron-peak elements in accreted and in situ born Galactic halo stars},} \aap, 682, A116, \dodoi{10.1051/0004-6361/202348392}

\bibitem[{J.~E. {Norris} {et~al.}(2013){Norris}, {Bessell}, {Yong}, {Christlieb}, {Barklem}, {Asplund}, {Murphy}, {Beers}, {Frebel}, \& {Ryan}}]{Norris2013}
{Norris}, J.~E., {Bessell}, M.~S., {Yong}, D., {et~al.} 2013, \bibinfo{title}{{The Most Metal-poor Stars. I. Discovery, Data, and Atmospheric Parameters},} \apj, 762, 25, \dodoi{10.1088/0004-637X/762/1/25}

\bibitem[{{\"O}. {{\"O}nal Ta{\c{s}}} {et~al.}(2018){{\"O}nal Ta{\c{s}}}, {Bilir}, \& {Plevne}}]{Onal-tas2018}
{{\"O}nal Ta{\c{s}}}, {\"O}., {Bilir}, S., \& {Plevne}, O. 2018, \bibinfo{title}{{Local stellar kinematics from RAVE data{\textemdash}VIII. Effects of the Galactic disc perturbations on stellar orbits of red clump stars},} \apss, 363, 35, \dodoi{10.1007/s10509-018-3248-7}

\bibitem[{C.~B. Pereira {et~al.}(2017)Pereira, Smith, Drake, Roig, Hasselquist, Cunha, \& Jilinski}]{Pereira2019}
Pereira, C.~B., Smith, V.~V., Drake, N.~A., {et~al.} 2017, \bibinfo{title}{Chemical abundances and kinematics of TYC 5619-109-1,} Monthly Notices of the Royal Astronomical Society, 469, 774, \dodoi{10.1093/mnras/stx786}

\bibitem[{P. {Petit} {et~al.}(2014){Petit}, {Louge}, {Th{\'e}ado}, {Paletou}, {Manset}, {Morin}, {Marsden}, \& {Jeffers}}]{Petit2014}
{Petit}, P., {Louge}, T., {Th{\'e}ado}, S., {et~al.} 2014, \bibinfo{title}{{PolarBase: A Database of High-Resolution Spectropolarimetric Stellar Observations},} \pasp, 126, 469, \dodoi{10.1086/676976}

\bibitem[{O. {Plevne} {et~al.}(2015){Plevne}, {AK}, {Karaali}, {Bilir}, {Ak}, \& {Bostanci}}]{Plevne2015}
{Plevne}, O., {AK}, T., {Karaali}, S., {et~al.} 2015, \bibinfo{title}{{Local Stellar Kinematics from RAVE Data - VI. Metallicity Gradients Based on the F-G Main-Sequence Stars},} \pasa, 32, e043, \dodoi{10.1017/pasa.2015.44}

\bibitem[{A. {Recio-Blanco} {et~al.}(2016){Recio-Blanco}, {de Laverny}, {Allende Prieto}, {Fustes}, {Manteiga}, {Arcay}, {Bijaoui}, {Dafonte}, {Ordenovic}, \& {Ordo{\~n}ez Blanco}}]{Recio-Blanco2016}
{Recio-Blanco}, A., {de Laverny}, P., {Allende Prieto}, C., {et~al.} 2016, \bibinfo{title}{{Stellar parametrization from Gaia RVS spectra},} \aap, 585, A93, \dodoi{10.1051/0004-6361/201425030}

\bibitem[{B.~E. {Reddy} {et~al.}(2003){Reddy}, {Tomkin}, {Lambert}, \& {Allende Prieto}}]{Reddy2003}
{Reddy}, B.~E., {Tomkin}, J., {Lambert}, D.~L., \& {Allende Prieto}, C. 2003, \bibinfo{title}{{The chemical compositions of Galactic disc F and G dwarfs},} \mnras, 340, 304, \dodoi{10.1046/j.1365-8711.2003.06305.x}

\bibitem[{G.~R. {Ricker} {et~al.}(2014){Ricker}, {Winn}, {Vanderspek}, {Latham}, {Bakos}, {Bean}, {Berta-Thompson}, {Brown}, {Buchhave}, {Butler}, {Butler}, {Chaplin}, {Charbonneau}, {Christensen-Dalsgaard}, {Clampin}, {Deming}, {Doty}, {De Lee}, {Dressing}, {Dunham}, {Endl}, {Fressin}, {Ge}, {Henning}, {Holman}, {Howard}, {Ida}, {Jenkins}, {Jernigan}, {Johnson}, {Kaltenegger}, {Kawai}, {Kjeldsen}, {Laughlin}, {Levine}, {Lin}, {Lissauer}, {MacQueen}, {Marcy}, {McCullough}, {Morton}, {Narita}, {Paegert}, {Palle}, {Pepe}, {Pepper}, {Quirrenbach}, {Rinehart}, {Sasselov}, {Sato}, {Seager}, {Sozzetti}, {Stassun}, {Sullivan}, {Szentgyorgyi}, {Torres}, {Udry}, \& {Villasenor}}]{Ricker2014}
{Ricker}, G.~R., {Winn}, J.~N., {Vanderspek}, R., {et~al.} 2014, in Society of Photo-Optical Instrumentation Engineers (SPIE) Conference Series, Vol. 9143, Space Telescopes and Instrumentation 2014: Optical, Infrared, and Millimeter Wave, ed. J.~{Oschmann}, Jacobus~M., M.~{Clampin}, G.~G. {Fazio}, \& H.~A. {MacEwen}, 914320, \dodoi{10.1117/12.2063489}

\bibitem[{I.~U. {Roederer} {et~al.}(2014){Roederer}, {Preston}, {Thompson}, {Shectman}, {Sneden}, {Burley}, \& {Kelson}}]{Roederer2014}
{Roederer}, I.~U., {Preston}, G.~W., {Thompson}, I.~B., {et~al.} 2014, \bibinfo{title}{{A Search for Stars of Very Low Metal Abundance. VI. Detailed Abundances of 313 Metal-poor Stars},} \aj, 147, 136, \dodoi{10.1088/0004-6256/147/6/136}

\bibitem[{R. {Ro{\v{s}}kar} {et~al.}(2008){Ro{\v{s}}kar}, {Debattista}, {Quinn}, {Stinson}, \& {Wadsley}}]{Roskar2008}
{Ro{\v{s}}kar}, R., {Debattista}, V.~P., {Quinn}, T.~R., {Stinson}, G.~S., \& {Wadsley}, J. 2008, \bibinfo{title}{{Riding the Spiral Waves: Implications of Stellar Migration for the Properties of Galactic Disks},} \apjl, 684, L79, \dodoi{10.1086/592231}

\bibitem[{T. {\c{S}ahin} {et~al.}(2023){\c{S}ahin}, {Marismak}, {Cinar}, \& {Bilir}}]{Sahin2023a}
{\c{S}ahin}, T., {Marismak}, M., {Cinar}, N., \& {Bilir}, S. 2023, \bibinfo{title}{{An Updated Line List for Spectroscopic Investigation of G Stars- I: Redetermination of the Abundances in the Solar Photosphere},} Physics and Astronomy Reports, 1, 54, \dodoi{10.26650/PAR.2023.00007}

\bibitem[{S. {Salvadori} {et~al.}(2019){Salvadori}, {Bonifacio}, {Caffau}, {Korotin}, {Andreevsky}, {Spite}, \& {Sk{\'u}lad{\'o}ttir}}]{Salvadori2019}
{Salvadori}, S., {Bonifacio}, P., {Caffau}, E., {et~al.} 2019, \bibinfo{title}{{Probing the existence of very massive first stars},} \mnras, 487, 4261, \dodoi{10.1093/mnras/stz1464}

\bibitem[{S. {Salvadori} \& A. {Ferrara}(2009){Salvadori} \& {Ferrara}}]{Salvadori2009}
{Salvadori}, S., \& {Ferrara}, A. 2009, \bibinfo{title}{{Ultra faint dwarfs: probing early cosmic star formation},} \mnras, 395, L6, \dodoi{10.1111/j.1745-3933.2009.00627.x}

\bibitem[{P. {Santos-Peral} {et~al.}(2023){Santos-Peral}, {S{\'a}nchez-Bl{\'a}zquez}, {Vazdekis}, \& {Palicio}}]{Santos-Peral2023}
{Santos-Peral}, P., {S{\'a}nchez-Bl{\'a}zquez}, P., {Vazdekis}, A., \& {Palicio}, P.~A. 2023, \bibinfo{title}{{Chemical characterisation of the X-shooter Spectral Library (XSL): [Mg/Fe] and [Ca/Fe] abundances},} \aap, 672, A166, \dodoi{10.1051/0004-6361/202245606}

\bibitem[{M.~F. {Skrutskie} {et~al.}(2006){Skrutskie}, {Cutri}, {Stiening}, {Weinberg}, {Schneider}, {Carpenter}, {Beichman}, {Capps}, {Chester}, {Elias}, {Huchra}, {Liebert}, {Lonsdale}, {Monet}, {Price}, {Seitzer}, {Jarrett}, {Kirkpatrick}, {Gizis}, {Howard}, {Evans}, {Fowler}, {Fullmer}, {Hurt}, {Light}, {Kopan}, {Marsh}, {McCallon}, {Tam}, {Van Dyk}, \& {Wheelock}}]{Skrutskie2006}
{Skrutskie}, M.~F., {Cutri}, R.~M., {Stiening}, R., {et~al.} 2006, \bibinfo{title}{{The Two Micron All Sky Survey (2MASS)},} \aj, 131, 1163, \dodoi{10.1086/498708}

\bibitem[{C. {Sneden}(1974){Sneden}}]{Sneden1974}
{Sneden}, C. 1974, \bibinfo{title}{{Carbon and nitrogen abundances in metal-poor stars.},} \apj, 189, 493, \dodoi{10.1086/152828}

\bibitem[{C. {Soubiran} {et~al.}(2003){Soubiran}, {Bienaym{\'e}}, \& {Siebert}}]{Soubiran2003}
{Soubiran}, C., {Bienaym{\'e}}, O., \& {Siebert}, A. 2003, \bibinfo{title}{{Vertical distribution of Galactic disk stars. I. Kinematics and metallicity},} \aap, 398, 141, \dodoi{10.1051/0004-6361:20021615}

\bibitem[{C. {Soubiran} {et~al.}(2016){Soubiran}, {Le Campion}, {Brouillet}, \& {Chemin}}]{Soubiran2016}
{Soubiran}, C., {Le Campion}, J.-F., {Brouillet}, N., \& {Chemin}, L. 2016, \bibinfo{title}{{The PASTEL catalogue: 2016 version},} \aap, 591, A118, \dodoi{10.1051/0004-6361/201628497}

\bibitem[{C. {Soubiran} {et~al.}(2024){Soubiran}, {Creevey}, {Lagarde}, {Brouillet}, {Jofr{\'e}}, {Casamiquela}, {Heiter}, {Aguilera-G{\'o}mez}, {Vitali}, {Worley}, \& {de Brito Silva}}]{Soubiran2024}
{Soubiran}, C., {Creevey}, O.~L., {Lagarde}, N., {et~al.} 2024, \bibinfo{title}{{Gaia FGK benchmark stars: Fundamental T$_{eff}$ and log g of the third version},} \aap, 682, A145, \dodoi{10.1051/0004-6361/202347136}

\bibitem[{S. {Ta{\c{s}}demir} \& D.~C. {{\c{C}}{\i}nar}(2025){Ta{\c{s}}demir} \& {{\c{C}}{\i}nar}}]{Tasdemir2025}
{Ta{\c{s}}demir}, S., \& {{\c{C}}{\i}nar}, D.~C. 2025, \bibinfo{title}{{A Detailed Analysis of Close Binary OCs},} arXiv e-prints, arXiv:2501.17235, \dodoi{10.48550/arXiv.2501.17235}

\bibitem[{S. {Ta{\c{s}}demir} \& T. {Yontan}(2023){Ta{\c{s}}demir} \& {Yontan}}]{Tasdemir2023}
{Ta{\c{s}}demir}, S., \& {Yontan}, T. 2023, \bibinfo{title}{{Analysis of the Young Open Cluster Trumpler 2 Using Gaia DR3 Data},} Physics and Astronomy Reports, 1, 1, \dodoi{10.26650/PAR.2023.00001}

\bibitem[{G. {Tautvai{\v{s}}ien{\.{e}}} {et~al.}(2021){Tautvai{\v{s}}ien{\.{e}}}, {Viscasillas V{\'a}zquez}, {Mikolaitis}, {Stonkut{\.{e}}}, {Minkevi{\v{c}}i{\={u}}t{\.{e}}}, {Drazdauskas}, \& {Bagdonas}}]{Tautvaivsien2021}
{Tautvai{\v{s}}ien{\.{e}}}, G., {Viscasillas V{\'a}zquez}, C., {Mikolaitis}, {\v{S}}., {et~al.} 2021, \bibinfo{title}{{Abundances of neutron-capture elements in thin- and thick-disc stars in the solar neighbourhood},} \aap, 649, A126, \dodoi{10.1051/0004-6361/202039979}

\bibitem[{J. {Tumlinson}(2010){Tumlinson}}]{Tumlinson2010}
{Tumlinson}, J. 2010, \bibinfo{title}{{Chemical Evolution in Hierarchical Models of Cosmic Structure. II. The Formation of the Milky Way Stellar Halo and the Distribution of the Oldest Stars},} \apj, 708, 1398, \dodoi{10.1088/0004-637X/708/2/1398}

\bibitem[{S. {Tun{\c{c}}el G{\"u}{\c{c}}tekin} {et~al.}(2019){Tun{\c{c}}el G{\"u}{\c{c}}tekin}, {Bilir}, {Karaali}, {Plevne}, \& {Ak}}]{Guctekin2019}
{Tun{\c{c}}el G{\"u}{\c{c}}tekin}, S., {Bilir}, S., {Karaali}, S., {Plevne}, O., \& {Ak}, S. 2019, \bibinfo{title}{{Vertical and radial metallicity gradients in high latitude galactic fields with SDSS},} Advances in Space Research, 63, 1360, \dodoi{10.1016/j.asr.2018.10.041}

\bibitem[{C. {Usher} {et~al.}(2019){Usher}, {Brodie}, {Forbes}, {Romanowsky}, {Strader}, {Pfeffer}, \& {Bastian}}]{Usher2019}
{Usher}, C., {Brodie}, J.~P., {Forbes}, D.~A., {et~al.} 2019, \bibinfo{title}{{The SLUGGS survey: measuring globular cluster ages using both photometry and spectroscopy},} \mnras, 490, 491, \dodoi{10.1093/mnras/stz2596}

\bibitem[{D. {Valcin} {et~al.}(2020){Valcin}, {Bernal}, {Jimenez}, {Verde}, \& {Wandelt}}]{Valcin2020}
{Valcin}, D., {Bernal}, J.~L., {Jimenez}, R., {Verde}, L., \& {Wandelt}, B.~D. 2020, \bibinfo{title}{{Inferring the age of the universe with globular clusters},} \jcap, 2020, 002, \dodoi{10.1088/1475-7516/2020/12/002}

\bibitem[{E. {Vasiliev} \& H. {Baumgardt}(2021){Vasiliev} \& {Baumgardt}}]{Vasiliev2021}
{Vasiliev}, E., \& {Baumgardt}, H. 2021, \bibinfo{title}{{Gaia EDR3 view on galactic globular clusters},} \mnras, 505, 5978, \dodoi{10.1093/mnras/stab1475}

\bibitem[{K.~A. {Venn} {et~al.}(2020){Venn}, {Kielty}, {Sestito}, {Starkenburg}, {Martin}, {Aguado}, {Arentsen}, {Bonifacio}, {Caffau}, {Hill}, {Jablonka}, {Lardo}, {Mashonkina}, {Navarro}, {Sneden}, {Thomas}, {Youakim}, {Gonz{\'a}lez-Hern{\'a}ndez}, {S{\'a}nchez Janssen}, {Carlberg}, \& {Malhan}}]{Venn2020}
{Venn}, K.~A., {Kielty}, C.~L., {Sestito}, F., {et~al.} 2020, \bibinfo{title}{{The Pristine survey - IX. CFHT ESPaDOnS spectroscopic analysis of 115 bright metal-poor candidate stars},} \mnras, 492, 3241, \dodoi{10.1093/mnras/stz3546}

\bibitem[{J.~I. {Vines} \& J.~S. {Jenkins}(2022){Vines} \& {Jenkins}}]{Vines2022}
{Vines}, J.~I., \& {Jenkins}, J.~S. 2022, \bibinfo{title}{{ARIADNE: measuring accurate and precise stellar parameters through SED fitting},} \mnras, 513, 2719, \dodoi{10.1093/mnras/stac956}

\bibitem[{P. Virtanen {et~al.}(2020)Virtanen, Gommers, Oliphant, Haberland, Reddy, Cournapeau, Burovski, Peterson, Weckesser, Bright, {van der Walt}, Brett, Wilson, Millman, Mayorov, Nelson, Jones, Kern, Larson, Carey, Polat, Feng, Moore, {VanderPlas}, Laxalde, Perktold, Cimrman, Henriksen, Quintero, Harris, Archibald, Ribeiro, Pedregosa, {van Mulbregt}, \& {SciPy 1.0 Contributors}}]{Scipy}
Virtanen, P., Gommers, R., Oliphant, T.~E., {et~al.} 2020, \bibinfo{title}{{{SciPy} 1.0: Fundamental Algorithms for Scientific Computing in Python},} Nature Methods, 17, 261, \dodoi{10.1038/s41592-019-0686-2}

\bibitem[{A. {Viswanathan} {et~al.}(2025){Viswanathan}, {Yuan}, {Ardern-Arentsen}, {Starkenburg}, {Martin}, {Youakim}, {Ibata}, {Sestito}, {Matsuno}, {Allende Prieto}, {Barwell}, {Bayer}, {Doliva-Dolinsky}, {Fern{\'a}ndez-Alvar}, {Anta}, {Jhass}, {Longeard}, {Arroyo-Polonio}, {Massana}, {Montelius}, {Rusterucci}, {Santos-Torres}, {Thomas}, {Vitali}, {Wu}, {Yarker}, {Ye}, {Aguado}, {Gran}, \& {Navarro}}]{Viswanathan2025}
{Viswanathan}, A., {Yuan}, Z., {Ardern-Arentsen}, A., {et~al.} 2025, \bibinfo{title}{{The Pristine survey: XXV. The very metal-poor Galaxy: Chemodynamics through the follow-up of the Pristine-Gaia synthetic catalogue},} \aap, 695, A112, \dodoi{10.1051/0004-6361/202450819}

\bibitem[{N.~C. {Weatherford} {et~al.}(2023){Weatherford}, {K{\i}ro{\u{g}}lu}, {Fragione}, {Chatterjee}, {Kremer}, \& {Rasio}}]{Weatherford2023}
{Weatherford}, N.~C., {K{\i}ro{\u{g}}lu}, F., {Fragione}, G., {et~al.} 2023, \bibinfo{title}{{Stellar Escape from Globular Clusters. I. Escape Mechanisms and Properties at Ejection},} \apj, 946, 104, \dodoi{10.3847/1538-4357/acbcc1}

\bibitem[{N.~C. {Weatherford} {et~al.}(2024){Weatherford}, {Rasio}, {Chatterjee}, {Fragione}, {K{\i}ro{\u{g}}lu}, \& {Kremer}}]{Weatherford2024}
{Weatherford}, N.~C., {Rasio}, F.~A., {Chatterjee}, S., {et~al.} 2024, \bibinfo{title}{{Stellar Escape from Globular Clusters. II. Clusters May Eat Their Own Tails},} \apj, 967, 42, \dodoi{10.3847/1538-4357/ad39df}

\bibitem[{M. {Wenger} {et~al.}(2000){Wenger}, {Ochsenbein}, {Egret}, {Dubois}, {Bonnarel}, {Borde}, {Genova}, {Jasniewicz}, {Lalo{\"e}}, {Lesteven}, \& {Monier}}]{Wenger2000}
{Wenger}, M., {Ochsenbein}, F., {Egret}, D., {et~al.} 2000, \bibinfo{title}{{The SIMBAD astronomical database. The CDS reference database for astronomical objects},} \aaps, 143, 9, \dodoi{10.1051/aas:2000332}

\bibitem[{E.~L. {Wright} {et~al.}(2010){Wright}, {Eisenhardt}, {Mainzer}, {Ressler}, {Cutri}, {Jarrett}, {Kirkpatrick}, {Padgett}, {McMillan}, {Skrutskie}, {Stanford}, {Cohen}, {Walker}, {Mather}, {Leisawitz}, {Gautier}, {McLean}, {Benford}, {Lonsdale}, {Blain}, {Mendez}, {Irace}, {Duval}, {Liu}, {Royer}, {Heinrichsen}, {Howard}, {Shannon}, {Kendall}, {Walsh}, {Larsen}, {Cardon}, {Schick}, {Schwalm}, {Abid}, {Fabinsky}, {Naes}, \& {Tsai}}]{Wright2010}
{Wright}, E.~L., {Eisenhardt}, P. R.~M., {Mainzer}, A.~K., {et~al.} 2010, \bibinfo{title}{{The Wide-field Infrared Survey Explorer (WISE): Mission Description and Initial On-orbit Performance},} \aj, 140, 1868, \dodoi{10.1088/0004-6256/140/6/1868}

\bibitem[{D. {Yong} {et~al.}(2013){Yong}, {Norris}, {Bessell}, {Christlieb}, {Asplund}, {Beers}, {Barklem}, {Frebel}, \& {Ryan}}]{Yong2013}
{Yong}, D., {Norris}, J.~E., {Bessell}, M.~S., {et~al.} 2013, \bibinfo{title}{{The Most Metal-poor Stars. II. Chemical Abundances of 190 Metal-poor Stars Including 10 New Stars with [Fe/H] <= -3.5},} \apj, 762, 26, \dodoi{10.1088/0004-637X/762/1/26}

\bibitem[{T. {Yontan}(2023){Yontan}}]{Yontan2023a}
{Yontan}, T. 2023, \bibinfo{title}{{An Investigation of Open Clusters Berkeley 68 and Stock 20 Using CCD UBV and Gaia DR3 Data},} \aj, 165, 79, \dodoi{10.3847/1538-3881/aca6f0}

\bibitem[{T. {Yontan} \& R. {Canbay}(2023){Yontan} \& {Canbay}}]{Yontan2023b}
{Yontan}, T., \& {Canbay}, R. 2023, \bibinfo{title}{{Comprehensive Analysis of the Open Cluster Collinder 74},} Physics and Astronomy Reports, 1, 65, \dodoi{10.26650/PAR.2023.00008}

\bibitem[{T. {Yontan} {et~al.}(2022){Yontan}, {{\c{C}}akmak}, {Bilir}, {Banks}, {Ra{\'u}l}, {Canbay}, {Ko{\c{c}}}, {Ta{\c{s}}demir}, {Er{\c{c}}ay}, {Tan{\i}k Ozt{\"u}rk}, \& {Dursun}}]{Yontan_2022}
{Yontan}, T., {{\c{C}}akmak}, T., {Bilir}, S., {et~al.} 2022, \bibinfo{title}{{A Study of the NGC 1193 and NGC 1798 Open Clusters Using CCD UBV Photometric and Gaia EDR3 Data},} The Revista Mexicana de Astronomía y Astrofísica, 58, 333, \dodoi{10.22201/ia.01851101p.2022.58.02.14}

\bibitem[{G. {Yucel} {et~al.}(2024){Yucel}, {Canbay}, \& {Bakis}}]{Yucel2024}
{Yucel}, G., {Canbay}, R., \& {Bakis}, V. 2024, \bibinfo{title}{{The Fundamental Parameters and Evolutionary Status of V454 Aurigae},} Physics and Astronomy Reports, 2, 18, \dodoi{10.26650/PAR.2024.00003}

\bibitem[{Y. {Zhang} {et~al.}(2012){Zhang}, {Han}, {Liu}, {Zhang}, \& {Kang}}]{Zhang2012}
{Zhang}, Y., {Han}, Z., {Liu}, J., {Zhang}, F., \& {Kang}, X. 2012, \bibinfo{title}{{Testing three derivative methods of stellar population synthesis models},} \mnras, 421, 1678, \dodoi{10.1111/j.1365-2966.2012.20430.x}

\end{thebibliography}
